\documentclass[a4paper,11pt]{article}
\usepackage{jcappub} 

\usepackage{setspace}
\usepackage[utf8x]{inputenc}

\DeclareUnicodeCharacter{2212}{-}

\usepackage{changepage} 
\usepackage{adjustbox}

\usepackage{amsmath,latexsym,amssymb,amsfonts}
\usepackage{float} \usepackage{graphicx}
\usepackage{epstopdf}
\usepackage{graphics}
\usepackage{caption}
\usepackage{subfigure}
\usepackage{bm}
\usepackage{color}

\usepackage{hyperref}
\hypersetup{colorlinks=true,linkcolor=blue,filecolor=magneta,urlcolor=blue}

\usepackage[dvips]{psfrag}
\usepackage{ulem}
\usepackage{epsfig}
\usepackage{cases}
\usepackage{tensor}
\usepackage{multirow,array}
\usepackage{booktabs} 

\usepackage[dvipsnames]{xcolor}

\newcommand{\ine}{\text{inert}}
\newcommand{\cro}{\text{cr}}

\newcommand{\m}{\text{m}}
\newcommand{\TA}{\text{A}}
\newcommand{\TB}{\text{B}}

\newcommand{\tc}{\tilde{c}}
\newcommand{\ttc}{\tilde{\tilde{c}}}
\newcommand{\tG}{\tilde{G}}

\newcommand{\ta}{\tilde{a}}

\newcommand{\tRR}{\tilde{R}}
\newcommand{\tSS}{\tilde{S}}
\newcommand{\tM}{\text{M}}
\newcommand{\TM}{\text{M}}
\newcommand{\TL}{\text{L}}
\newcommand{\TN}{\text{N}}
\newcommand{\SD}{\text{SD}}
\newcommand{\GR}{\text{GR}}
\newcommand{\EM}{\text{EM}}

\newcommand{\rs}{\text{rs}}
\newcommand{\cm}{\text{cm}}
\newcommand{\Tp}{\text{p}}
\newcommand{\Teq}{\text{eq}}
\newcommand{\tgam}{\tilde{\gamma}}

\newcommand{\tbeta}{\tilde{\beta}}

\newcommand{\tkapp}{\tilde{\kappa}}

\newcommand{\Tn}{\text{n}}
\newcommand{\Tm}{\text{m}}
\newcommand{\Tr}{\text{r}}
\newcommand{\Tre}{\text{re}}
\newcommand{\Tc}{\text{c}}
\newcommand{\Tb}{\text{b}}
\newcommand{\Td}{\text{d}}
\newcommand{\Te}{\text{e}}
\newcommand{\Tw}{\text{w}}
\newcommand{\TW}{\text{W}}
\newcommand{\TT}{\text{T}}
\newcommand{\TtH}{\text{H}}
\newcommand{\TD}{\text{D}}
\newcommand{\Tde}{\text{de}}
\newcommand{\bbc}{\bar{c}}
\newcommand{\tih}{\tilde{h}}
\newcommand{\Tdrag}{\text{drag}}
\newcommand{\tnu}{\tilde{\nu}}
\newcommand{\tmu}{\tilde{\mu}}

\newcommand{\tepsilon}{\tilde{\epsilon}}

\newcommand{\Ch}{\text{Ch}}

\newcommand{\TMpc}{\text{Mpc}}
\newcommand{\Temis}{\text{emis}}
\newcommand{\Tchirp}{\text{chirp}}
\newcommand{\Tgw}{\text{gw}}
\newcommand{\Ts}{\text{s}}
\newcommand{\Tret}{\text{ret}}
\newcommand{\Teff}{\text{eff}}
\newcommand{\Tobs}{\text{obs}}

\newcommand{\Tls}{\text{ls}}
\newcommand{\Tso}{\text{so}}
\newcommand{\YPBBN}{Y_{\text{P}}^{(\text{BBN})}}
\newcommand{\thbar}{\tilde{\hbar}}
\newcommand{\hTT}{\tilde{h}} 

\title{The minimally extended Varying Speed of Light (meVSL)}

\author[a]{Seokcheon Lee}
\affiliation[a]{Department of Physics, Institute of Basic Science, Sungkyunkwan University, Suwon 16419, Korea}

\emailAdd{skylee@skku.edu}

\abstract{
Even though there have been various models of the time-varying speed of light (VSL), they remain out of the mainstream because of their possible violation of physics laws built into fundamental physics. For VSL to remain a viable theory, it must at least build upon the successes of special relativity, including Maxwell's equations and thermodynamics. For this purpose, we adopt the assumption that the speed of light, $\tc$, (\textit{i.e.}, $\tc[a]$), varies for the scale factor, $a$. This model naturally arises if there are no additional assumptions about cosmological time dilation in the FLRW metric. The background FLRW universe can be defined by the constant cosmic time hypersurface using physical quantities such as temperature, density, $\tc$, etc. Because they evolve in cosmic time, the homogeneity of the Universe demands that they must be equal at the same cosmic time. The variation of $\tc$ accompanies the joint variations of all related physical constants to satisfy the Lorentz invariance, thermodynamics, Bianchi identity, etc. We dub this VSL model as a ``minimally extended VSL (meVSL)''. We derive cosmological observables of meVSL and obtain the constraints on the variation of $\tc$ by using current cosmological observations. Interestingly, the cosmological redshift $z$ and all geometrical distances except the luminosity distance of the meVSL model are the same as those of general relativity. However, the Hubble parameter of meVSL is rescaled as $H(z) = (1+z)^{-b/4} H^{(\GR)}(z)$, where $H^{(\GR)}(z)$ denotes the Hubble parameter obtained from general relativity. Therefore, it can be used as an alternative solution to alleviate tensions in the measurements of the Hubble parameter. This manuscript discusses the principal impacts of the meVSL model on a range of cosmological observations, encompassing BBN, CMB, SZE, BAO, SNe, GWs, H, SL, and $\Delta \alpha$. The meVSL might provide alternative solutions for various late-time problems of the standard $\Lambda$CDM model. It is the main motivation for proposing the meVSL model.
}

\begin{document}
\maketitle
\flushbottom

\section{Introduction}
\label{sec:intro}

Although special relativity (SR) and general relativity (GR) attributed to Einstein have withstood numerous tests, no one can be certain that they hold everywhere and under all circumstances. SR is an inseparable part of quantum field theory that describes the interactions of elementary particles with an almost incredible degree of accuracy. Tests for SR on a wide range of experiments have not revealed any violations of Lorentz invariance (LI). People believe it to be locally exact. However, the local LI has to be replaced by GR at cosmological scales. Therefore, arguing whether SR is typically held and testable across cosmological distances and time scales is pointless.

SR contains only one parameter, $c$, the speed of light in a vacuum. We demonstrate in Sec.~\ref{sec:SR} that the Minkowski spacetime single postulate, or the universal Lorentz covariance, is sufficient to satisfy the SR \cite{Das93,Schutz97}. Thus, it is possible to make the LI varying speed of light (VSL) model as long as $c$ is locally constant and changes at cosmological scales. To avoid trivial rescaling of units, one must test the simultaneous variation of $c$ and Newton's gravitational constant $G$ because $c$ and $G$ enter as the combination $G/c^4$ in the Einstein action \cite{Barrow:1998eh}.

Sometimes, researchers have suggested different VSLs as potential explanations for inconclusive observational results based on GR. Einstein postulated the very first VSL, stating that $c = \nu \lambda$ with constant frequency $\nu$ causes a decreased speed of light at a shorter wavelength $\lambda$. He assumed that a gravitational field makes the clock run slower by $\nu_1 = \nu_2 (1 + GM/rc^2)$ \cite{Einstein:1911}. Dicke assumed that the wavelength and the frequency vary by defining a refractive index $n \equiv c/c_0 = 1 + 2GM/rc^2$ \cite{Dicke:1957}. He considered a cosmology with an alternative description to the cosmological redshift using a decreasing $c$ in time. The proponents of early VSL cosmology have proposed it to address the horizon problem of the Big Bang model and provide an alternative to cosmic inflation \cite{Barrow:1998eh,Petit:1988,Petit:1988-2,Petit:1989,Midy:1989,Moffat:1992ud,Petit:1995ass,Albrecht:1998ir,Barrow:1998he,Clayton:1998hv,Barrow:1999jq,Barrow:1999st,Clayton:1999zs,Brandenberger:1999bi,Bassett:2000wj,Clayton:2000xt,Gopakumar:2000kp,Jacobson:2000xp,Magueijo:2000zt,Magueijo:2000au,Magueijo:2003gj,Magueijo:2007gf,Petit:2008eb,Roshan:2009yb,Sanejouand:2009,Nassif:2012dr,Moffat:2014poa,Ravanpak:2017kdg,Costa:2017abc,Nassif:2018pdu}. A VSL model that proposed the change of the speed of light only without allowing the variations of other physical constants is called minimal VSL (mVSL). Petit suggested including the joint variations of all associated physical constants if one permits the time variation of c. These variations should be based on the consistency of all physical equations and measurements, ensuring that these constants remain in line with the laws of physics throughout the evolution of the Universe. It may be possible to derive a universal gauge relationship and the time variation of the parameters regarded as constants \cite{Petit:1995ass,Petit:2008eb}
\begin{align}
G &= G_0 a^{-1} \,, \quad m = m_{0} a \,, \quad c = c_0 a^{\frac{1}{2}} \,, \quad h = h_0 a^{\frac{3}{2}} \,, \quad e = e_{0} a^{\frac{1}{2}} \,, \quad \mu = \mu_0 a \label{Petitconst} \,.
\end{align}

Despite the success of standard cosmology based on GR and FLRW metrics, there have been several shortcomings of standard cosmology. Thus, one of the primary motivations for the proposal of VSL models is to look for explanations for some unusual properties of the Universe and to prevail over some of the limitations of standard cosmology \cite{Avelino:2002jn,Pedram:2007mj,Salzano:2016pny,Leszczynska:2018juk}. Also, the VSL theory provides a solution to the cosmological constant problem. Both theoretical and empirical research have explored VSL dynamics \cite{Avelino:1999is,Belinchon:1999kq,Drummond:1999ut,Alexander:1999cb,Avelino:2000ph,Szydlowski:2002kz,Shojaie:2004sq,Shojaie:2004xw,Balcerzak:2013kha,Balcerzak:2014rga,Franzmann:2017nsc,Hanimeli:2019wrt,Skara:2019usd,Bhattacharjee:2020fgl,Gupta:2020anq}.

However, we know that if someone proposes a varying $c$, they need to rewrite the related physics to replace the current system that depends on the assumption of the constant $c$. This is because the LI builds into fundamental physics \cite{Ellis:2003pw,Ellis:2007ah}. Therefore, one cannot simply change one or two arbitrary equations by replacing the constant $c$ with the time-varying $c$ and then disregard the rest of physics. Any viable VSL theory has to provide an integrated viable replacement to the entire set of physical equations and consequent effects (kinematical and dynamical) dependent on $c$. The speed of light in Einstein's relativity is related both to the metric tensor and to Maxwell's equations. The former establishes the geometry of null geodesics and specifies measurements in spacetime, while the latter determines the paths taken by light rays in spacetime. The properties of wavelike solutions of Maxwell's equations are null geodesics and it is determined by the metric tensor. Thus, they are related to each other. Light rays ({\it i.e.}, the paths of photons or other massless particles in spacetime) are solutions to the geodesic equation. It might include the redefinition of distance measurements, the validity of LI, the modification of Maxwell's equations, and consistencies concerning all other physical theories.

For this purpose, one should investigate the observational status of variations of fundamental constants \cite{Uzan:2002vq}. A dimensionless physical constant is a constant that is a pure number having no units attached to it. Thus, its numerical value is independent of the used system of units. Sometimes, one uses the terminology of the fundamental physical constant to refer to universal dimensionless constants like the fine-structure constant, $\alpha$. One might restrict the fundamental physical constants to the dimensionless universal physical ones. Thus, one cannot derive them from any other source \cite{Rich:2013oxa,Duff:2001ba,Duff:2014mva}. However, the universal dimensioned physical constants, such as the speed of light $c$, the gravitational constant $G$, the Planck constant $h$, and the vacuum permittivity $\epsilon_0$, also have been referred to as the fundamental physical constants \cite{Mohr:2015ccw}. One denotes the physical constant as the notion of a physical quantity subject to experimental measurement which is independent of the time or location of the experiment. The constancy of any physical constant is thus verified by the experiment. One cannot derive fundamental physical constants and they have to be measured. The current precision measurements of cosmology might be used to constrain any time variation of fundamental constants.

Dirac made the large numbers hypothesis (LNH) by relating ratios of size scales in the Universe to that of force scales \cite{Dirac37}. One obtains big dimensionless numbers from these ratios. From this hypothesis, he interprets that the apparent similarity of these ratios could imply a cosmology with several unusual features. For example, he proposes that the gravitational constant representing the strength of gravity is inversely proportional to the age of the Universe, $G \propto 1/t$. Also, he suggests that physical constants are not constant but depend on the age of the Universe. A distaste for Einstein's General Relativity (GR) led to the proposal of a time-varying $G$ cosmology \cite{Milne35}. It is represented by $G=\left(c^{3} / M_{\text{U}}\right)t$ to satisfy Einstein's conclusions, where $M_{\text{U}}$ and $t$ are the mass and the age of the Universe, respectively. There are recent reviews and applications in \cite{Ray:2007cc,Ray:2019lxv}. The constancy of fundamental physical constants is a primary foundation of the laws of physics. Any deviation from the established physical constants suggests the existence of a new law of physics. It concerns the speed of light, the gravitational constant, the fine structure constant, the proton-to-electron mass ratio, etc. There have been ongoing efforts to improve the accuracies of experiments on the time-dependence of these constants \cite{Uzan:2010pm,Chiba:2011bz,Will:2014kxa,Martins:2017yxk,Lopez-Honorez:2020lno}. From this point of view, the known values of physical constants are just an accident of the current epoch when they happen to be measured. A time variation of $\alpha$, the fine-structure constant, was announced \cite{Webb:2000mn} based on quasar observations, while no variation was seen \cite{13081496} based on observations using CH molecules. The time variation of $\alpha$ is still up for debate. However, it matters because, since $\alpha = e^2/(4 \pi \epsilon_0 \hbar c)$, its time variation is equivalent to the time variation of one or more of the vacuum permittivity, Planck constant, speed of light, and elementary charge. Thus, there have been updates for the limits on the time variation of $\alpha$ \cite{Murphy:2016yqp,Kotus:2016xxb,Levshakov:2017ivg,Murphy:2017xaz,Kanekar:2018mxs}.

Time variation of $\alpha$ affects various cosmological observables. Big Bang nucleosynthesis (BBN) refers to the formation of nuclei other than those of the lightest isotope of hydrogen ($^{1}$H) during the early phases of the Universe roughly at a temperature of about 0.1 MeV, which corresponds to a redshift $z \approx 10^{9}$ \cite{Alpher:1948pr}. If the primordial Helium mass fraction, $Y_{\text{P}}^{(\text{BBN})} = 4 n_{{}^{4}\text{He}}/n_{b}$ are changed, then they induce changes in the details of nucleosynthesis \cite{Steigman:2007xt}. In GR, the expansion rate of the Universe is well known. With this information and for the given value of the photon-baryon ratio, the process of standard BBN is well established and provides an accurate prediction of the values of $\YPBBN$. However, values of $\YPBBN$ can be changed if one relaxes any BBN prior or gravity theory. The changes in the values of $\YPBBN$  might cause a change in the recombination history. It can modify both the last scattering epoch and the diffusion damping scale. These changes affect cosmic microwave background (CMB) anisotropies \cite{Steigman:2010pa}. Since the weak interaction rate is dependent on $\alpha$, changing it from the standard model results in a change in the freeze-out temperature, which in turn impacts the BBN.  \cite{Kolb:1985prd,Bergstrom:1999wm,Nollett:2002da,Ichikawa:2004ju}. Since the electromagnetic processes are what form the CMB, changes in $\alpha$ have an impact on the Thomson scattering cross-section. It causes the change in the ionization of the fraction of free electrons to modify the CMB power spectra \cite{Hannestad:1998xp,Kaplinghat:1998ry,Avelino:2000ea,Martins:2002iv,Ichikawa:2006nm,Nakashima:2008cb,Wang:2009mj,Menegoni:2009rg,Hart:2017ndk}. The Sunyaev-Zel$^{’}$dovich effect (SZE) is a slight distortion of the CMB spectrum as a result of the inverse Compton scattering of the CMB photons on hot electrons of the intra-cluster medium (ICM) of galaxy clusters, which preserves the number of photons. Still, it allows photons to gain energy. It thus generates an increment of the photon temperature in the Wien region while a decrement of the temperature in the Rayleigh-Jeans part of the black-body spectrum. Therefore, the SZE is the imprint on the CMB frequency spectrum of the X-ray of clusters mainly due to bremsstrahlung. People use two parameters to describe two physical processes associated with SZE. One is the integrated Comptonization parameter $Y_{\text{SZ}}$, and the other is its X-ray counterpart, the $Y_X$ parameter. The dependency of the ratio of these two quantities on the fine structure constant is given by $\propto \alpha^{3.5}$. Thus, researchers use it to investigate the time variation of $\alpha$ \cite{Galli:2012bf,Holanda:2015oda,Holanda:2016iud,Colaco:2019fvl,Holanda:2019vmh,Goncalves:2019xtc}. Investigations have also been conducted on the impact of time-varying $\alpha$ on other cosmological observables such as white dwarfs (WDs), strong lensing (SL), etc \cite{Colaco:2020ndf,Hu:2020zeq,Milakovic:2020tvq,Bora:2020sws}. The spectrum of a distant galaxy can put an upper bound of the change in the proton-to-electron mass ratio that gives $10^{-16} \text{yr}^{-1}$ \cite{Bagdonaite12,Levshakov:2020ule,Hanneke2020}.

Due to the weakness of the gravitational interaction, the gravitational constant is difficult to measure with high precision. There have been conflicting measurements in the 2000s, and thus there have been controversial suggestions of a periodic variation of its value \cite{Anderson:2015bva}. Under the assumption that the physics in type Ia supernovae (SNe Ia) is universal, one might put an upper bound on $G$ of less than $10^{-10}$ per year for the gravitational constant over the last nine billion years \cite{Mould:2014iga}. The selection of units may affect the dimensional quantity's value and potential variations. The gravitational constant is a dimensional quantity. Thus, one might need to compare it with a non-gravitational force to provide a meaningful test on its time variation. For example, the ratio of the gravitational force to the electrostatic force between two electrons can give the dimensionless quantity related to the dimensionless fine-structure constant. From a theoretical point of view, one can establish gravity theories with a time-varying gravitational constant that satisfies the weak equivalence principle (WEP) but does not satisfy the strong equivalence principle (SEP) \cite{Will:2014kxa}. Most SEP-violating theories of gravity predict the locally time-varying gravitational constant. A variation of the gravitational constant is a pure gravitational phenomenon. Thus, it does not affect the local physics, such as the atomic transitions or nuclear physics. Most constraints on the time variation of the gravitational constant are obtained from systems where gravity is non-negligible. These include the motion of bodies of the Solar system, astrophysical systems, and cosmological ones. Again, one obtains this by comparing a gravitational time scale to a non-gravitational one. One can easily use Kepler's third law to encode a time variation of $G$ into an anomalous evolution of the orbital periods of astronomical bodies as shown in \cite{Merkowitz:2010lr}. The Lunar Laser Ranging (LLR) experiment has provided measurements of the relative position of the Moon for the Earth with an accuracy of the order of $1$ cm over three decades since the pioneering work done in 1978 
\cite{McElhinny:1978na,Muller:1991aj,Damour:1994bd,Williams:1996prl,Williams:1996prd,Williams:2004qba,Williams:2005rv,Battat:2007uh,Merkowitz:2010lr,Williams:2012nc,Murphy:2013qya,Bourgoin:2016ynf,Marki:2020ajaa,Cheng:2019aa,Viswanathan:2020oor}. When calculating the time variation of the period in pulsar timing, one cannot ignore the dependency of the gravitational binding energy, unlike in the case of the solar system \cite{Eardley:1975apj,Heintzmann:1975plA,Mansfield:1976na,Damour:1988prl,Nordtvedt:1990prl,Goldman:1990mnras,Damour:1991apjl,Kaspi:1994apj,Verbiest:2008gy,Zhu:2018etc,Ding:2020sig}. From the Poisson equation, one can interpret a change in the gravitational constant as a change in the star density. Thus, one can constrain the possible value of G from the stellar evolution \cite{Teller:1948pr,Gamow:1967prl,Demarque:1994,Guenther:1995apj,Gracia-Berro:1995mnras,DeglInnocenti:1995hbi,Thorsett:1996prl,Guenther:1998apj,Gaztanaga:2001fh,Ricci:2002ra,Biesiada:2003sr,Benvenuto:2004bs,GarciaBerro:2005yw,Bambi:2007an,Ketchum:2008fb,Althaus:2011ca,GarciaBerro:2011wc,Garcia-Berro:2013lea}. Cosmological constraints on the time variation of $G$ come from an extension of GR and require modifying all equations describing the background evolution and the perturbations. Its influence is affected at the BBN via a speed-up factor, $H/H^{(\GR)}$. The BBN limits on $\dot{G}/G$ for specific models have been considered 
\cite{Barrow:1978mnras,Rothman:1982apj,Kolb:1986prd,Arai:1987AA,Accetta:1990plb,Damour:1991prd,Casas:1992mpla,Santiago:1997prd,Damour:1998ae,Copi:2003xd,Olive:2004kq,Cyburt:2004yc,Umezu:2005ee,Clifton:2005xr,Coc:2006rt,Coc:2006sx,Acquaviva:2007mm,Iocco:2008va,Coc:2008yu,Nesseris:2009jf,Will:2014kxa,Alvey:2019ctk,Fields:2019pfx}. One can also investigate the time-dependent $G$ from CMB. It causes the modification of the Friedmann equation to change the sound horizon. This means that the time variation of $G$ can be constrained by the shift in angular scales and the adjustment of damping scales.  
\cite{Bond:1997wr,Chen:1999qh,Nagata:2003qn,Boucher:2004ta,Chan:2007fe,Wu:2009za,Wu:2009zb,Galli:2009pr,Brax:2011ta,Galli:2011dsa,Li:2013nwa,Ade:2014zfo,Li:2015aug,Ooba:2016slp,Ooba:2017gyn,Lin:2018nxe,Yadav:2019fjx,Wang:2020bjk}. The time variation of $G$ modifies the absolute magnitude of SNe and thus provides a modified magnitude vs redshift relation \cite{Riazuelo:2001mg,Zhao:2018gwk,Kazantzidis:2018jtb}. The time-varying $G$ causes the difference in the propagation of gravitational waves (GWs) between GR and the given gravity theories. Due to this discrepancy, the luminosity distance for GWs deviates from that for electromagnetic signals \cite{Zhao:2018gwk,Belgacem:2017ihm,Dalang:2019fma,Wolf:2019hun,Noller:2020afd,Vijaykumar:2020nzc}. 

One can also investigate the effect of the variation of fundamental constants on gravitational observables, such as black holes and WDs \cite{Davies:2002,MacGibbon:2007qh,Landau:2020vkr}. Or one can also investigate the time variation of other fundamental constants related to particle physics, like the Fermi constant $G_{F}$ \cite{Ferrero:2010ab,Karpikov:2015sra}.

Observational bounds on $\dot{\tc}/ \tc$ is $\left( 0 \pm 2 \right) \times 10^{-12} \text{yr}^{-1}$ obtained from the time variation of the radius of Mercury $c = c_0 e^{\psi}$ with the modification on the Hilbert-Einstein action \cite{Racker:2007hj} 
\begin{align}
I = \int d^4 x \sqrt{-g} \left( e^{a \psi} \left( R - 2 \Lambda - \kappa \nabla_{\mu} \psi \nabla^{\mu} \psi \right) + \frac{16 \pi G}{c_0^4} e^{b\psi} \mathcal L_{m} \right) \label{IRacker} \,.
\end{align}
In VSL models, various cosmological observables are affected by changing $c$ at different epochs. As a result, studies have looked into how changes in fundamental constants, such as $c$, affect a variety of cosmological observables, including BBN~\cite{Bergstrom:1999wm,Ichikawa:2002bt,Muller:2004gu,Coc:2006sx,Coc:2012xk,Coc:2013gc,Clara:2020efx}, CMB \cite{Avelino:2000ph,Ade:2014zfo,Martins:2010gu,Qi:2014zja,Hart:2019dxi}, baryonic acoustic oscillations (BAO)~\cite{Salzano:2014lra,Salzano:2015mxk,Salzano:2016hce,Salzano:2017nvc,Wang:2019tdn}, SNe~\cite{Wang:2019tdn,Izadi:2017phr,Cao:2018rzc}, GWs~\cite{Moffat:2014poa,Csaki:2000dm,Izadi:2009tr,Tahura:2019dgr}, Hubble parameter~\cite{Cai:2016vmn}, strong lensing (SL)~\cite{Cao:2020vra}, and others~\cite{Cao:2016dgw,Landau:2000mx,Calcagni:2013yra}. 

There have been various attempts to explain mechanisms for VSL models:  the hard breaking of Lorentz invariance~\cite{Barrow:1998he,Albrecht:1998ir,Barrow:1998eh,Magueijo:2000zt,Bassett:2000wj},  bimetric VSL theories~\cite{Clayton:1998hv,Drummond:1999ut,Clayton:1999zs,
Clayton:2000xt}, color-dependent speed of light~\cite{Amelino-Camelia:1996bln,Amelino-Camelia:1997ieq,Ellis:1999sd,Amelino-Camelia:2000bxx,Amelino-Camelia:2000cpa,Kowalski-Glikman:2001vvk,Bruno:2001mw,Magueijo:2001cr,Magueijo:2002am,Ellis:2002in,Amelino-Camelia:2002uql}, locally Lorentz invariant VSL~\cite{Moffat:1992ud,Barrow:1999st,Jacobson:2000xp}, extra dimensions induced VSL~\cite{Randall:1999ee,Randall:1999vf,Kiritsis:1999tx,Kaelbermann:1999jw,Chung:1999xg,Alexander:1999cb,Ishihara:2000nf,Csaki:2000dm}, etc. We extend the Einstein Hilbert action by allowing both $c$ and $G$ to vary for the cosmic time and thus can write the action as shown in Eq \eqref{tSHmp}. It is possible to have the cosmic time-varying speed of light model when we adopt the unconventional cosmic time dilation. However, we focus on the phenomenological effects of this model on various late-time cosmological observations and look for possible solutions for known problems in this manuscript.

We briefly review previous VSL models and shortly introduce the minimally extended VSL (meVSL) model in the next section~\ref{sec:PreVSL}. In Sec.~\ref{sec:SR}, we investigate the LI of SR to obtain the cosmological evolutions of fundamental constants in meVSL. We also probe any modification of meVSL compared to GR. We investigate any modification of the geodesic equation and the deviation of it in meVSL in section~\ref{sec:geodesics}. In section~\ref{sec:cosmology}, we derive Friedmann equations of meVSL and show that the cosmological redshift of meVSL is the same as that of GR. We investigate modifications of cosmological observables in meVSL and try to obtain the constraint of the variation of the speed of light based on the current observations. We conclude in section~\ref{sec:conclusion}.
\section{Previous VSL}
\label{sec:PreVSL}

When one describes the background Friedmann-Lema\^{i}tre-Robertson-Walker (FLRW) universe, one can define the constant-time hypersurface by using physical quantities such as temperature or density. It is because the temperature and density evolve in time, and the homogeneity of the Universe demands that they must be equal at the same cosmic time. Thus, the speed of light is also constant at a given time, even though it can evolve through cosmic time, $\tc [a]$. In other words, the speed of light is a function of the scale factor, $a$. This fact makes it possible to construct the LI VSL models on each hypersurface.

Even though GR has been a successful theory to describe the Universe, there exist some drawbacks to standard cosmology based on GR. Thus, it is worth trying a new minimally extended theory to overcome those shortcomings while keeping the success of GR. VSL can be a candidate among these kinds of minimally extended theories.

As an alternative to cosmic inflation, the early VSL models focus on solving the horizon problem of the Big Bang model~\cite{Barrow:1998eh,Petit:1988,Petit:1988-2,Petit:1989,Midy:1989,Moffat:1992ud,Petit:1995ass,Albrecht:1998ir,Barrow:1998he,Clayton:1998hv,Barrow:1999jq,Clayton:1999zs,Brandenberger:1999bi,Bassett:2000wj,Clayton:2000xt,Gopakumar:2000kp,Magueijo:2000zt,Magueijo:2000au,Magueijo:2003gj,Magueijo:2007gf,Petit:2008eb,Roshan:2009yb,Sanejouand:2009,Nassif:2012dr,Moffat:2014poa,Ravanpak:2017kdg,Costa:2017abc,Nassif:2018pdu}. Petit also proposed that the variation of $c$ accompanies the joint variations of all physical constants as given in Eq.~\eqref{Petitconst}~\cite{Petit:1995ass,Petit:2008eb}. Researchers have studied the theoretical and empirical aspects of the dynamics of the VSL model~\cite{Avelino:2002jn,Pedram:2007mj,Salzano:2016pny,Leszczynska:2018juk,Avelino:1999is,Belinchon:1999kq,Drummond:1999ut,Alexander:1999cb,Avelino:2000ph,Szydlowski:2002kz,Shojaie:2004sq,Shojaie:2004xw,Balcerzak:2013kha,Balcerzak:2014rga,Franzmann:2017nsc,Hanimeli:2019wrt,Skara:2019usd,Bhattacharjee:2020fgl,Gupta:2020anq}.

Recently, there have been various investigations of mVSL model effects on cosmological observables, such as CMB~\cite{Qi:2014zja}, BAO~\cite{Salzano:2014lra,Salzano:2015mxk,Salzano:2016hce,Salzano:2017nvc,Wang:2019tdn}, SNe~\cite{Izadi:2017phr,Cao:2018rzc}, GWs~\cite{Izadi:2009tr}, H~\cite{Cai:2016vmn}, and SL~\cite{Cao:2020vra}. However, the mVSL model only considers the variation of $c$, a dimensional constant. By changing units, one can obtain time dependence of dimensional constants. Thus, time-varying dimensional constants are not invariant statements. Since dimensional parameters are defined as dimensionless ratios between the parameter and the unit, they become invariant once one fixes the units.

However, one needs to rewrite modern physics for varying $c$ to propose an integrated viable alternative to the whole set of physical equations and consequent effects dependent on $c$~\cite{Ellis:2003pw,Ellis:2007ah}. In addition to the geometry of null geodesics, both the temporal and the spatial measurements are affected by the speed of light. Thus, any VSL model may require the validity of LI and the redefinition of distance measurements. Also, it may cause the modification of Maxwell’s equations. One also needs to investigate the consistencies of all other physical theories.

In the next section, we show that we can obtain an extended theory satisfying both the LI and the law of energy conservation even when the speed of light varies as a function of the cosmic time, $c[a]$. We obtain the cosmological evolutions on other physical constants to satisfy LI, electromagnetism, and thermodynamics. We compare the cosmological evolutions on both physical constants and quantities between different VSL models in table~\ref{tab:tabVSL}. The results in the last column come from meVSL, and we derive these relations in section~\ref{sec:SR}.

\begin{table}[h!]
    \caption{Cosmological evolutions on both physical quantities and constants of various VSL models. NC means not considered. }
		\label{tab:tabVSL}
             \begin{adjustbox}{width=\columnwidth,center}
		\begin{tabular}{|c|c|c|c|c|c|c|c|c|c|c|} 
			\hline
			 $c$ & $G$ & $\hbar$ &$\lambda$ & $\nu$  &  $m$ & $k_{\TB}$ & $T$ & $e$  & $\alpha$   &reference \\ \hline
			 $c_0 a^{-1/2}$  & $G_0 a^{-1}$ & $\hbar_0 a^{3/2}$ &$\lambda_0 a$ &$\nu_0 a^{-3/2}$ & $m_0 a$ & NC & NC & $e_0 a^{1/2}$ & const & \cite{Petit:1988,Petit:1995ass,Petit:2008eb} \\
			 $c_0 a^n$ & const & $\hbar_0 a^{n}$ &const & $\nu_0 a^n$ & const & const & $T_0 a^{2n}$ & const & $\alpha_0 a^{-2n}$  & \cite{Barrow:1998eh,Albrecht:1998ir} \\
			 $c_0 a^{-1/4}$ & const & NC & $\lambda_0 a$ & $\nu_0 a^{-5/4}$ & $\m_0 a^{1/2}$ & const & $T_0 a^{-5/4}$ & NC & NC  & \cite{Shojaie:2004sq,Shojaie:2004xw} \\
			 $c_0 a^{n}$ & const & NC  & NC  & NC & NC & NC & NC & NC & NC  & \cite{Balcerzak:2013kha,Balcerzak:2014rga,Qi:2014zja} \\
			  $c_0 a^{b/4}$ & $G_0 a^{b}$ & $\hbar_0 a^{-b/4}$ & $\lambda_0 a$ & $\nu_0 a^{-1+b/4}$ & $m_0 a^{-b/2}$ & const & $T_0 a^{-1}$ & $e_0 a^{-b/4}$ & $\alpha_0 a^{-b/4}$  & meVSL \\
			\hline
		\end{tabular}
         \end{adjustbox}
 \end{table}

\section{Special Relativity}
\label{sec:SR}

When quantum and gravitational effects are minimal, SR is the most accurate theory of motion at any speed, as demonstrated by experiments. Experiments have verified SR in a wide range of cases. Thus, the meVSL should inherit the success of SR. In this section, we revisit SR and modifications to physical laws associated with SR in the meVSL. It provides us with modifications of additional physical constants in the meVSL. 

 Two postulates originally formed the basis of SR. One is that the laws of physics are the same (invariant) in all inertial frames of reference (\textit{i.e.}, non-accelerating frames of reference), and the other is that the vacuum speed of light is the same for all observers, regardless of the motion of the light source or observer. However, if the spacetime transition between inertial frames follows Lorentz transformations, all observers within inertial frames will have an equally finite maximum speed. Therefore, the SR is sufficiently satisfied by the single postulate of Minkowski spacetime or the single postulate of universal Lorentz covariance \cite{Das93,Schutz97}.

 In relativistic physics, SR implies that the laws of physics are the same for all observers moving for one another within an inertial frame. It provides an equivalence of observation or observational symmetry called Lorentz symmetry. If a physical quantity transforms under a given representation of the Lorentz group, then we call it the Lorentz covariant. One can build Lorentz covariant quantities from scalars, four vectors, four tensors, and spinors. In particular, a Lorentz invariance (LI) defines a Lorentz covariant scalar that remains the same under Lorentz transformations. One also calls an equation Lorentz covariant if one writes it in terms of Lorentz covariant quantities. In all other inertial frames, Lorentz covariant quantities hold in the same way if they hold in one inertial frame. It follows from the fact that if all the components of a tensor vanish in one frame, they also vanish in every frame. It is a required condition based on the principle of relativity (\textit{i.e.}, all non-gravitational laws must make the same predictions for identical experiments taking place at the same spacetime event in two different inertial frames of reference). In GR, local Lorentz covariance means that Lorentz covariance is applied only within infinitesimal regions of spacetime.

 In Einstein's theory of relativity, a point in Minkowski's space is an assembly of one temporal and three spatial positions called an event, or sometimes the position four-vector described in one reference frame by a set of four coordinates. One can define the path of an object moving relative to a particular frame of reference by this position four-vector ({\it i.e.}, four-position), $x^{\mu} = (ct, x^{i}) = (c \tau, 0)$, where $\mu$ is a spacetime index which takes the value $0$ for the timelike component, $i = 1, 2, 3$ for the spacelike coordinates, and $\tau$ is the so-called proper time measured at the instantaneous rest frame. One should emphasize that the speed of light of meVSL is a function of cosmic time. Thus, it is better to express the time variation of the speed of light as $c[a]$ as a function of the scale factor, $a$
\begin{align}
x^{0}(\tau) = c \left[ a \left[ \tau \right] \right] \tau \quad , \quad x^{0}(t) = c \left[ a \left[ t \right] \right] t \,. \label{x0}
\end{align}
Then the differentials of $x^{0}$ in different coordinates are given by
\begin{align}
d x^{0} &= \begin{cases} \left( 1 + \frac{d \ln c}{d \ln a} \frac{d \ln a}{d \ln \tau} \right) c (a[\tau]) \, d \tau \equiv \ttc (a[\tau]) \, d \tau \\
\left( 1 + \frac{d \ln c}{d \ln a} H t  \right) c [a[t]] \, d t \equiv \tc (a[t]) \, d t  \end{cases} \label{dx0ps} \,,
\end{align}
where $H = d \ln a / dt$ is the Hubble parameter. We introduce new definitions of the speed of light, $\tc$ and $\ttc$.

Time dilation (TD) is a difference in the elapsed time measured between two events, as measured by two observers moving relative to each other. From TD, the relation between the differentials in both coordinate time $t$ and proper time $\tau$ can be parameterized by
\begin{align}
dt &\equiv \gamma(v, \tc, \ttc) d \tau \label{gamma} \,,
\end{align}
where $\gamma (v, \tc, \ttc)$ is the Lorentz factor in the meVSL model. It might depend on both $\tc$ and $\ttc$ compared to the conventional Lorentz factor that depends on only the relative motion between two frames, $v$.

The conservation of the line element, $ds = \sqrt{\eta_{\mu\nu}dx^{\mu} dx^{\nu}}$, allows us to obtain the relation between the Lorentz factor of the meVSL model and that of SR. For Lorentzian coordinate $\eta_{00} = -1$ and $\eta_{ii} =1$ and the line-element square is given by
\begin{align}
ds^2 &= \eta_{\mu\nu} dx^{\mu} dx^{\nu} = - \ttc^2(\tau) d \tau^2 = - \tc^2(t) dt^2 + \sum_{i=1}^{3} \left(dx^{i}\right)^2  \,.\label{ds2Lor}
\end{align}
We define the norm of the three tangent vector $v^{i} = dx^{i}/dt$ as $v = \sqrt{\sum_{i} (v^{i}})^2$. From Eq.~\eqref{ds2Lor}, one obtains
\begin{align}
\left( \frac{\ttc(\tau)}{\tc(t)} \frac{d\tau}{d t} \right)^2 &= 1 - \frac{v^2}{\tc(t)^2} \equiv 1 - \tbeta^2 \equiv \tgam^{-2} \, \Longrightarrow \,
\gamma (v, \tc, \ttc) = \breve{r} (\tc, \ttc) \tgam(v, \tc)  \,,  \text{where} \,\,  \breve{r} \equiv \frac{\ttc}{\tc} \,, \label{gamma2}
\end{align}
where $\tgam$ equals the Lorentz factor of SR when the speed of light is $\tc$. From Eq.~\eqref{gamma2}, one can interpret that the time delay in the meVSL model arises from two effects. One is from the relative motion of each frame $\tgam$, and the other comes from the differences of the time-varying speed of light at two frames, $\breve{r}$.
In the local inertial frame (LIF), $v = 0$, as $\tgam = 1$ and $\breve{r} = 1$, so $\gamma = 1$. It also means that $\ttc = \tc$ in the LIF. However, this relation alone cannot specifically determine the conditions of $\tc$ and $\ttc$. Thus, we investigate the other relation specifying the relation between $\tc$ and $\ttc$. Furthermore, one should notice that the expression for $\tc[a]$ in Eq.~\eqref{dx0ps} is a function of the scale factor $a[t]$. Thus, at the given constant time hypersurface (\textit{i.e.}, at the given cosmic epoch), $\tc$ is a constant value \footnote{Based on Minkowski spacetime, SR's universal Lorentz covariance sufficiently satisfies its tenets. On the other hand, an IF in GR denotes a freely falling one. Although LI spacetime intervals between events can be established, the lack of a universal IF in GR makes it difficult to define a global time. 
Meanwhile, a global time that satisfies the cosmological principle for the universe can be designated, allowing it to be divided into non-intersecting 3-dimensional surfaces. This is the universe described by the RW metric \cite{Islam01,Narlikar02,Hobson06,Roos15}. If $c$ varies on cosmic time but is locally constant (that is, at each given epoch), then the LI VSL model is possible. In other words, if the speed of light in an expanding universe is expressed as a function of the scale factor, $c[a]$, then even though its value changes at different epochs such as $a_1$ and $a_2$, it will have a constant local value at each epoch. It ensures LI, thereby maintaining the validity of quantum mechanics and electromagnetism that satisfy special relativity at each epoch. The RW metric assumes that all galaxies exist on the same hypersurface. This hypersurface coincides with the simultaneity surface of each galaxy's local Lorentz frame. It allows us to visualize the hypersurface as a smooth conjunction of the Lorentz frames of all galaxies, with each galaxy's 4-velocity being perpendicular to this hypersurface, a concept known as Weyl's postulate. 
This series of hypersurfaces can be given a parameter $t$, which becomes the appropriate time for every galaxy and creates a global time standard. The measurement made by a comoving observer, who sees the cosmos expanding uniformly around her, corresponds to this cosmic time. 
Consequently, the proper time is the cosmic time in the RW metric. The standard RW metric's constant speed of light assumption is based on particular hypotheses about cosmological time dilation rather than on rules that can be immediately drawn from the metric's foundations \cite{Lee:2023FoP,Lee:2024part}. In the meVSL model, the value of $c$ remains constant on each hypersurface, satisfying Lorentz invariance on each hypersurface related to SR.}. Later, we investigate the Einstein field equation (EFE) to obtain the specific form of $\tc$ from the Friedmann-Lema\^{i}tre-Robertson-Walker (FLRW) universe.

\subsection{Four velocity and four acceleration}
\label{subsec:4vel4accel}

The four-velocity is the rate of change of the four-position for the proper time along the curve. Whereas the velocity denotes the rate of change of the position in the three-dimensional space of the object, as seen by an observer, for the observer's time. The value of the magnitude square of a four-velocity, $| {\bf U} |^2 = {\bf U} \cdot {\bf U} = \eta_{\mu\nu} U^{\mu} U^{\nu}$, is always equal to $- \ttc^2$, where $\ttc$ is the speed of light in the inertial frame. For an object at rest, the direction of its four-velocity is parallel to that of the time coordinate with $U^0 = \ttc$. Thus, a four-velocity is a contravariant vector with the normalized future-directed timelike tangent vector to a world line. Even though the four-velocity is a vector, the addition of two of them does not yield another four-velocity. It means that the space of four-velocity is not itself a vector space but the tangent four-vector of a timeline worldline. Thus, four-velocity $U^{\mu}$ at any point is defined as
\begin{align}
U^{\mu} &\equiv \frac{d x^{\mu}}{d \tau} = \begin{cases} \left( \ttc \,, 0 \right) \\
\gamma \left( \tc \,, v^{i} \right) = \tgam \left( \ttc \,, \frac{\ttc}{\tc} v^{i} \right)  \end{cases} \label{Umu} \,.
\end{align}
When described in the particular slice of the flat spacetime, the three spacelike components of four-velocity define a traveling object's proper velocity $\gamma \vec {v} = d \vec {x} /d\tau$. One can obtain the magnitude of four-velocity from Eq.~\eqref{Umu}
\begin{align}
U^{\mu} U_{\mu} &= \gamma^2 \left( -\tc^2 + v^2 \right) = - \breve{r} ^2 \tc^2 = - \ttc^2 \label{UU} \,.
\end{align}
Similarly, the four-acceleration, $A^{\mu}$, is defined as the rate of change in four-velocity for the particle's proper time along its worldline. Thus, one can obtain $A^{\mu}$ from Eq.~\eqref{Umu}
\begin{align}
\resizebox{\textwidth}{!}{
$A^{\mu} \equiv \frac{d U^{\mu}}{d \tau} = \gamma^2 \left[ \frac{\dot{\gamma}}{\gamma}  \left( \, \tc  , v^{i} \, \right) + \left( \, \dot{\tc}  , a^{i} \, \right) \right] = \gamma^2 \tgam^2  \left( \frac{\vec{v} \cdot \vec{a}}{\tc} - \tbeta^2 \dot{\tc} + \tgam^{-2} \frac{\dot{\ttc}}{\ttc} \tc  \,, \vec{a} + \frac{\vec{v} \times \left( \vec{v} \times \vec{a} \right)}{\tc^2} - \frac{\dot{\tc}}{\tc} \vec{v}  + \tgam^{-2} \frac{\dot{\ttc}}{\ttc} \vec{v} \right) \,, \label{4accel}$
}
\end{align}
where dots denote the derivatives for the coordinate time, $t$. We show the detailed derivation of the above Eq.~\eqref{4accel} in appendix~\ref{subsec:FourAccelerationApp}. Geometrically, four-acceleration is a curvature vector of a worldline.

In an instantaneously co-moving inertial reference frame (\textit{i.e.}, $\vec{v} = 0$, $\tgam = 1$, $\dot{\gamma } = 0$, and $\dot{\tc} = 0$), the four-acceleration in Eq.~\eqref{4accel} becomes
\begin{align}
A_{(\ine)}^{\mu} &= \left( \frac{\dot{\ttc}}{\ttc} \tc  \,, \vec{a} \right) = \left( \dot{\ttc}  \,, \vec{a} \right) = \left( 0 \,, \vec{a} \right) \,, \label{4acceline}
\end{align}
where we use $\ttc = \tc$ in the second equality. This result is identical to the result of special relativity. We want to establish a VSL model that takes over the success of SR. Thus, we make a VSL model that satisfies all three equivalence principles. In other words, the result of any local experiment (gravitational or not) on a freely falling observer is independent of the observer's velocity and location in spacetime. An inertial frame of reference in SR possesses the property that the acceleration of an object with zero net force acting upon it is zero in this frame of reference. It means that the object is at rest or moving at a constant velocity. The core concept in the equivalence principles is locality. Thus, if one adopts that $\tc$ (equally $\ttc$) depends on the cosmic time only ({\it i.e.}, $c = c[a[t]]$), then one can establish the constant $\tc$ at the given cosmic time. We are already familiar with this concept when we consider the temperature of the cosmic microwave background radiation (CMB). The temperature of the cosmic photon $T_{\gamma}[a[t]]$ is expressed as $T_{\gamma 0} (a[t]/a[t_0])^{-1}$, where $t_0$ is the cosmic time at the present epoch (\textit{i.e.}, the current age of the universe), $a[t_0] \equiv a_0$ is the present scale factor which will later be set to $1$, and $T_{\gamma 0} = T[a_0]$ is the present value of the CMB temperature. $T_{\gamma}$ at the given cosmic epoch is constant and one considers the cosmic evolution of $T_{\gamma}$ as a function of the scale factor only. This kind of VSL model is called the minimally varying speed of light, ``mVSL''. Thus, the four-velocity and four-acceleration of mVSL in an instantaneously co-moving inertial frame become
\begin{align}
U_{(\ine)}^{\mu} &= \left( \ttc \,, \vec{0} \right) = \left( \tc \,, \vec{0} \right) \quad , \quad
A_{(\ine)}^{\mu} = \left( \frac{\dot{\ttc}}{\ttc} \tc  \,, \vec{a} \right) = \left( 0 \,, \vec{a} \right)  \label{4velaccinert} \, .
\end{align}
These are the same as those of SR. We also investigate the scalar product of a particle's four-velocity and its four-acceleration $U^{\mu} A_{\mu}$ which is given by
\begin{align}
U^{\mu} A_{\mu} & = - \gamma^3 \tgam^{-2} \frac{d\tau}{dt} \frac{d \ttc}{d \tau} \frac{1}{\ttc} \tc^2  = - \ttc \frac{d \ttc}{d \tau} \,. \label{UmuAmu}
\end{align}
The detailed derivation of the above Eq.~\eqref{UmuAmu} is given in appendix~\ref{subsec:FourAccelerationApp}.
To satisfy $U^{\mu} A_{\mu} = 0$ in the inertial frame,  $\ttc = \tc =$ const is required and this is satisfied in the mVSL. Now, one needs to investigate other consequences of the mVSL.

\subsection{Four momentum}
\label{subsec:4momentum} 

One can generalize the classical three-dimensional momentum to four-momentum in the four-dimensional spacetime. Classical momentum is a three-dimensional vector, so the four-momentum in spacetime is a four-vector. The contravariant four-momentum of a massive particle is given by the particle's rest mass, $m_{\rs}$ multiplied by the particle's four-velocity
\begin{align}
P^{\mu} &= m_{\rs} U^{u} = m_{\rs} \gamma \left( \tc \,, \vec{v} \right) = m_{\rs} \tgam \left( \ttc \,, \frac{\ttc}{\tc} \vec{v} \right) \equiv \left( \frac{E}{\tc} \,, \vec{p} \right) \label{Pmu} \,,
\end{align}
where we use Eq.~\eqref{Umu}. Thus, the relativistic energy $E$ and three-momentum $\vec{p} = \gamma m_{\rs} \vec{v}$, where $\vec{v}$ is the particle's three-velocity and $\gamma$ is the Lorentz factor, are given by
\begin{align}
E &= m_{\rs} \gamma \tc ^2  \quad , \quad \vec{p} = m_{\rs} \gamma \vec{v}  \,. \label{Evecp}
\end{align}
The energy-momentum relation (relativistic dispersion relation) is obtained from the squaring of the four-momentum
\begin{align}
P^{\mu}P_{\mu}&= - m_{\rs}^2 \tc^2 = - \frac{E^2}{\tc^2} + p^2  \quad \rightarrow \quad E^2 = m_{\rs}^2 \tc^4 + p^2 \tc^2 \label{Pmu2} \,.
\end{align}
As expected, the dispersion relation of the massive particle of mVSL in Eq.~\eqref{Pmu2} is the same as that of SR. One can also recover the classical mechanics for the non-relativistic limit, $v \ll \tc$
\begin{align}
E &\approx m_{\rs} \tc^2 \left( 1 + \frac{1}{2} \left( \frac{m_{\rs} v }{m_{\rs} \tc} \right)^2 + \cdots \right) = m_{\rs} \tc^2 + \frac{1}{2} m_{\rs} v^2 + \cdots \label{Eclass} \,,
\end{align}
where the second term is the classical kinetic energy.

For massless particles, one needs to redefine energy and momentum as
\begin{align}
E &\equiv \hbar \, \omega = h \, \nu \quad , \quad \vec{p} \equiv  \hbar \, \vec{k} = \frac{h}{\lambda} \hat{n} \label{Epm0} \,,
\end{align}
where $h (\hbar)$ is the (reduced) Planck constant, $\nu (\omega)$ is the (angular) frequency, $\vec{k}$ is the wavevector with a magnitude $|\vec{k}| = k$, equals to the wavenumber, $\lambda$ is the wavelength, and $\hat{n}$ is the unit vector. Thus, the energy-momentum relation in Eq.~\eqref{Pmu2} becomes
\begin{align}
E^2 &= p^2 \tc^2 \quad \Longrightarrow \quad \lambda \tnu = \tc  \label{masslessEvecp} \,.
\end{align}
One might wonder why we are reiterating seemingly obvious results that appear to be equivalent to those of SR.  However, one should be careful $\tc$ in equations \eqref{Pmu2} and \eqref{masslessEvecp}. In section~\ref{sec:cosmology}, we obtain the explicit form of $\tc$ as a function of the scale factor, $a$, and we may investigate any deviation of dispersion relation in cosmic time.

For the matter waves, one can use the de Broglie relations for energy and momentum of matter to obtain
\begin{align}
\left( \hbar \omega \right)^2 = \left( \hbar k \tc \right)^2 + m_{\rs} ^2\tc^4 \quad \Longrightarrow \left( \frac{\omega}{\tc} \right)^2 = k^2  + \left( \frac{m_{\rs} \tc}{\hbar} \right)^2 \label{EPrel}  \,.
\end{align}
When we consider the cosmological evolution of $\tc$, we also obtain the evolution equations for $\hbar$, $\nu$, and $m_{\rs}$ from the conservations of energy and number densities. As a result, the interpretation of the dispersion relation in Eq.~\eqref{EPrel} may vary depending on the epoch. The last energy relation used in cosmology is the relation between the microscopic energy $E$ and the macroscopic temperature $T$ given by
\begin{align}
E = k_{\TB} T \,, \label{kBT}
\end{align}
where $k_{B}$ is the Boltzmann constant. In cosmology, we consider the thermal equilibrium when calculating the relic density of particles. In subsection~\ref{subsec:TEQ}, we show that both $k_{B}$ and $T$ are not affected by variation in the speed of light.

\subsection{Electromagnetism}
\label{subsec:EM}

Consistent analysis of VSL theories can lead to significant violations of charge conservation \cite{Landau:2000mx}. However, VSL theories can be resolved by adjusting units according to the scale of dynamics. Time and space units can be redefined as $dt \rightarrow [f(x)]^a dt$ and $dx^{i} \rightarrow [f(x)]^b dx^{i}$, where $f$ is a function and $a$ and $b$ are constants. To maintain the local LI of the line element, it is necessary that $c(x) \propto [f(x)]^{b-a}$. This approach forms anisotropic multi-scaling by scaling the spatial and temporal variables differently. When $b = 0$, the time coordinate can be redefined by reabsorbing $c$ into the $x^{0}$ coordinate \cite{Calcagni:2013yra}
\begin{align}
x^0 = \int dt c(x) \label{x0Calcagni} \,,
\end{align}
With this choice of coordinates, specific conditions allow all equations to become generally covariant and gauge-invariant \footnote{When we think of an event in four-dimensional spacetime as a composition of a temporal point and a three-dimensional hypersurface at a given epoch, we may talk about this problem. However, in a VSL model that satisfies the LI, the time variation of the speed of light is due to actual changes in physical wavelength and frequency caused by the Universe's expansion but not from the redefinition of time. It refers to the functional change in the cosmological redshift of the speed of light that occurs when the cosmological TD does follow the relation $T_0 = T (1+z)^{n}$ with $n \neq 1$ \cite{Lee:2023FoP,Lee:2024part}.}.

In this subsection, we review Maxwell's equations in 4-dimensional spacetime to investigate the effect of meVSL on Maxwell's equation. We adopt the speed of electromagnetic waves propagating in vacuum is related to the distributed capacitance and inductance of vacuum, $\tc^{-2} = \tepsilon \tmu$ where $\tepsilon$ and $\tmu$ represent the permittivity and the permeability of vacuum, respectively. One consequence of meVSL is that both $\tepsilon$ and $\tmu$ can also vary as a function of the scale factor, $a$ (\textit{i.e.}, as a cosmic time). In section~\ref{sec:cosmology}, we adopt $\tc = \tc_{0} a^{b/4}$ and as a result we obtain $\tepsilon = \tepsilon_{0} a^{-b/4}$ and $\tmu = \tmu_{0} a^{-b/4}$ where subscripts $0$ represent the values of the present Universe instead of vacuum. Therefore, in the meVSL, the Maxwell's equations can be modified. The electromagnetic field is fully described by a vector field called the 4-potential $A^{\alpha}$ that is given by
\begin{align}
A^{\alpha} (t \,, \vec{x}) &\equiv \left( \frac{\phi}{\tc} \,, \vec{A} \right) \quad , \quad A_{\alpha} (t \,, \vec{x}) \equiv \left( - \frac{\phi}{\tc} \,, \vec{A} \right) \label{Aalpha} \,,
\end{align}
where $\phi$ is the electrostatic scalar potential, $\vec{A}$ is the vector potential, and $\tc$ is speed of light given in Eq.~\eqref{dx0ps}. The Lagrangian of a charged particle and an electromagnetic field is given by
\begin{align}
L_{\text{EM}} &\equiv \int \mathcal L d^3 x = - \int \rho_{m} \tc \sqrt{ U_{\alpha} U^{\alpha}} d^3 x - \int \frac{1}{4 \mu} F_{\alpha\beta} F^{\alpha\beta} d^3 x + \int j_{\alpha} A^{\alpha} d^3 x \label{LEM} \,, 
\end{align}
where $F^{\alpha\beta} = \partial^{\alpha} A^{\beta} - \partial^{\beta} A^{\alpha}$ is an electromagnetic field strength tensor, $j^{\alpha} = \rho_{\EM \rs} U^{\alpha} = \rho_{\EM \rs} \gamma \left( \tc \,, \vec{v} \right)$ is a four-current density, and $\rho_{\EM \rs} = q_{\EM \rs} \delta (\vec{r} - \vec{s} )$. $\rho_{\EM \rs}$ is the rest charge density \textit{i.e.}, the charge density for a comoving observer (an observer moving at the constant speed $\vec{v}$).

The Euler-Lagrange equations for the electromagnetic field provide 
\begin{align}
\partial^{\alpha} F_{\alpha\beta} &= - \mu j_{\beta} + \frac{\partial^{\alpha} \mu}{\mu} F_{\alpha\beta} \,, \label{ELEqofEM} \\
\epsilon^{\alpha\beta\gamma\delta} \partial_{\gamma} F_{\alpha\beta} &= 0 \,, \label{ELEqofEM2} 
\end{align}
where Eq.~\eqref{ELEqofEM} are inhomogeneous Maxwell's equations (Gauss's law and Amp\`{e}re's law) and Eq.~\eqref{ELEqofEM2} are Bianchi identity (Gauss's law for magnetism and Maxwell-Faraday equation). One can refer to the appendix~\ref{subsec:ElectromagnetismApp} for the detailed derivation. We adopt $E_{i} = - \tc F_{0i}$ and $B_i = 1/2 \epsilon_{ijk} F^{jk}$ to obtain 
\begin{align}
\vec{\nabla} \cdot \vec{E} (t) = - \tmu \tc j_{0} + \vec{\nabla} (\ln \tmu) \cdot \vec{E} = \tmu \tc^2 \rho_{\EM \rs}  = \frac{\rho_{\EM \rs}}{\tepsilon} \quad , \quad \text{for} \quad \beta = 0 \,, \label{Gausslaw}
\end{align}
where we use $\tc^2 = 1/(\tmu \tepsilon)$, $\tmu$, and $\tepsilon$ are functions of the cosmic time only. In the last equality we use $\tc = \tc_0 a^{b/4}$ and $\tepsilon = \tepsilon_0 a^{-b/4}$ and $\tmu = \tmu_0 a^{-b/4}$. 
\begin{align}
-\frac{1}{\tc^2} \frac{d \vec{E}(t)}{d t} + \vec{\nabla} \times \vec{B} = \tmu \vec{j} (t) - \frac{ d \ln[\tc \tmu]}{d t} \frac{\vec{E}(t)}{\tc} = \tmu \vec{j} (t) - \frac{ d \ln[\tc_0 \tmu_0]}{d t} \frac{\vec{E}(t)}{\tc} = \tmu \vec{j} (t) \quad , \quad \text{for} \quad \beta = k \,, \label{Amerelaw}
\end{align}
where we use the fact that the present values of $\tc_0$ and $\tmu_0$ are given constant values at the present hypersurface. Thus, Amp\`{e}re's law is also the same as that of SR. 

In SR, charge conservation is that the Lorentz invariant divergence of $J^{\alpha}$ is zero
\begin{align}
\partial_{\alpha} J^{\alpha} =  \frac{d}{\tc d t} \left( \rho_{\EM} \tc \right) + \vec{\nabla} \cdot \vec{j} = 0 \label{charveconserv} \,.
\end{align}
In the LIF, one can conclude $\rho_{\EM} \tc =$ constant, where both $\rho_{\EM}(t_{\ast})$ and $\tc(t_{\ast})$ are constants at the local time $t_{\ast}$ in the absence of the local current. Thus, this is consistent with the conservation of charge in the LIF. Similarly, the continuity equation in GR with FLRW metric is written as
\begin{align}
\nabla_{\alpha} J^{\alpha} = \frac{d}{\tc d t} \left( \rho_{\EM} \tc \right) + \vec{\nabla} \cdot \vec{j} + 3 \frac{H}{\tc} \rho_{\EM} \tc = 0 \,.
\label{charcon}
\end{align}
When $\vec{j} = 0$, the above continuity equation gives the solution as
\begin{align}
&\rho_{\EM} = \rho_{\EM 0} a^{-3 - \frac{b}{4}} \label{rhoEM2} \,.
\end{align}
By using the above Eq.~\eqref{rhoEM2}, the Gauss's law in Eq.~\eqref{Gausslaw} becomes
\begin{align}
\nabla \cdot \vec{E}(t) = \frac{\rho_{\text{EM}}}{\tepsilon} = \frac{\rho_{\text{EM}0} a^{-b/4}}{\tepsilon_0 a^{-b/4}} = \frac{\rho_{\text{EM}0}}{\tepsilon_0} \label{Gausslaw0} \,.
\end{align}
Therefore, Gauss's law holds in the same form even if the speed of light changes with the expansion of the universe.

\subsection{Thermal Equilibrium}
\label{subsec:TEQ} 

From the perfect blackbody spectrum of the CMB, we know that the early Universe was in the local thermal equilibrium. We need to use statistical mechanics to turn microscopic laws into behaviors of macroscopic laws. It is convenient to describe the system in phase space, where the gas of weakly interacting particles is described by the positions and momenta of all particles. The density of momentum eigenstates of particles in momentum space is volume divided by $\tih^3$ and the state density in position and momentum phase space is $\tih^{-3}$. Thus, if the particle has $g$ internal degrees of freedom (\textit{e.g.}, spin), then the density of states becomes $g/(2 \pi \thbar)^3$. One needs to know the phase space distribution function, $f(\vec{x}, \vec{p}, t)$ to obtain macroscopic quantities (\textit{e.g.}, number density, energy density, etc). If we adopt the cosmological principle (CP), then the homogeneity requires $f$ to be independent of the position, $\vec{x}$ and the isotropy makes $f$ a function of the magnitude of momentum $p =|\vec{p}|$. Thus, the local number density of particles  in real space is given by
\begin{align}
n = \frac{g}{(2 \pi \thbar)^3} \int_{0}^{\infty} d^3 p f(p) \label{nthermal} \,.
\end{align}

For weakly interacting particles, one can ignore the interaction energies between particles and thus the energy-momentum relation given in Eq.~\eqref{Pmu2} can be used to give the energy of particles. Then, the energy density and the pressure are defined by
\begin{align}
\rho \tc^2 &= \frac{g}{(2 \pi \thbar)^3} \int_{0}^{\infty} d^3 p f(p) E(p) \quad , \quad
P = \frac{g}{(2 \pi \thbar)^3} \int_{0}^{\infty} d^3 p f(p) \frac{p^2}{3 E} \label{rhoPthermal}  \,.
\end{align}
If the particles exchange energy and momentum efficiently, then a system of particles is in kinetic equilibrium, and distribution functions are given by the Fermi-Dirac or Bose-Einstein distributions for fermions and bosons, respectively. In the early universe, the chemical potentials of all particles were so small that one could neglect them, and thus the distribution functions are given by
\begin{align}
f(p) = \frac{1}{\exp [E/(k_{\TB} T)] \pm 1} \label{fp} \,,
\end{align}
where the $+$ sign and the $-$ one are for fermions and bosons, respectively. From the above equations~\eqref{nthermal} -~\eqref{fp}, one can obtain the number densities, energy densities, and the pressures of relativistic and non-relativistic particles
\begin{align}
n &= \begin{cases} \frac{g}{\pi^2} \left( \frac{k_{\TB} T}{\thbar \tc} \right)^3 \frac{3}{4} \zeta(3) & \text{fermion} \\
\frac{g}{\pi^2} \left( \frac{k_{\TB} T}{\thbar \tc} \right)^3 \zeta(3) & \text{boson} \\
g \left( \frac{1}{2 \pi} \frac{m_{\rs} \tc^2}{\thbar \tc} \frac{k_{\TB}T}{\thbar \tc} \right)^{\frac{3}{2}}  e^{-\frac{m_{\rs} \tc^2}{k_{\TB} T}} & \text{non-relativistic} \end{cases} \label{nth} \,, \\
\rho \tc^2 &= \begin{cases} \frac{g \pi^2}{30} \frac{ \left( k_{\TB} T \right)^4}{\left( \thbar \tc \right)^3} \frac{7}{8} & \text{fermion} \\
\frac{g \pi^2}{30} \frac{ \left( k_{\TB} T \right)^4}{\left( \thbar \tc \right)^3}  & \text{boson} \\
n \left( m_{\rs} \tc^2 + \frac{3}{2} k_{\TB} T \right) \approx n m_{\rs} \tc^2 & \text{non-relativistic} \end{cases} \,, \quad
P = \begin{cases} \frac{1}{3} \rho \tc^2 & \text{fermion} \\
\frac{1}{3} \rho \tc^2  & \text{boson} \\
n T \approx 0 & \text{non-relativistic} \end{cases} \label{rhoc2Pth} \,.
\end{align}

They are local quantities whose cosmological evolution information is determined by $T$ and $\tc$. As we mentioned, $\tc = \tc_0 a^{b/4}$, $T = T_{0} a^{-1}$, and $k_{\TB}$ is a constant. Also, the number density is the total number of particles, $N$, divided by the volume, $V$. In the expanding universe, it is given by $n = N / V = N / (V_{\cm} a^{3})$ where $V_{\cm}$ means the comoving volume. It is most natural to assume that the total number of particles and energy remain constant as the Universe expands. The conservations of them provide the cosmological evolutions of other physical constants (quantities) as
\begin{align}
\thbar = \thbar_{0} a^{-\frac{b}{4}} \quad , \quad m_{\rs} = m_{\rs 0} a^{-\frac{b}{2}} \label{hbarmrs} \,,
\end{align}
where $\thbar_{0}$ and $m_{\rs 0}$ denote the present values of the reduced Planck constant and the rest mass, respectively. Consequently, we also obtain the mass density is redshifted as $\rho \propto a^{-3(1+\omega_i) - b/2}$ from Eq.~\eqref{rhoc2Pth}. We obtain consistent results when considering cosmology in Section~\ref{sec:cosmology}. We emphasize that the relations in Eq.~\eqref{hbarmrs} depend on our assumptions on conservations of the total number of particles and their energy. We call these requirements the \textbf{m}inimal \textbf{e}xtension of VSL and dubbed this model as the \textit{meVSL}. Thus, if one chooses other conditions as the required physical principle, one may obtain other cosmological redshift relations for $\thbar$ and $m_{\rs}$.

\subsection{Lorentz transformation and Lorentz covariance}
\label{subsec:LT} 

We briefly review the LT in this subsection. Thus, this subsection seems to be an unnecessary repetition. However, we want to emphasize that the equality of the local speed of light is a condition for satisfying LI. Thus, any model with the cosmic time-varying speed of light is safe from violating the LI. From the translational symmetry of space and time, a transformation of the coordinates $x$ and $t$ from one inertial reference frame $\mathcal{O}$ to $x'$ and $t'$ in another reference frame $\mathcal{O}'$ should be linear functions. This fact can be expressed as
\begin{align}
\begin{pmatrix} t' \\ x' \end{pmatrix} &= \begin{pmatrix} A & B \\ C & D \end{pmatrix} \begin{pmatrix} t \\ x \end{pmatrix} \label{Lor1} \,.
\end{align}
If one chooses that $x' = 0$ is the origin of $\mathcal{O}'$ and it moves with velocity $v$ relative to $\mathcal{O}$ so that $x = vt$, then one obtains $C = - v D$. One can also choose $x = 0$ as the origin of $\mathcal{O}$, and it moves with velocity $-v$ relative to $\mathcal{O}'$ so that $x' = -vt'$, then one obtains $t' = D t = At $ and thus $A = D$. With these relations, one can rewrite $t'= A (t + F x)$ where $F = B/A$. If one changes the notation $A = \gamma$, then one has
\begin{align}
\begin{pmatrix} t' \\ x' \end{pmatrix} &= \gamma[v] \begin{pmatrix} 1 & F[v] \\ - v & 1 \end{pmatrix} \begin{pmatrix} t \\ x \end{pmatrix} \label{Lor6} \,.
\end{align}
A combination of two Lorentz transformations also must be a Lorentz transformation (form a group). If a reference frame $\mathcal{O}'$ moving relative to $\mathcal{O}$ with velocity $v_1$ and a reference frame $\mathcal{O}''$ moving relative to $\mathcal{O}'$ with velocity $v_2$ then
\begin{align}
\begin{pmatrix} t'' \\ x'' \end{pmatrix}
&= \gamma[v_2] \gamma[v_1] \begin{pmatrix} 1 - F[v_2] v_1 & F[v_1] + F[v_2] \\ -v_2 - v_1 & 1 - F[v_1] v_2 \end{pmatrix} \begin{pmatrix} t \\ x \end{pmatrix}  \,.\label{Lor7Double}
\end{align}
One can compare the coefficients in Eqs.~\eqref{Lor6} and ~\eqref{Lor7Double} to obtain
\begin{align}
1 - F[v_2] v_1 = 1 - F[v_1] v_2 \Rightarrow \frac{F[v_1]}{v_1} = \frac{F[v_2]}{v_2} \equiv \alpha^{-1} = \text{constant} \label{Lor8alpha} \,.
\end{align}
By inserting Eq.~\eqref{Lor8alpha} into Eq.~\eqref{Lor6}, one obtains
\begin{align}
\begin{pmatrix} t' \\ x' \end{pmatrix} &= \gamma[v] \begin{pmatrix} 1 & \frac{v}{\alpha} \\ - v & 1 \end{pmatrix} \begin{pmatrix} t \\ x \end{pmatrix} \,. \label{Lor9}
\end{align}
If one performs a LT from the reference frame $\mathcal{O}$ to $\mathcal{O}'$ and then transforms back from $\mathcal{O}'$ to $\mathcal{O}$, one obtains the relation $1 + v^2/\alpha \equiv \gamma[v]^{-2}$. Finally, if one put $\alpha = -\tc^2$, then the LT is given by
\begin{align}
\begin{pmatrix} \tc t' \\ x' \end{pmatrix} &=  \frac{1}{\sqrt{1 - \tbeta^2}} \begin{pmatrix} 1 & - \tbeta \\ -\tbeta & 1 \end{pmatrix} \begin{pmatrix} \tc t \\ x \end{pmatrix} \label{Lor12} \,,
\end{align}
where $\tbeta \equiv v/\tc$. In the meVSL model, the local value of the speed of light is constant, so thus the LT is well established in the meVSL model.

Due to the Lorentz symmetry, the laws of physics are the same for all inertial observers. Thus, experimental results are independent of the orientation or the magnitude of the observer's velocity. As we mentioned, Lorentz covariance means that a Lorentz covariant scalar stays the same under LTs. It is also said to be a LI. An equation is Lorentz covariant if stated in terms of Lorentz covariant quantities. Lorentz covariance holds in any inertial frame if they hold in one inertial frame. Local Lorentz covariance, which follows from GR, refers to Lorentz covariance applying only locally in an infinitesimally small region of spacetime at every point. And the meVSL satisfies Lorentz covariance as shown in this section~\ref{sec:SR}.
\section{Geodesics}
\label{sec:geodesics}

Now, we extend the meVSL model in the curved spacetime. A geodesic is a generalized idea of a ``straight line" in curved spacetime in GR. It is the shortest path that a physical object follows when moving freely, given by the metric of the given spacetime. It means that a freely moving or falling particle always follows a geodesic. In this section, we investigate both the geodesic equation and the geodesic deviation equation in the meVSL model.

\subsection{Geodesic equation}
\label{subsec:geodesicEq}

We adopt the equivalence principle in the meVSL model, and the derivation of the geodesic equation is directly from it. A free-falling particle does not accelerate in the neighborhood of a point-event for a freely falling coordinate system, $X^{\mu}$. Setting $X^0 \equiv c \tau$, one has the following equation that is locally applicable in free fall
\begin{align}
\frac{d^2 x^{\lambda}}{ d \tau^2} &= - \Gamma^{\lambda}_{\alpha\beta} \frac{d x^{\alpha}}{d \tau} \frac{d x^{\beta}}{d \tau}  + \frac{d \ln \ttc}{d\tau} \frac{dx^{\lambda}}{d\tau} \quad , \quad \text{where} \quad \Gamma^{\lambda}_{\alpha\beta} = \frac{\partial^2 X^{\mu}}{\partial x^{\alpha} \partial x^{\beta}} \frac{ \partial x^{\lambda}}{\partial X^{\mu}} \label{geod22} \,.
\end{align}
One can rewrite Eq.~\eqref{geod22} in terms of the time coordinate $t$
\begin{align}
\frac{d^2 x^{\lambda}}{d t^2} &= -  \Gamma^{\lambda}_{\alpha\beta} \frac{d x^{\alpha}}{d t} \frac{d x^{\beta}}{d t} - \frac{d \ln \gamma}{d t} \frac{d x^{\lambda}}{d t} + \frac{d \ln \ttc}{dt} \frac{dx^{\lambda}}{dt} = -  \Gamma^{\lambda}_{\alpha\beta} \frac{d x^{\alpha}}{d t} \frac{d x^{\beta}}{d t} + \frac{d \ln [\tc/\tgam]}{d t} \frac{d x^{\lambda}}{d t} \label{geod3} \,,
\end{align}
where we use $\gamma/ \ttc = \tgam/\tc$ from Eq.~\eqref{gamma}. By expressing the last term with four coordinate, one obtains
\begin{align}
\frac{d^2 x^{\lambda}}{d t^2} &= -  \Gamma^{\lambda}_{\alpha\beta} \frac{d x^{\alpha}}{d t} \frac{d x^{\beta}}{d t} + \Gamma^{0}_{\alpha\beta} \frac{d x^{\alpha}}{d t} \frac{d x^{\beta}}{d t}  \frac{d x^{\lambda}}{d t} + \frac{d \ln \tc}{dt} \frac{dx^{\lambda}}{dt} \,. \label{geod4}
\end{align}
Thus, compared to the GR, the meVSL has the correction term due to $d \ln \tc/dt$. To estimate the effect of this contribution, we apply the geodesic equation to the Newtonian limit. Because the particle in the Newtonian limit is moving slowly, the time component dominates the spatial components, and every term containing one or two spatial four-velocity components will be dwarfed by the term containing two time components. Therefore, we can take the approximation
\begin{align}
\frac{d^2 x^{\lambda}}{d t^2} &\approx -  \Gamma^{\lambda}_{00} \tc^2 + \frac{d \ln \tc}{dt} \frac{dx^{\lambda}}{dt} \label{geod2New} \,.
\end{align}
If the gravitational field is weak enough, then spacetime will be only slightly deformed from the gravity-free Minkowski space of SR, and we can consider the spacetime metric as a small perturbation from the Minkowski metric $\eta_{\mu\nu}$
\begin{align}
g_{\mu\nu} &= \eta_{\mu} + h_{\mu\nu} \quad , \quad |h_{\mu\nu} | \ll 1 \quad , \quad
g_{00,i} = h_{00,i} \label{Minko2} \,.
\end{align}
Because we are interested in the Newtonian $3$-D space, we can then replace the four-dimensional index $\lambda$ in Eq.~\eqref{geod2New} by the spatial component, $i$
\begin{align}
\frac{d^2 \vec{x}}{dt^2} = - \vec{\nabla} \Phi + \vec{v} \frac{d \ln \tc}{dt}  \label{geod2New3}  \,,
\end{align}
where we define $2 \Phi = h_{00} \tc^2$. Now we can estimate the magnitudes of each term in the right-hand side of Eq.~\eqref{geod2New3}
\begin{align}
h_{00} &= 2 \frac{\Phi}{\tc^2} = \frac{2}{\tc^2} \frac{G M_{\text{Earth}}}{R_{\text{Earth}}} \approx 1.39 \times 10^{-9}  \label {estiNew} \,, \\
\left| \nabla \Phi \right| &\approx \frac{G M_{\text{Earth}}}{R_{\text{Earth}}^2} \sim 10 [\text{m}/\text{s}^2] \quad , \quad \left| v \frac{d \ln \tc}{dt} \right|  = \left| v b H \right| \sim  10^{-15}  [\text{m}/\text{s}^2] \label{extiNew2} \,,
\end{align}
where we use $v \sim 100$ km/s and $\tc/\tc_0 = a^{b/4}$ with $b \sim 10^{-2}$ which will be obtained later. Thus, the geodesic equation of the meVSL model deviates from that of GR, but that effect is negligible. However, we should emphasize that the local variation of $\tc$ is ignored in the meVSL model, and the correction term in Eq.~\eqref{geod3} should not be considered for the local observer.

\subsection{Geodesic deviation equation}
\label{subsec:geodesicdeviation}

Now, we consider how the evolution of the separation measured between two adjacent geodesics, also known as geodesic deviation can be modified in the meVSL model. We consider two particles following two very close geodesics. We denote their respective path as $x^{\mu}(\tau)$ (reference particle) and $y^{\mu}(\tau) = x^{\mu}(\tau)+ \xi^{\mu}(\tau)$ (second particle) where $\xi^{\mu}$ refers to the deviation four-vector joining one particle to the other at each given time $\tau$ ($\xi^{\mu} \ll x^{\mu}$). The relative acceleration $A^{\mu}$ of the two objects is defined, roughly, as the second derivative of the separation vector $\xi^{\mu}$ as the objects advance along their respective geodesics. As each particle follows a geodesic as in Eq.~\eqref{geod22}, the equations of their respective coordinates are given by
\begin{align}
&\frac{d^2 x^{\alpha}}{d \tau^2} + \Gamma^{\alpha}_{\mu\nu} (x^{\alpha}) \frac{d x^{\mu}}{d \tau} \frac{d x^{\nu}}{d \tau} = \frac{d \ln \ttc}{d \tau} \frac{d x^{\alpha}}{d \tau} \label{d2xaodt2} \,, \\
&\frac{d^2 x^{\mu}}{d \tau^2} + \frac{d^2 \xi^{\mu}}{d \tau^2} + \left( \Gamma^{\alpha}_{\mu\nu} (x^{\alpha} )  + \partial_{\sigma} \Gamma^{\alpha}_{\mu\nu} \xi^{\sigma} \right) \left( \frac{d x^{\mu}}{d \tau} + \frac{d \xi^{\mu}}{d \tau}\right) \left( \frac{d x^{\nu}}{d \tau} + \frac{d \xi^{\nu}}{d \tau}\right) = \frac{d \ln \ttc}{d \tau} \left( \frac{d x^{\alpha}}{d \tau} + \frac{d \xi^{\alpha}}{d \tau}\right)  \label{d2xxiaodt2} \,,
\end{align}
If one subtracts Eq.~\eqref{d2xaodt2} from Eq.~\eqref{d2xxiaodt2}, then one obtains equation for $\xi$ upto the linear order of $\xi$ ({\it i.e.}, $\mathcal{O}(\xi)$)
\begin{align}
\frac{d^2 \xi^{\alpha}}{d \tau^2} + 2 \Gamma^{\alpha}_{\mu\nu} U^{\mu} \frac{d \xi^{\nu}}{d \tau} + \partial_{\sigma} \Gamma^{\alpha}_{\mu\nu} U^{\mu} U^{\nu} \xi^{\sigma} = \frac{d \ln \ttc}{d \tau} \frac{d \xi^{\alpha}}{d \tau} \label{d2xiodt2} \,,
\end{align}
where we use the torsion free condition $\Gamma^{\alpha}_{\mu\nu} = \Gamma^{\alpha}_{\nu\mu}$. We now have an expression for $d \xi^{\mu}/ d \tau$, but this is not the total derivative of the four-vector $\xi^{\mu}$ since its derivative could also get a contribution from the change of the basis vectors $e_{\alpha}$ as the object moves along its geodesic. To get the total derivative, we have
\begin{align}
\left( \frac{d \xi}{d \tau} \right)^{\alpha} &= \frac{d \xi^{\alpha}}{d \tau} + \Gamma^{\alpha}_{\mu\sigma} U^{\mu} \xi^{\sigma} \quad , \quad \text{where} \quad \frac{d e_{\alpha}}{d \tau} \equiv \Gamma^{\sigma}_{\mu\alpha} U^{\mu} e_{\sigma}  \label{dxiodtaualpha} \,.
\end{align}
Since ${\bf \xi}$ is a four-vector, its derivative for proper time is also a four-vector, so we can find the second absolute derivative by using the same development as for the first-order derivative
\begin{align}
\left( \frac{d^2 \xi}{d \tau^2} \right)^{\alpha}  &= - \left( \partial_{\sigma} \Gamma^{\alpha}_{\mu\nu} - \partial_{\nu} \Gamma^{\alpha}_{\mu\sigma} + \Gamma^{\alpha}_{\gamma\sigma} \Gamma^{\gamma}_{\nu\mu} - \Gamma^{\alpha}_{\nu\sigma} \Gamma^{\gamma}_{\mu\sigma} \right) U^{\mu} U^{\nu} \xi^{\sigma} + \left( \frac{d \ln \ttc}{d \tau} \right)^2 \xi^{\alpha} \nonumber \\
&\equiv - \tensor{R}{^\alpha_\mu_\sigma_\nu} U^{\mu} U^{\nu} \xi^{\sigma} + \left( \frac{d \ln \ttc}{d \tau} \right)^2 \xi^{\alpha}  \label{d2xiodt2alpha} \,,
\end{align}
where we use Eqs.~\eqref{d2xiodt2} and ~\eqref{dxiodtaualpha} in the third equality. Eq.~\eqref{d2xiodt2alpha} is the geodesic deviation equation of the VSL model. Compared to the GR, we obtain the additional term related to $d \ttc/ d \tau$. However, for the local observer, this modification term is ignored, and the geodesic deviation equation is the same as that of the GR. Thus, the meVSL predicts the same polarization of gravitational waves (GWs) as the GR. In other words, if one breaks the equivalence principle, one needs to consider the effect of VSL in the GW polarization detections.
\section{Cosmology}
\label{sec:cosmology}

We now investigate the cosmology in the meVSL model. Thus, the Einstein-Hilbert action based on the meVSL model is rewritten as
\begin{align}
\tSS &= \int \Biggl[ \frac{\tc^4}{16 \pi \tG} \left( \tRR - 2 \Lambda \right)  + \mathcal{L}_{\text{m}} \Biggr] \sqrt{-g} dt d^3x \equiv \int \Biggl[ \frac{1}{2 \tkapp} \left( \tRR - 2 \Lambda \right)  + \mathcal{L}_{\text{m}} \Biggr] \sqrt{-g} dt d^3x \label{tSHmp} \,,
\end{align}
where $g = \det (g_{\mu\nu})$ is the determinant of the metric tensor, $\tRR$ is the Ricci scalar, $\tG$ is the time-varying Newton's gravitational constant, and $\tc$ is the time-varying speed of light. We show that as we allow the speed of light to change with time, so does $\tG$ to obtain the consistent theory. This becomes obvious when we consider the field equation. We obtain the field equations by using the fact that the variation of the action for the inverse metric must be zero to recover Einstein's field equation. By doing this, we obtain
\begin{align}
0 &= \delta \tSS = \int \frac{\sqrt{-g}}{2 \tkapp} \left[ \tRR_{\sigma\nu} - \frac{1}{2} g_{\sigma\nu} \left( \tRR - 2 \Lambda \right) - \tkapp T_{\sigma\nu} \right] \delta g^{\sigma\nu} dtd^3 x + \int \frac{\sqrt{-g}}{2 \tkapp} \left[ \nabla_{\sigma} \nabla_{\nu} - g_{\sigma\nu} \Box \right] \delta g^{\sigma\nu} dtd^3 x \label{deltatSHmp2} \,,
\end{align}
where $T_{\mu\nu}$ is the stress-energy tensor and the second term on the right-hand side is the so-called Palatini identity term. If one uses the integrate by part, then this term gives the contributions such as $\nabla_{\mu} \nabla_{\nu} (\tkapp^{-1})$. This is the case for the Brans-Dicke theory when there exists a coupling between the Ricci scalar and the scalar field. Thus, if we want to avoid these additional unexpected dynamical contributions, then we should have the constraint on the meVSL model as
\begin{align}
\frac{d}{\tc dt} \frac{1}{\tkapp} = 0 \quad \Longrightarrow \quad \tkapp [a] = \text{const} \quad \Longrightarrow \quad \tc[a]^4 \propto \tG[a] \label{tkappaconstmp}  \,,
\end{align}
where $t$ is the cosmic time. This is the main constraint of the meVSL model. We specify this constraint equation by using scale factor $a$ when we adopt the energy conservation of matter ({\it i.e.}, Bianchi identity) later. One can also proceed with other general kinds of VSL models without this constraint. In that case, one should include the terms from this Palatini identity term. However, we adopt this minimal model in this manuscript in order not to spoil the success of the GR.

\subsection{FLRW solution}
\label{subsec:FLRWsol}

We now investigate the cosmology of the meVSL model for the FLRW metric given by
\begin{align}
g_{\mu\nu} = \text{diag} \left(-1 , \frac{a^2}{1-kr^2} , a^2 r^2, a^2 r^2 \sin^2 \theta \right)
\,. \label{FLRWmetric}
\end{align}
The line element is written as
\begin{align}
ds^2 = - \tc dt^2 + a^2 \gamma_{\ij} dx^{i} dx^{j} \label{dsFRW} \,.
\end{align}
Then, Riemann curvature tensors, Ricci tensors, and Ricci scalar curvature are given by
\begin{align}
&\tensor{\tRR}{^0_i_0_j} = \frac{g_{ij}}{\tc^2} \left( \frac{\ddot{a}}{a} - H^2 \frac{d \ln \tc}{d \ln a} \right) , \quad
\tensor{\tRR}{^i_0_0_j} = \frac{\delta^{i}_{j}}{\tc^2} \left( \frac{\ddot{a}}{a} - H^2 \frac{d \ln \tc}{d \ln a} \right) , \nonumber \\ 
&\tensor{\tRR}{^i_j_k_m} =  \left( \frac{H^2}{\tc^2} + \frac{k}{a^2} \right) \left( \delta^{i}_{k} g_{jm} - \delta^{i}_{m} g_{jk} \right) \label{tRijkmApp} \,, \\
& \tRR_{00} = - \frac{3}{\tc^2} \left( \frac{\ddot{a}}{a} - H^2 \frac{d \ln \tc}{d \ln a} \right)  \quad , \quad
\tRR_{ii} = \frac{g_{ii}}{\tc^2} \left( 2 \frac{\dot{a}^2}{a^2} + \frac{\ddot{a}}{a} + 2 k \frac{\tc^2}{a^2} - H^2 \frac{d \ln \tc}{d \ln a} \right) \label{tR00mp} \,, \\
& \tRR = \frac{6}{\tc^2} \left( \frac{\ddot{a}}{a} + \frac{\dot{a}^2}{a^2} + k \frac{\tc^2}{a^2} - H^2 \frac{d \ln \tc}{d \ln a} \right) \label{tRmp} \,.
\end{align}

The stress-energy tensor of a perfect fluid in thermodynamic equilibrium is given by
\begin{align}
T_{\mu\nu} = \left( \rho + \frac{P}{\tc^2} \right) U_{\mu} U_{\nu} + P g_{\mu\nu} \label{Tmunump} \,.
\end{align}
In an inertial frame of reference comoving with the fluid, the fluid's four-velocity becomes $ U^{\mu} = \left(\tc\,,\vec{0} \right)$. Thus,
the energy-momentum tensor is given by
\begin{align}
T_{\mu}^{\nu} = \text{diag} \left(-\rho \tc^2, P, P, P \right) \label{Tmunumpdia} \,.
\end{align}
One needs to investigate Bianchi's identity to provide the energy conservation given by
\begin{align}
&\rho_i \tc^{2} = \rho_{i0} \tc_0^2 a^{-3 (1 + \omega_i)} \label{rhotc2mp} \,,
\end{align}
where $\tc_0$ is the present value of the speed of light, $\rho_{i0}$ is the present value of mass density of the i-component, and we use $a_0 = 1$.

We obtain EFEs including the cosmological constant by using Eqs.~\eqref{tR00mp}-~\eqref{rhotc2mp}
\begin{align}
& \frac{\dot{a}^2}{a^2} + k \frac{\tc^2}{a^2}  -\frac{ \Lambda \tc^2}{3} = \frac{8 \pi \tG}{3} \sum_{i} \rho_i \label{tG00mp} \,, \\ 
& \frac{\dot{a}^2}{a^2} + 2 \frac{\ddot{a}}{a} +  k \frac{\tc^2}{a^2} - \Lambda \tc^2 - 2 H^2 \frac{d \ln \tc}{d \ln a}  = -8 \pi \tG \sum_{i} \frac{P_i}{\tc^2} = -8 \pi \tG \sum_{i} \omega_{i} \rho_{i} \label{tG11mp} \,. 
\end{align}
One substracts Eq.~\eqref{tG00mp} from Eq.~\eqref{tG11mp} to obtain 
\begin{align}
& \frac{\ddot{a}}{a} = -\frac{4\pi \tG}{3} \sum_{i} \left( 1 + 3 \omega_i \right) \rho_i  + \frac{\Lambda \tc^2}{3} + H^2 \frac{d \ln \tc}{d \ln a} \label{tG11mG00mp} \,. 
\end{align} 
From Eqs.~\eqref{tG00mp} and \eqref{tG11mG00mp}, one can understand that the expansion velocity of the Universe does depend on not only the speed of light $\tc$ but also both on $\tG$  and on $\rho$. Also, so does the acceleration of the expansion of the Universe. 

\begin{figure*}
\centering
\vspace{1cm}
\begin{tabular}{cc}
\includegraphics[width=0.5\linewidth]{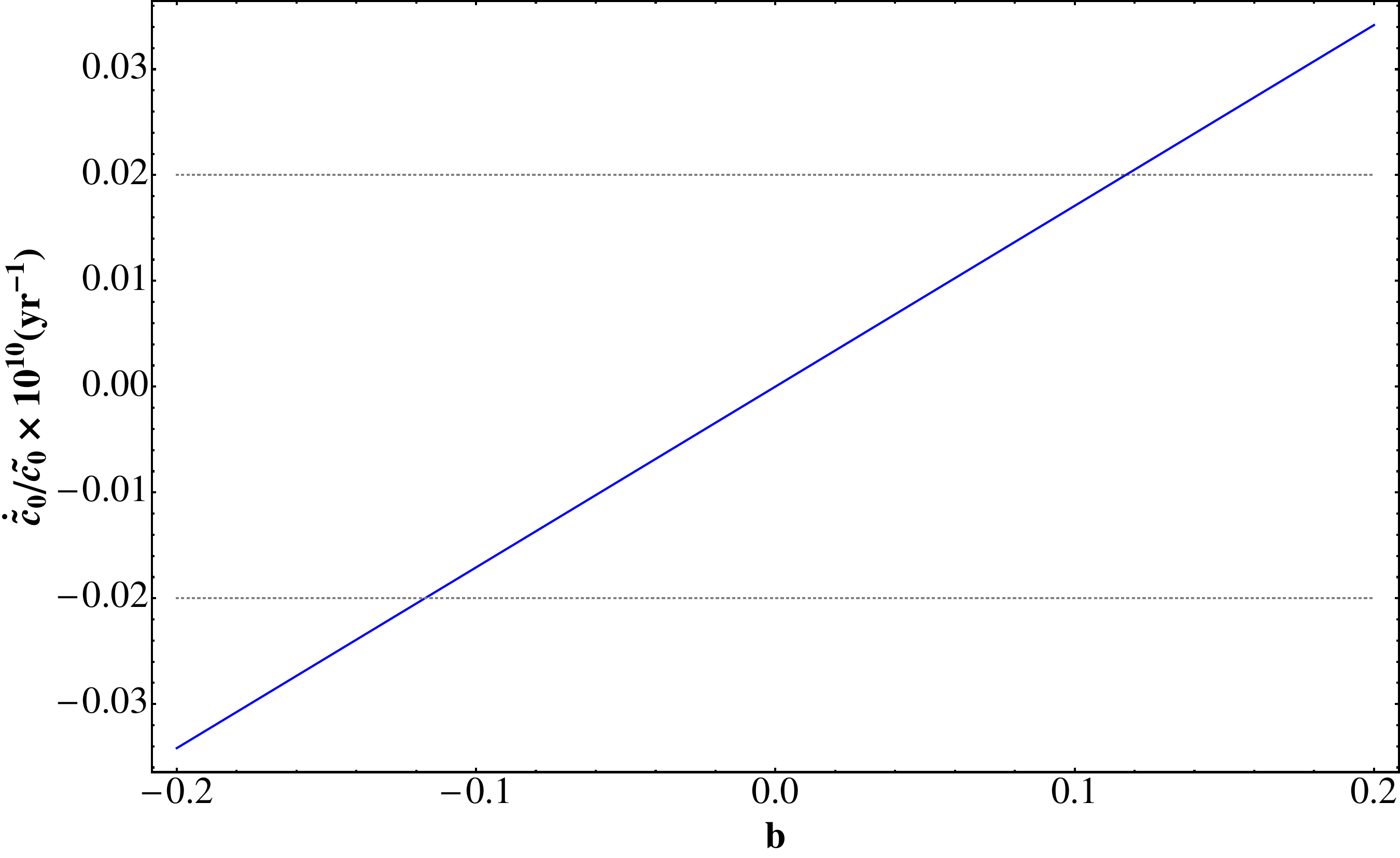} &
\includegraphics[width=0.49\linewidth]{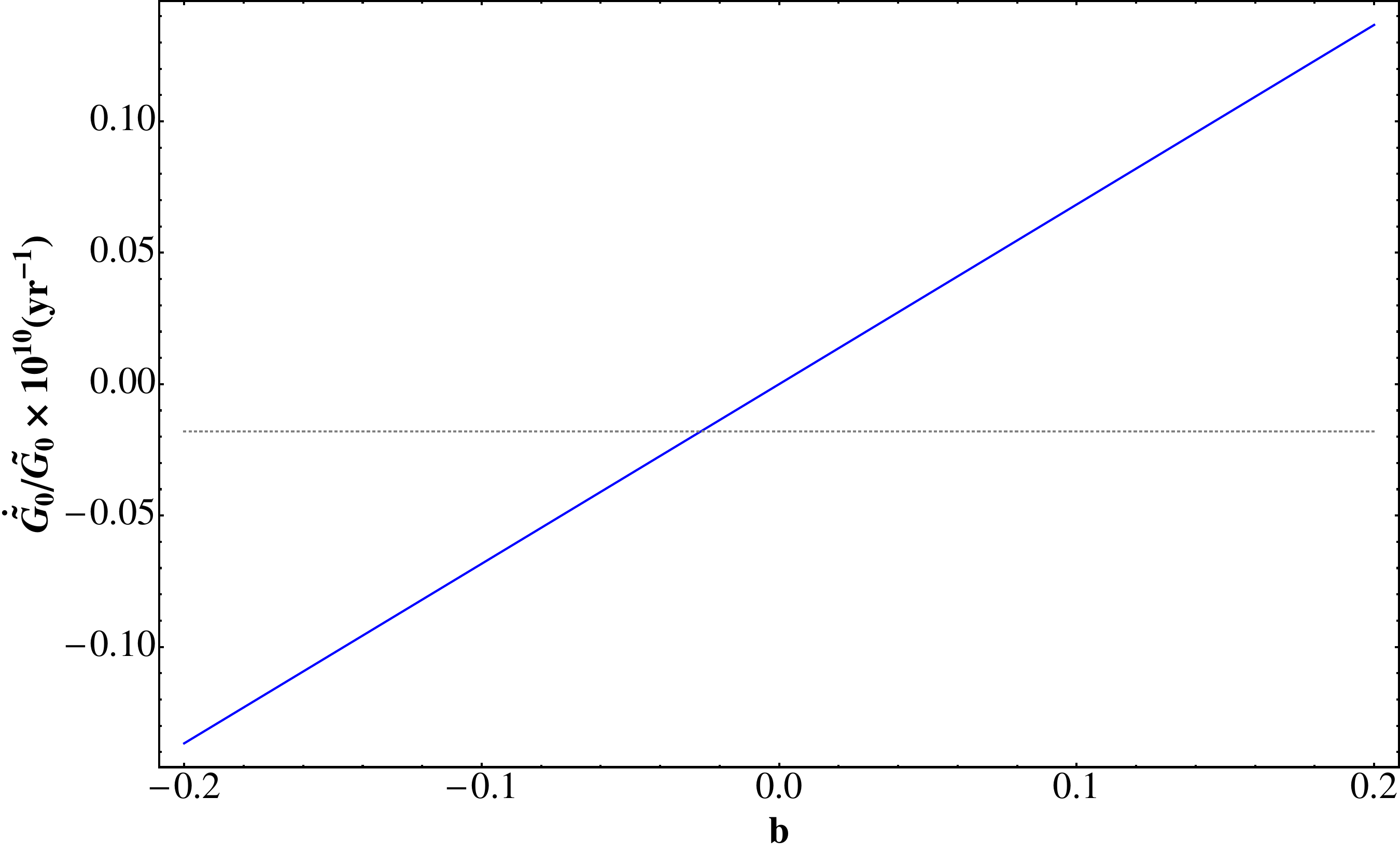}
\end{tabular}
\vspace{-0.5cm}
\caption{The present values of time variation of physical constants as a function of $b$. a) The values of $\dot{\tc}_0/\tc_0$ multiplied by $10^{10}$ for the different values of $b$ in the units $yr^{-1}$. a) The values of $\dot{\tG}_0/\tG_0$ multiplied by $10^{10}$ for the different values of $b$ in the units $yr^{-1}$.} \label{fig-1}
\vspace{1cm}
\end{figure*}

Eq.~\eqref{tG11mG00mp} also should be obtained by differentiating Eq.~\eqref{tG00mp} with respect to the cosmic time, $t$ and using Eq.~\eqref{rhotc2mp}. This provides the relation between $\tG$ and $\tc$ as
\begin{align}
\frac{d \ln \tG}{d \ln a} = 4 \frac{d \ln \tc}{d \ln a} \equiv b = \text{const.} \quad \Longrightarrow \quad \frac{\tG}{\tG_0} = \left( \frac{\tc}{\tc_0} \right)^4 = \left( \frac{a}{a_0} \right)^{b} = a^{b} \label{dotGoGmp} \,.
\end{align}
From the above Eq.~\eqref{dotGoGmp}, one can obtain the expressions for the time variations of $\tc$ and $\tG$ as
\begin{align}
\frac{\dot{\tG}}{\tG} = b H \quad , \quad \frac{\dot{\tc}}{\tc} = \frac{b}{4} H \label{dotcoc} \,.
\end{align}
Thus, the present values of both the ratio of the time variation of the gravitational constant and that of the speed of light are given by
\begin{align}
\frac{\dot{\tG}_0}{\tG_0} = b H_0 \quad , \quad \frac{\dot{\tc}_0}{\tc_0} = \frac{b}{4} H_0 \label{dotcoc0} \,.
\end{align}
One way to obtain the limits on the time variations of both the speed of light and the gravitational constant is by using the evolution of the radius of Mercury \cite{Racker:2007hj}. The radius of a planet is determined by the hydrostatic equilibrium equation besides an equation of state and boundary conditions. The presence of the time-varying speed of light causes in general time variations in the radius of a planet. On the other hand, one uses different topographical observations to estimate the actual change in the size of several bodies of the Solar system. There exists a stringent bound for the radius of Mercury. It has not changed more than 1 km in the last $3.9 \times 10^9$ years \cite{McElhinny:1978na}. This fact provides a bound for the temporal variation of the speed of light. The hydrostatic equilibrium equation is equivalent to another equation in which the temporal dependence exists only in $G$. The present bound on the time variation of $\tc$ is $\dot{\tc}_0/\tc_0 = 0 \pm 2 \times 10^{-12} \text{yr}^{-1}$ \cite{Racker:2007hj}. This provides the bound on $b$ as  $-0.11 \leq b \leq 0.11$. The present bound on the time variation of $\tG$ is $\dot{\tG}_0/\tG_0 \leq 1.8 \times 10^{-12} \text{yr}^{-1}$ \cite{GarciaBerro:2011wc}. This gives the bound $-0.026 \leq b$. We show this in the figure.~\ref{fig-1}. In the left panel of Fig.~\ref{fig-1}, the present values of $\dot{\tc}/\tc$ for the different values of $b$ are depicted. The value of $\dot{\tc}_0/\tc_0$ is proportional to the present value of the Hubble parameter $H_0 (\sim 6.83 \times 10^{-11}) \text{yr}^{-1}$ as shown in Eq.~\eqref{dotcoc0}. The horizontal dotted lines indicate the bound on $\dot{\tc}_0/\tc_0$ in the reference \cite{Racker:2007hj}. The sign of $b$ can be determined if $\dot{\tc}_0/\tc_0$ is obtained. We also show the behavior of $\dot{\tG}_{0} / \tG_{0}$ as a function of $b$ in the right panel of Fig.~\ref{fig-1}. Because it is proportional to $H_0$ as for the time variation of the speed of light, this behavior is the same as that of $\dot{\tc}_0/\tc_0$ except the slope is increased by factor $4$.

Eq.~\eqref{dotGoGmp} is consistent with Eq.~\eqref{tkappaconstmp} and this guarantees the consistency of the theory of meVSL. The above equation~\eqref{dotGoGmp} is one of the main properties of meVSL from which the cosmological evolutions of other quantities are obtained. One adopts Eq.~\eqref{dotGoGmp} into Eq.~\eqref{rhotc2mp} to obtain
\begin{align}
\rho_{i} = \rho_{i0} a^{-3(1+\omega_i)-\frac{b}{2}} \equiv \rho_{i \rs} a^{-3(1+\omega_i)} \label{rhoimp} \,,
\end{align}
where we define $\rho_{i \rs} \equiv \rho_{i 0} a^{-b/2}$ as the rest-mass density of the $i$-component. Thus, the mass density of $i$-component redshifts slower (faster) than that of the GR for a negative (positive) value of $b$. Or, one can interpret this equation as that the rest mass cosmologically evolves as $a^{-b/2}$. For later use, it is convenient to rewrite the equations ~\eqref{tG00mp} and ~\eqref{tG11mG00mp} by using Eqs.~\eqref{rhotc2mp} and~\eqref{dotGoGmp}
\begin{align}
H^2 &= \left( \frac{8 \pi \tG_0}{3} \sum_{i} \rho_{i0} a^{-3(1+\omega_i)} - k \frac{\tc_0^2}{a^2} \right) a^{\frac{b}{2}} 
\equiv \left( \frac{8 \pi \tG_0}{3} \rho_{\cro} - k \frac{\tc_0^2}{a^2} \right) a^{\frac{b}{2}}  \equiv H^{(\GR)2} a^{\frac{b}{2}}  \label{H2mp} \,, \\
\frac{\ddot{a}}{a} &= \left( -\frac{4\pi \tG_0}{3} \sum_{i} \left( 1 + 3 \omega_i \right) \rho_{i0} a^{-3(1+\omega_i)}   \right) a^{\frac{b}{2}}  + \frac{b}{4} H^2 \nonumber \\
	&= \left( -\frac{4\pi \tG_0}{3} \sum_{i} \left( 1 + 3 \omega_i \right) \rho_{i0} a^{-3(1+\omega_i)} + \frac{b}{4} H^{(\GR)2} \right) a^{\frac{b}{2}} \equiv \left( \left( \frac{\ddot{a}}{a} \right)^{(\GR)} + \frac{b}{4} H^{(\GR)2} \right) a^{\frac{b}{2}} \label{ddotaoa} \,,
\end{align}
where we denote $H^{(\GR)}$ is the Hubble parameter of GR, $\Lambda \tc_0^2 = 8 \pi \tG_0 \rho_{\Lambda}$, the equation of state (e.o.s) of the cosmological constant $\omega_{\Lambda} = -1$, and $\rho_{\cro}$ is the critical density to have a flat Universe. 

Henceforth, we limit ourselves to the flat universe (\textit{i.e.}, $k=0$) only. In this case, the deceleration parameter, $q$ can be written as
\begin{align}
q \equiv - \frac{\ddot{a}}{a H^2} = \frac{1}{2} \sum_{i} \left( 1 + 3 \omega_{i} \right) \Omega_{i} - \frac{b}{4} \equiv q^{(\GR)} - \frac{b}{4} \label{q} \,,
\end{align}
where $\Omega_{i} \equiv \rho_i / \rho_{\cro}$ is the mass density contrasts of i-component.
Because both $H^2$ and $\ddot{a}/a$ of meVSL are modified by factor $a^{b/2}$ as shown in Eqs.~\eqref{H2mp} and~\eqref{ddotaoa}, the deceleration parameter does not include an extra scale factor. However, it still has the meVSL effect as $-b/4$. Thus, the value of the deceleration parameter of the meVSL model decreases (increases) by $-b/4$ compared to GR when $b$ is positive (negative). This difference is independent of the cosmic time ({\it i.e.}, the scale factor $a$) and thus gives crucial information when combined with other observational quantities that depend on the cosmic time.

\begin{figure*}
\centering
\vspace{1cm}
\begin{tabular}{cc}
\includegraphics[width=0.5\linewidth]{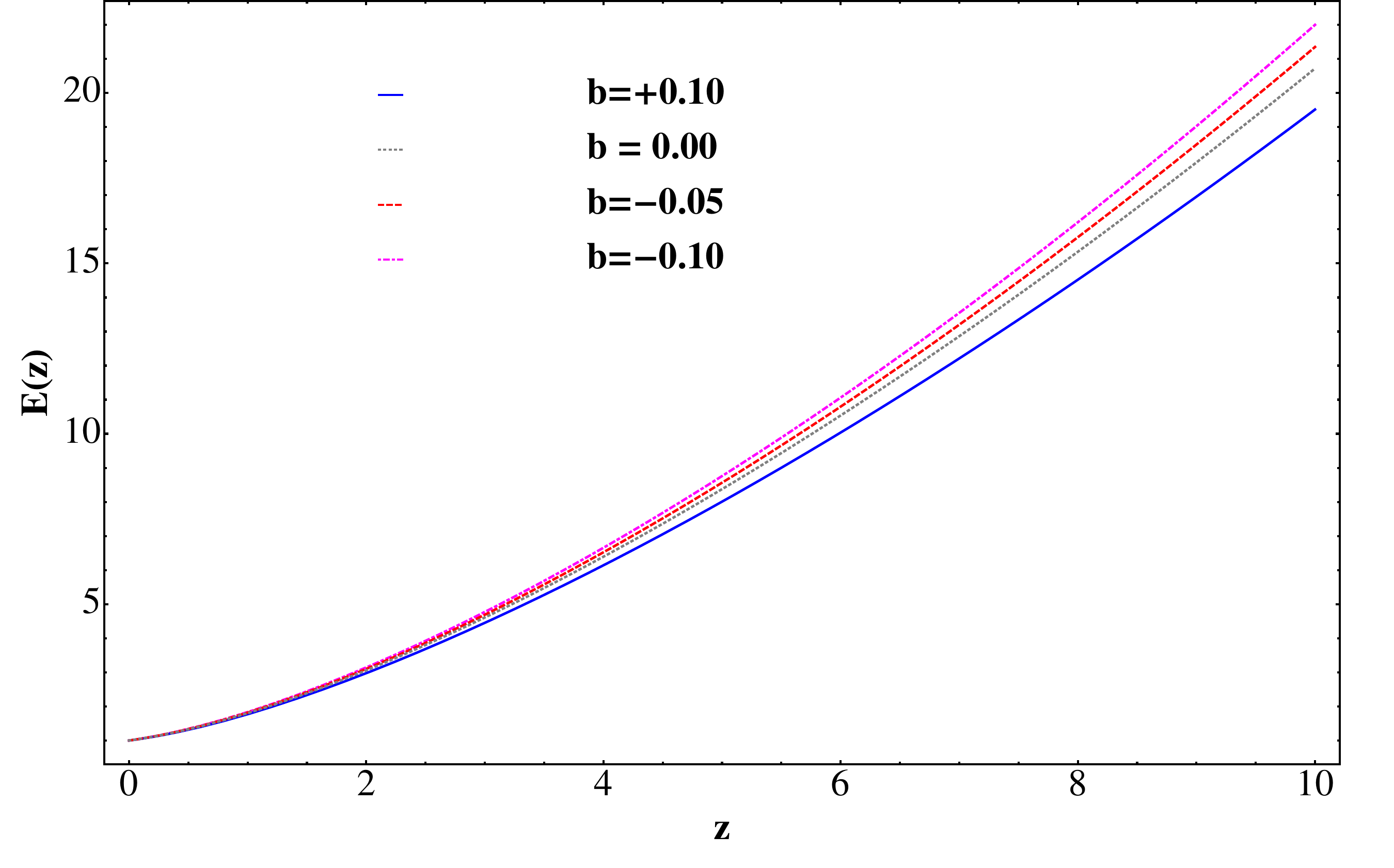} &
\includegraphics[width=0.49\linewidth]{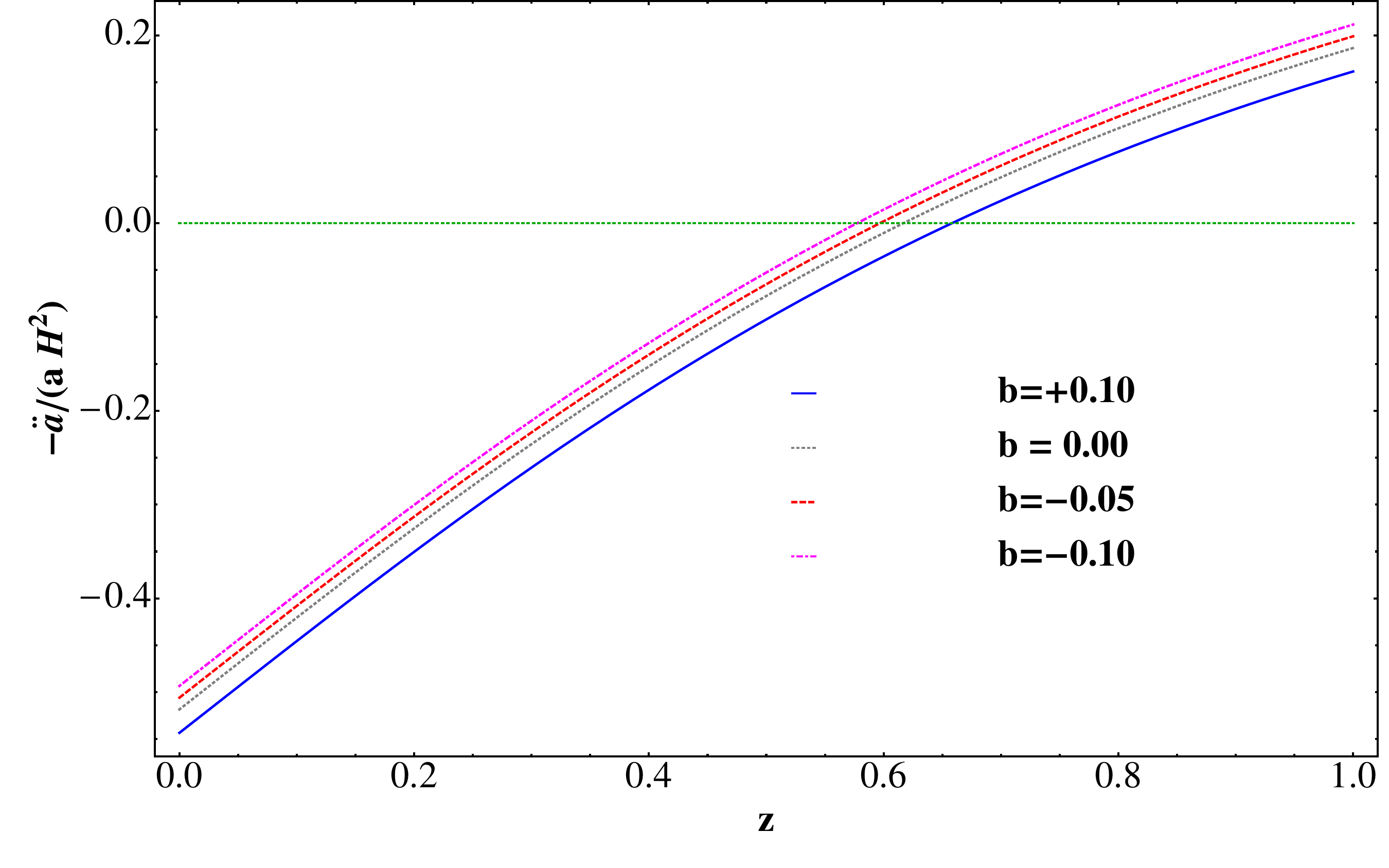}
\end{tabular}
\vspace{-0.5cm}
\caption{Cosmological evolutions of $E(z)$ and $q(z)$ for the different values of $b$. a) The $E(z)$ for $b = 0.1$ (solid), 0 (dotted), -0.05 (dashed), and -0.1 (dot-dashed), respectively. b) $q(z)$ with the same notation as $E(z)$.} \label{fig-2}
\vspace{1cm}
\end{figure*}

One can rewrite the above equations~\eqref{H2mp} and~\eqref{ddotaoa} by dividing them with $H_0^2$
\begin{align}
\frac{H^2}{H_0^2} &\equiv E^2 = \left( \sum_{i} \Omega_{i0} a^{-3(1+\omega_i)} \right) a^{\frac{b}{2}} \equiv E^{(\GR)2} a^{\frac{b}{2}}  \label{H2oH02mp} \,, \\
\frac{\ddot{a}}{a} / H_{0}^2 &= \left( -\frac{1}{2} \sum_{i} \left( 1 + 3 \omega_i \right) \Omega_{i0} a^{-3(1+\omega_i)}  \right) a^{\frac{b}{2}}  + \frac{b}{4}  E^{(\SD)2} a^{\frac{b}{2}} \equiv \left[ \left( \frac{\ddot{a}}{a} / H_0^2 \right)^{(\GR)} + \frac{b}{4}  E^{(\GR)2} \right] a^{\frac{b}{2}} \label{ddotaoaoH02} \,.
\end{align}

Thus, one can find that the present values both $E_0$ and $E_0^{(\GR)}$ are equal to one. However, the present value of the deceleration parameter of meVSL is modified as $q_0 \equiv -\ddot{a}_0/(a_0 H_0^2) = q_0^{(\GR)} + b/4$. Thus, the magnitude of the $q_0$-value depends on the sign of $b$ too in addition to cosmological parameters. These facts are shown in figure~\ref{fig-2}. The values of $E(z)$ for the different values of $b$ are shown in the left panel of Fig.~\ref{fig-2}. Through this manuscript, we adopt best fit cosmological parameters based on Planck $2018$ TT $+$ lowE data Planck \cite{Aghanim:2018eyx}. The ratio of $E(z)$ of meVSL to that of GR is $(1+z)^{-b/4}$. Thus, $E(z)$-values of meVSL are smaller (bigger) than those of GR for the positive (negative) values of $b$. The dot-dashed, dashed, dotted, and solid lines correspond $b = -0.1, -0.05, 0$, and $0.1$, respectively. The percent differences between $E(z)$ and $E^{(\GR)}$ at $z = 10$ (\textit{i.e.}, $\Delta E (z=10) = (E - E^{(\GR)})/E^{(\GR)} \times 100 (\%)$) $= 6.2, 3.0$, and $-5.8$ for $b = -0.1, -0.05$, and $0.1$, respectively. The cosmological evolutions of values of the deceleration parameter, $q$ of meVSL for different values of $b$ are depicted in the right panel of Fig~\ref{fig-2}. As shown in Eq.~\eqref{q}, the deceleration parameter of meVSL is shifted by $-b/4$ compared to that of GR. Thus, the value of $q$ at the given redshift $z$ decreases as the value of $b$ increases. This induces the delay of late-time acceleration of the Universe as the value of $b$ decreases. Again, the dot-dashed, dashed, dotted, and solid lines correspond $b = −0.1, −0.05, 0$, and $0.1$, respectively. One can define the accelerating redshift, $z_{\text{acc}}$ as $q(z_{\text{acc}}) = 0$ and  $z_{\text{acc}} = 0.577, 0.597, 0.617,$ and $0.658$ for $b = -0.1, -0.05, 0$, and $0.1$, respectively. 

One of the main motivations of previous VSL models is providing the model alternative to cosmic inflation by shrinking the so-called comoving Hubble radius in time ({\it i.e.}, $d (c/aH)/dt < 0$). However, one can obtain the comoving Hubble radius of meVSL by using Eqs.~\eqref{dotGoGmp} and \eqref{H2mp} 
\begin{align}
\frac{\tc}{a H} = \frac{\tc_0}{a H^{(\GR)}} \label{HR} \,.
\end{align}
As shown in the above equation~\eqref{HR}, the Hubble radius of meVSL is the same as that of GR and thus meVSL cannot replace the early inflation. 

Now, one can obtain an explicit form of $\tc$ by using Eqs.~\eqref{dx0ps} and~\eqref{H2mp}. If the Universe is dominated by the $i$-component, then Eq.~\eqref{H2mp} gives
\begin{align}
H t = \frac{2}{3(1+\omega_i) -b/2} \label{Ht} \,.
\end{align} 
By combining Eqs.~\eqref{dx0ps} and ~\eqref{Ht}, one obtains
\begin{align}
\tc &\equiv \tc_0 a^{\frac{b}{4}} = \left( \frac{3(1+\omega_i)}{3(1+\omega_i) -b/2} \right) c_0 a^{\frac{b}{4}} \label{tca} \,, 
\end{align}
where we assume $c = c_0 a^{b/4}$. Thus, there exists the upper limit on $b$ as $b < 6(1+\omega_i)$. 

\subsection{Redshift}
\label{subsec:Redshift} 

The line element of the FLRW metric is provided by Eq.~\eqref{dsFRW}. 
The proper distance $D_{p}$ from our galaxy ($r=0$) to another galaxy at cosmic time $t$ is given by
\begin{align}
D_{\Tp} = a(t) \int_{0}^{r} \frac{dr'}{\sqrt{1 - kr^{'2}}} \equiv a(t) f(r) = \begin{cases} a(t) \sin^{-1} r & k = 1 \\ a(t) r & k =0 \\ a(t) \sinh^{-1} r & k = -1 \end{cases} \label{Dpmp} \,.
\end{align}
Now we consider a light reaching us, at $r = 0$, has been emitted from a galaxy at $r=r_1$. Also, we consider successive crests of light, emitted at times $t_1$ and $t_1 + \Delta t_1$ and received at times $t_0$ and $t_0 + \Delta t_0$. Since $ds^2 = 0$ and the light is traveling radially, one has for the first and the second crest of light
\begin{align}
\int_{r_1}^{0} \frac{dr}{\sqrt{1 - kr^2}} &=  \int_{t_1}^{t_0} \frac{\tc dt}{a} = \int_{t_1+\Delta t_1}^{t_0+\Delta t_0} \frac{\tc dt}{a} \label{t1t2}
\,. \end{align}
One can rewrite the above equation as
\begin{align}
\int_{t_0}^{t_0+\Delta t_0} \frac{\tc dt}{a} &= \int_{t_1}^{t_1+\Delta t_1} \frac{\tc dt}{a}   \quad \Longrightarrow \quad  \int_{t_0}^{t_0+\Delta t_0} \frac{\tc_0 dt}{a^{1-b/4}} = \int_{t_1}^{t_1+\Delta t_1} \frac{\tc_0 dt}{a^{1-b/4}} \label{Deltaa} \,.
\end{align}
Now if $\Delta t_1$ and $\Delta t_0$ are very small ({\it i.e.}, $\Delta t \ll t$) and then we may assume that $a(t)$ is constant over these intervals, then Eq.~\eqref{Deltaa} provides
\begin{align}
\frac{\Delta t_1}{a(t_1)^{1-b/4}} =  \frac{\Delta t_0}{a(t_0)^{1-b/4}} \label{redshiftmp} \,.
\end{align}
The relative difference between an object's received and emitted wavelengths defines cosmological redshift. The redshift, $z$, is a dimensionless quantity that represents this change. If $\lambda$ represents wavelength and $\nu$ represents frequency, then the emitted and observed wavelengths are given by
\begin{align}
\lambda_{\text{e}} \equiv \tc(t_1) \Delta t_1 = \tc_0 a_{1}^{b/4} \Delta t_1 \quad , \quad \lambda_{\text{o}} \equiv \tc(t_0) \Delta t_0 = \tc_0 a_{0}^{b/4} \Delta t_0 \label{lambdaeo} \,.
\end{align}
By using Eqs.~\eqref{redshiftmp} and~\eqref{lambdaeo}, we obtain 
\begin{align}
1 + z \equiv \frac{\lambda_{\text{o}}}{\lambda_{\text{e}}} = \frac{a_0}{a_1}  \label{redshiftmp2} \,.
\end{align}
As a result, the meVSL model's redshift, $z$, is equal to GR's. It is significant because, rather than utilizing cosmic time, one uses the redshift to express cosmological observations. If the $z$ of any VSL model differs from that of GR, then one needs to reinterpret the observational data using a new redshift obtained from the corresponding model. In meVSL, as $\tc$ changes as a function of time, so does the frequency. One can use $\tc = \tnu \lambda$ where $\tnu = \tnu_0 a^{b/4}$ by using Eq.~\eqref{redshiftmp}. It means the wavelength is not changed. Also, one can investigate the so-called redshift drift, $\Delta z$, the source redshift changes during the time interval between the first and the second crest of light \cite{Loeb:1998bu}
\begin{align}
\Delta z &\equiv \frac{a(t_0 + \Delta t_0)}{a(t_1 + \Delta t_1)} - \frac{a(t_0)}{a(t_1)} \approx \Delta t_0 \left[ H_0 (1+z) - H(t_1) \frac{\tc(t_0)}{\tc(t_1)} \right] = \Delta t_0 \left[ H_0 (1+z) - H(z)^{(\GR)} \right] \nonumber \\ &= \Delta z^{(\GR)} \label{DzLoeb} \,.
\end{align}
The redshift drift, $\Delta z$, in the meVSL model is the same as that of GR. This result is different from other VSL models \cite{Balcerzak:2013kha}.

\subsection{Distances}
\label{subsection:cosDistance}

There are a few different definitions of the distances between two objects or events in the Universe. They are used to relate some observable quantities (such as the redshift of a distant galaxy, the luminosity of a supernova, or the angular size of the acoustic peaks in the CMB power spectrum) to other quantities (like mass density contrast, equation of state of dark energy, and curvature constant) that are not directly observable. We often refer to background observables as cosmological observables related to a distance measurement. It is more practical to write these distances as functions of the redshift, $z$ rather than the cosmic time, $t$. The comoving distance is defined as the distance that is measured locally between two events today if those two points were locked into the Hubble flow.
As we show in Eq.~\eqref{t1t2}, the comoving distance is defined as
\begin{align}
D_{\text{C}}(z) \equiv \int_{0}^{r} \frac{dr'}{\sqrt{1-kr^2}} = \frac{\tc_0}{H_0} \int_{0}^{z} \frac{dz'}{E^{(\GR)}(z')} =  D_{\text{C}}^{(\GR)}(z)\label{Dcmp} \,,
\end{align}
where we use the fact that the Hubble radius is identical in both the meVSL model and GR as shown in Eq.~\eqref{HR}. Thus, the comoving distances both in GR and in meVSL are the same. The transverse comoving distance is defined as the ratio of the transverse velocity of an object to its proper motion and it is given as
\begin{align}
D_{\text{M}}(z) = D_{\text{M}}^{(\GR)}(z) = \begin{cases} \frac{\tc_0}{H_0} \frac{1}{\sqrt{\Omega_{K0}}} \sinh \left( \sqrt{\Omega_{k0}} \frac{H_0}{\tc_0} D_{\text{C}} \right) & \Omega_{k0} > 0 \, \\D_{\text{C}} & \Omega_{k0} = 0 \, \\ \frac{\tc_0}{H_0} \frac{1}{\sqrt{|\Omega_{K0}|}} \sin \left( \sqrt{|\Omega_{k0}|} \frac{H_0}{\tc_0} D_{\text{C}} \right) & \Omega_{k0} < 0 \, \end{cases} \label{DMmp} \,.
\end{align}
The luminosity distance, $D_{\text{L}}$ is defined by the relationship between bolometric flux and bolometric luminosity. Thus, the relation between the luminosity distance and the transverse comoving distance is modified to that of GR as derived in sec.~\ref{subsec:dLapp}. And the angular diameter distance, $D_{\TA}$ is defined as the ratio of an object's physical transverse size to its angular size. They are given by
\begin{align}
D_{\text{L}}(z) = (1+z)^{1-b/8} D_{\text{M}}(z)  =  (1+z)^{-\frac{b}{8}} D_{\text{L}}^{(\GR)}(z) \quad , \quad  D_{\text{A}}(z) = \frac{D_{\text{M}}}{(1+z)} = D_{\text{A}}^{(\GR)}(z) \, . \label{DLDA}
\end{align}
Thus, the so-called cosmic distance duality relation (DDR) of meVSL is different from that of GR and can be written as
\begin{align}
\frac{(1+z)^2 D_{\TA}}{D_{\TL}} = \left( 1 + z \right)^{-\frac{b}{8}} \label{ccdrmeVSL} \,.
\end{align}
The comoving volume $V_{\text{C}}$ is the volume measured in which the number densities of non-evolving objects locked into the Hubble flow are constant with redshift. Thus, the comoving volume element in solid angle $d \Omega$ and redshift interval $dz$ is given by
\begin{align}
\frac{dV_{\text{C}}}{d \Omega dz} = (1+z)^2 \frac{\tc_0}{H_0} \frac{D_{\text{A}}^2}{E^{(\GR)}(z)} = \left(  \frac{dV_{\text{C}}}{d \Omega dz}  \right)^{(\GR)} \label{dVC} \,.
\end{align}
Thus, all of the cosmological distances of meVSL are the same as those of GR except the luminosity distance. However, this does not imply that one obtains the same values of cosmological parameters extracted from the background evolution observables. This is because some physical constants and quantities also vary as a function of cosmic time due to a time variation in the speed of light. Thus, we need to investigate subsequent changes in relevant quantities, such as the fine structure constant, the Thomson cross-section, the decay rate of weak interaction, etc, related to physical processes.

\section{Observations}
\label{sec:observations}

Cosmological observations ought to confirm the viability of theories of gravity. In addition to background evolution observations, there have been various cosmological observations based on the thermal history of the Universe. These include Big Bang Nucleosynthesis (BBN), cosmic microwave background (CMB), baryon acoustic oscillation (BAO), type Ia supernova (SNe), Hubble parameter (H), and gravitational waves (GWs). Also, the time variation of the fine structure constant has been investigated as a possible probe for the time variations of fundamental physical constants. In this section, we examine how meVSL affects those cosmological observations. 

\subsection{BBN}
\label{subsec:BBN}

BBN is the formation of primordial light elements other than those of the lightest isotope of hydrogen during the early Universe. At temperatures higher than $1$MeV, photons, electrons, positrons, neutrinos, antineutrinos, protons, and neutrons formed the primordial plasma of the early Universe. At this epoch, neutrinos start being decoupled and then the number of neutrons begins to diminish through the $\beta$-decay. Neutrons are also captured by protons and form deuterium nuclei. The result of these reactions was to lock up most of the free neutrons into $\tensor[^4]{\text{He}}{}$ nuclei and to create trace amounts of D, $\tensor[^3]{\text{He}}{}$, $\tensor[^7]{\text{Li}}{}$, and $\tensor[^7]{\text{Be}}{}$. One can investigate the modification of meVSL at each mentioned step.

For $T > 1$ MeV,  a first stage during which the neutrons, protons, electrons, positrons, and neutrinos are kept in statistical equilibrium by the weak interaction. As long as the statistical equilibrium holds, the neutron-to-proton ratio is
\begin{align}
\left( \frac{n}{p} \right) = e^{-E_{\text{np}}/k_{\TB}T} = \left( \frac{n}{p} \right)^{(\GR)} \label{nop} \,,
\end{align}
where we use $E_{\text{np} } \equiv (m_{\Tn \, \text{rs}} - m_{\Tp\, \text{rs}}) \tc^2 = (m_{\Tn 0} - m_{\Tp 0}) \tc_0^2 =  E_{\text{np}}^{(\GR)} = 1.293$ MeV. Thus, the neutron-to-proton ratio in the meVSL model is the same as that of GR. The abundance of the other light element of the mass number $A$ and charge $Z$ is given by
\begin{align}
Y_{A} = g_{A} \left( \frac{\zeta(3)}{\sqrt{\pi}} \right)^{A-1} 2^{(3A-5)/2} A^{5/2} \left[ \frac{k_{\TB} T}{m_{\TN} \tc^2} \right]^{3(A-1)/2} \eta^{A-1} Y_{\Tp}^{Z} Y_{\Tn}^{A-Z} e^{B_{A}/k_{\TB} T} = Y_{A}^{(\GR)} \label{YA} \,,
\end{align}
where $g_{A}$ is the number of degrees of freedom of the nucleus ${}^{A}_{Z}X$, $m_{\TN}$ is the nucleon mass, $\eta$ is the baryon to photon ratio, and $B_{A} \equiv (Z m_{\Tp} + (A-Z) m_{\Tn} - m_{A})c^2 = B_{A}^{(\GR)}$ is the binding energy which is identical for both GR and meVSL. Thus, the abundances of light elements are the same in both models.

Around $T \sim 0.8$ MeV, the weak interactions freeze out at a temperature $T_{f} = T(z_{f})$ determined by the competition between the weak interaction rate and the expansion rate of the Universe. The total decay rate of neutrons is given by 
\begin{align}
\Gamma_{\Tw} = \frac{1}{4 \pi^3 \hbar} \left( \frac{g_{\Tw}}{2 M_{\TW} \tc^2} \right)^4 \left(m_e \tc^2 \right)^5 \left[ \frac{1}{15} \left(2 b^4 - 9b^2 - 8 \right) \sqrt{b^2-1} + b \ln \left[ b + \sqrt{b^2 -1} \right] \right] \equiv \Gamma_{\Tw}^{(\GR)} a^{\frac{b}{4}} \label{Gammaw} \,,
\end{align}
where $g_{\Tw}$ is the coupling constant of the weak interaction measured as 0.653, $M_{\TW}$ is the mass of the W-boson, and $b \equiv (m_{\Tn} - m_{\Tp})/m_{\Te} \approx 2.53$. $b$ of GR is the same as that of meVSL. In GR, $\Gamma_{\Tw}^{(\GR)} \approx 7.5876 \times 10^{-4} s^{-1}$. The decay rate of the neutron is modified from that of GR as shown in Eq.~\eqref{Gammaw}. However, this does not cause a change in the decoupling epoch of neutrons. The thermal equilibrium of neutrons is maintained so long as the timescale for the weak interaction is short compared with the timescale of the cosmic expansion. They begin to decouple from the primordial plasma when the condition $\Gamma_{\Tw} \sim H$ is reached. We can show that the decoupling condition of meVSL is equal to that of GR by using Eqs.~\eqref{H2mp} and~\eqref{Gammaw}
\begin{align}
\Gamma_{\Tw} (z_f) = H (z_f) \quad \Longrightarrow \quad \Gamma_{\Tw}^{(\GR)} (z_f) = H^{(\GR)} (z_f) \label{decoup} \,.
\end{align}
Thus, the neutrons are decoupled from other elements after $z_f$ which is the same for meVSL and GR. After $z_f$, the number of neutrons and protons change only through the neutron $\beta$-decay between $T_f$ to $T_{\TN} \sim 0.1$MeV when $p + n$ reactions proceed faster than their inverse dissociation.

For 0.05 MeV $< T < 0.6$MeV, only two-body reactions produce the synthesis of light elements. This requires two conditions. One is that the deuteron is to be synthesized ($p+n \rightarrow D$). The other is the very low photon density to neglect the photon-dissociation. This happens roughly when
\begin{align}
\left( \frac{n_{\Td}}{n_{\gamma}} \right) \sim \eta^2 \exp \left[ - \frac{B_{\TD}}{k_{\TB} T} \right] =  \left( \frac{n_{\Td}}{n_{\gamma}} \right)^{(\GR)} \label{ndongamma} \,.
\end{align}
The abundance of ${}^{4}$He by mass, $Y_{\Tp}$, is then well estimated by
\begin{align}
Y_{\Tp} \simeq 2 \frac{(n/p)_{\TN}}{1 + (n/p)_{\TN}} \quad \text{where} \quad  \left( \frac{n}{p} \right)_{\TN} =  \left( \frac{n}{p} \right)_{f} \exp \left[ -\frac{t_{\TN}}{\tau_{\Tn}} \right] \label{Yp} \,,
\end{align}
where $t_{\TN} = t_{\TN}^{(\GR)} (1+z_{\TN})^{b/4}$ and $\tau_{\Tn} = \tau_{\TN}^{(\GR)}  (1+z_{\Tn})^{b/4}$. This means $Y_{\Tp} \simeq Y_{\Tp}^{(\GR)}$.
Thus, unlike other VSL models, meVSL does not affect the BBN predictions compared to those of GR. Thus, if one wants to obtain the cosmological limit on the value of $b$, one should use other observations rather than BBN. The same cosmological parameters from BBN based on GR can be adopted in the meVSL model.

\subsection{CMB}
\label{subsec:CMB}

\begin{figure*}
\centering
\vspace{1cm}
\includegraphics[width=0.8\linewidth]{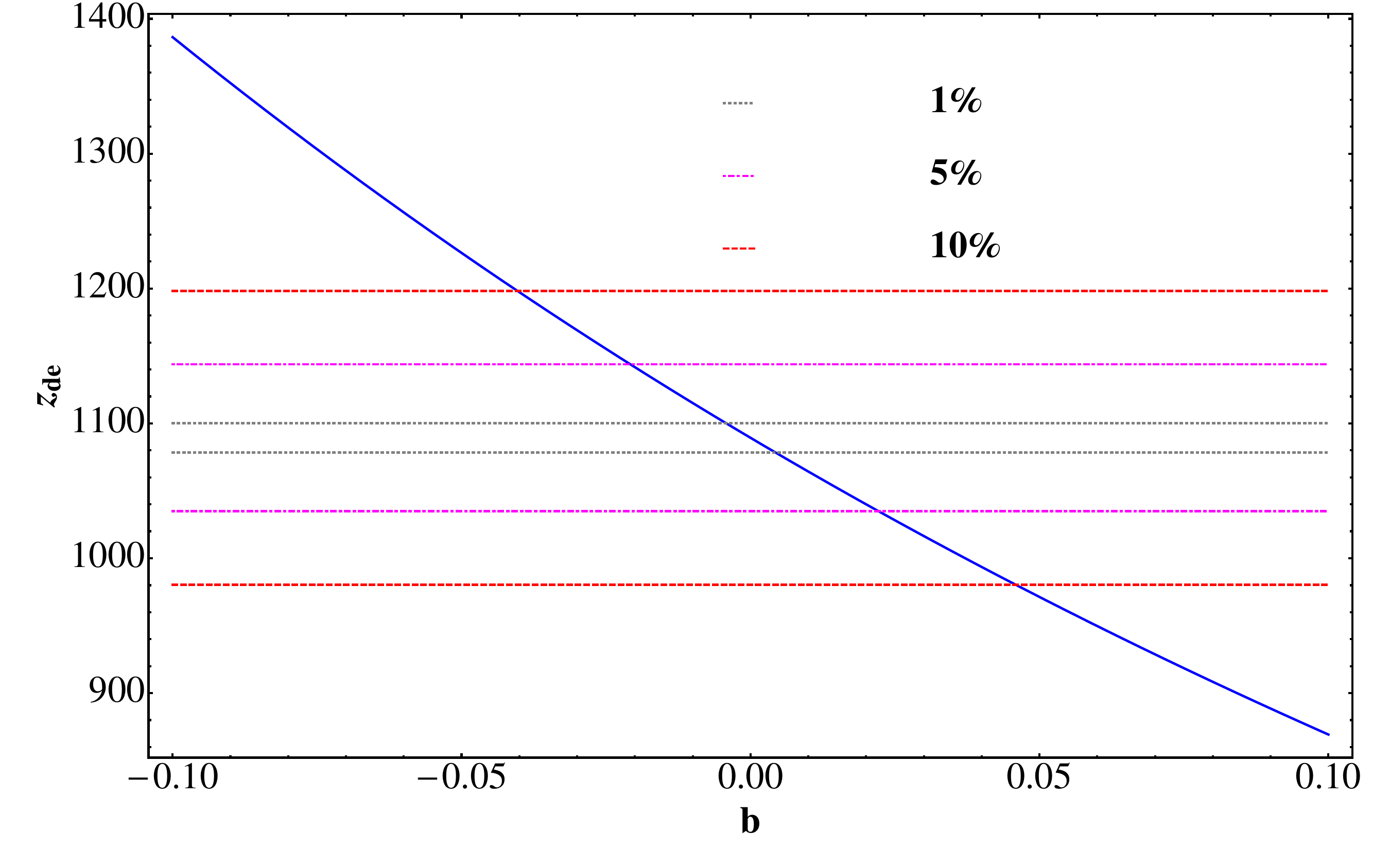} 
\vspace{-0.5cm}
\caption{The decoupling redshift as a function of $b$. The horizontal lines correspond to $1$ (dot-dashed), $5$ (dotted), and $10$ (dashed) \% error deviations, respectively.} \label{fig-3}
\vspace{1cm}
\end{figure*}

CMB is electromagnetic radiation as a remnant from an early Universe after it decouples from the primordial plasma. One can distinguish the recombination from the decoupling. Protons and electrons combine in a process called recombination to produce neutral hydrogen. At sufficiently low temperatures, photons were no longer able to ionize the hydrogen atoms, and thus free electron numbers dramatically dropped. Therefore, the number densities of protons and electrons determine the recombination epoch alone. The number densities of them are the same both in GR and in meVSL. Thus, the recombination epoch is not modified in the meVSL model compared to GR. However, photons interacted primarily with electrons through Thomson scattering. In this process, the electron oscillates due to the electromagnetic field of the photon, resulting in the emission of radiation at the same frequency as the incident wave, thus scattering the wave. An important feature of Thomson scattering is that it introduces polarization along the direction of motion of the electron. For Thomson scattering, the cross-section is 
\begin{align}
\sigma_{\TT} = \frac{8\pi}{3} \left( \frac{e^2}{4 \pi \tepsilon m_{\Te} \tc^2 } \right)^2 = \sigma_{\TT}^{(\GR)} a^{-\frac{b}{2}} \label{sigmaT} \,.
\end{align}
This process is tiny, and the Thomson scattering is most important in regions with a high density of free electrons, such as the early Universe or the dense interiors of stars. The scattering rate per photon, $\Gamma_{\TT}$, can be estimated as the speed of light divided by the mean free path for photons (the mean distance traveled between scatterings)
 \begin{align}
 \Gamma_{\TT} = n_{\Te} \sigma_{\TT} \tc = \Gamma_{\TT}^{(\GR)} a^{-\frac{b}{4}}  \label{GammaThomson} \,.
 \end{align}
 Thus, the decoupling epoch is determined by using Eqs.~\eqref{H2mp} and \eqref{GammaThomson}
 \begin{align}
 \Gamma_{\TT} = H \quad \Longrightarrow \quad \Gamma_{\TT}^{(\GR)} = H^{(\GR)} a^{\frac{b}{2}} \label{GammaTH} \,.
 \end{align} 
 Recombination is the process by which neutral hydrogen is formed through a combination of protons and electrons. Generally speaking, decoupling is the epoch at which photons cease interacting with unbound electrons. In this case, their mean free path becomes larger than the Hubble radius and we can detect them as CMB coming from the last scattering surface at the present epoch. We can estimate the deviation of the decoupling epoch in the meVSL model compared to GR. For this purpose, we assume that the Universe is dominated by radiation at that epoch. Then the Hubble parameter at that epoch is given by $H_{\Tde} = H_{0} (1 +z_{\Tde})^{2 - b/4}$ where $z_{\Tde}$ is the redshift at the decoupling epoch. Also, if we assume that the Universe is fully ionized at this epoch $n_{\Te}(z_{\Tde}) = X_{\Te} n_{\Tb}(z_{\Tde}) = X_{\Te} n_{\Tb 0} (1+z_{\Tde})^3$ where $X_{\Te}$ is the free electron fraction. Then the decoupling epoch defined to be $\Gamma_{\TT}(z_{\Tde}) = H(z_{\Tde})$ is estimated by
\begin{align}
\Gamma_{\TT} (z_{\Tde}) &= X_{\Te} n_{\Tb} \sigma_{\TT} \tc = \frac{3 H_0^2 \Omega_{\Tb 0}}{8 \pi \tG_{0} m_{\Tp \rs}} X_{\Te} \sigma_{\TT}^{(\GR)} \tc_0 \left( 1+ z_{\Tde} \right)^{3 + \frac{b}{4}} \label{Gammade} \,, \\
\sigma_{\TT} &= \frac{8\pi}{3} \left( \frac{e^2}{4 \pi \epsilon m_{\Te} \tc^2} \right)^2 = \frac{8\pi}{3} \left( \frac{e_{0}^2}{4 \pi \epsilon_{0} m_{\Te \rs} \tc_0^2} \right)^2 a^{-\frac{b}{2}} \equiv \sigma_{\TT}^{(\GR)} a^{-\frac{b}{2}} \label{sigmaT2} \,, \\
H (z_{\Tde})  &\simeq H_{0} \sqrt{\Omega_{\Tm 0}} \sqrt{ 1 + \frac{1+z_{\Tde}}{1 + z_{\Teq}} } (1+z_{\Tde})^{\frac{3}{2} - \frac{b}{4}} = H_{0} \sqrt{\Omega_{\Tm 0}} \sqrt{ 1 + \left( 1+z_{\Tde} \right) \frac{\Omega_{\Tr 0} h^2}{\Omega_{\Tm 0} h^2} } (1+z_{\Tde})^{\frac{3}{2} - \frac{b}{4}} \label{Hde} \,, \\
\frac{\Gamma_{\TT}(z_{\Tde})}{H(z_{\Tde})} &= \frac{3 \sigma_{\TT}^{(\GR)} \tc_0 H_0}{8 \pi \tG_{0} m_{\Tp \rs} h} X_{\Te}(z_{\Tde}) \frac{\Omega_{\Tb 0} h^2}{\sqrt{\Omega_{\Tm 0} h^2}} \left( 1 +  z_{\Tde} \right)^{\frac{3+b}{2}} \left(  1 + \left( 1+z_{\Tde} \right) \frac{\Omega_{\Tr 0} h^2}{\Omega_{\Tm 0} h^2} \right)^{-\frac{1}{2}} \label{GdeoHde} \, \\
&= 145 \times 33^{b} \times X_{\Te} (z_{\Tde}) \left( \frac{\Omega_{\Tb 0} h^2}{0.02212} \right) \left( \frac{0.1434}{\Omega_{\Tm 0} h^2} \right)^{\frac{1}{2}} \left( \frac{1+z_{\Tde}}{1090} \right)^{\frac{3+b}{2}} \left(  1 + \frac{1+z_{\Tde}}{3411} \frac{0.1434}{\Omega_{\Tm 0} h^2} \right)^{-\frac{1}{2}} \,,  \nonumber
\end{align}
where $\sigma_{\TT}^{(\GR)} = 6.635 \times 10^{-29} \text{m}^2$. Thus, the decoupling epoch of meVSL is earlier (later) than that of GR for the negative (positive) value of $b$. The observed decoupling epoch is $z_{\text{de}} = 1090$. We show the effect of $b$ on $z_{\Tde}$ in Fig.~\ref{fig-3}. The horizontal lines are depicted $1$\% (dotted), $5$\% (dot-dashed), and $10$\% (dashed) errors, respectively. As the value of $b$ increases, $z_{\Tde}$ decreases. If one allows the 1\% error in $z_{\text{de}}$, then the allowed range of $b$ is $-0.004 \leq b \leq 0.004$. For 5\% deviation of $z_{\Tde}$, $-0.021 \leq b \leq 0.022$ is obtained. For 10\% deviations of $z_{\Tde}$, the allowed regions for $b$ is $-0.048 \leq b \leq 0.045$.

The shift parameter, $R$, is presented in the CMB measurement as a practical means of assessing the cosmological models' likelihood quickly~\cite{Bond:1997wr}
\begin{align}
R = \sqrt{\Omega_{\Tm 0}} \int_{0}^{z_{\Td}} dz' \frac{\tc(z')}{E(z')} = \sqrt{\Omega_{\Tm 0}} \int_{0}^{z_{\Td}} dz' \frac{\tc_0(z')}{E^{(\GR)}(z')} = R^{(\GR)} \label{RCMB} \,.
\end{align}
This shift parameter is often used to investigate the VSL model. However, $R$ is the same both in GR and in the meVSL model. For this reason, $R$ cannot be used in the meVSL model to limit $b$.

\begin{figure*}
\centering
\vspace{1cm}
\begin{tabular}{cc}
\includegraphics[width=0.5\linewidth]{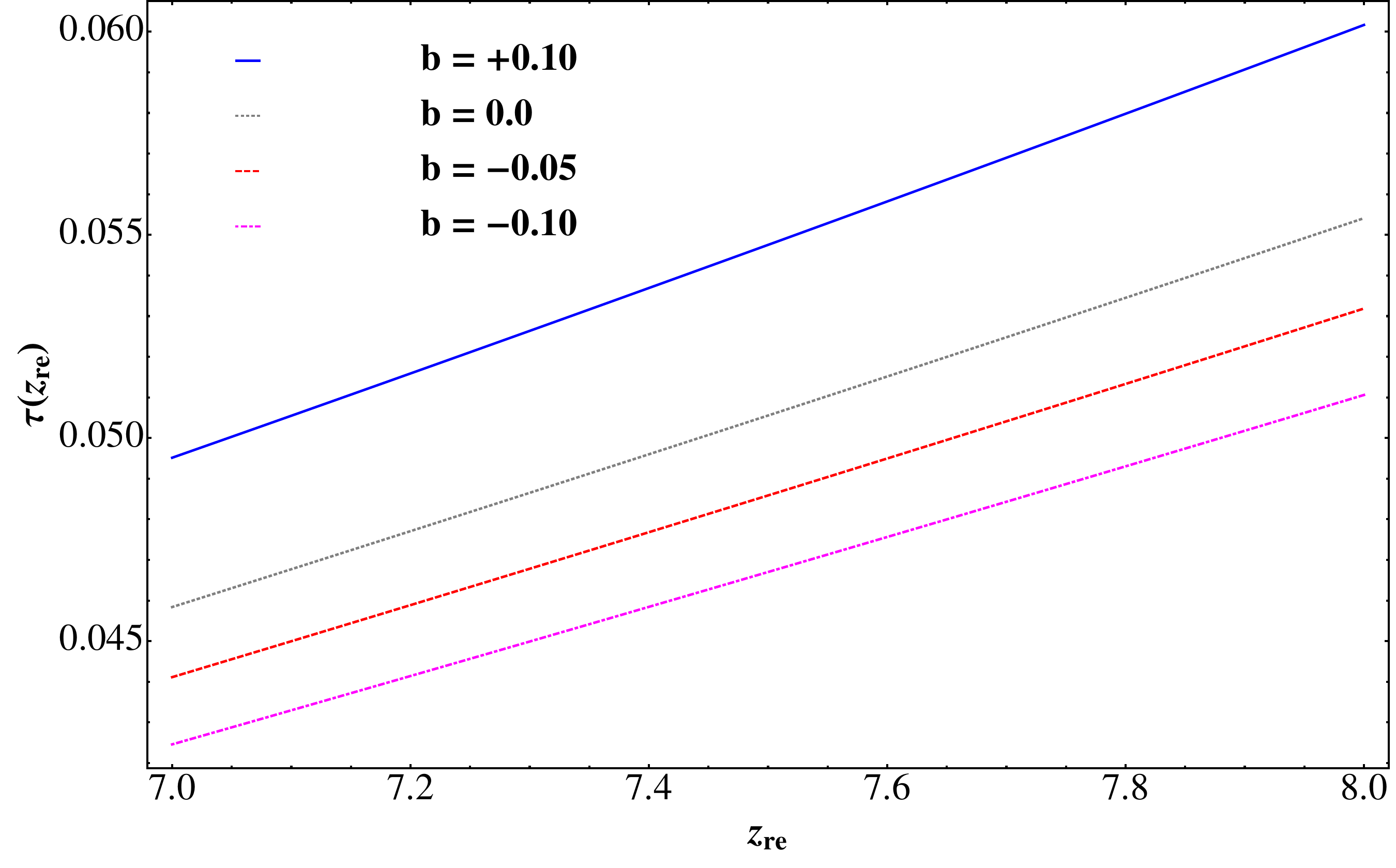} &
\includegraphics[width=0.49\linewidth]{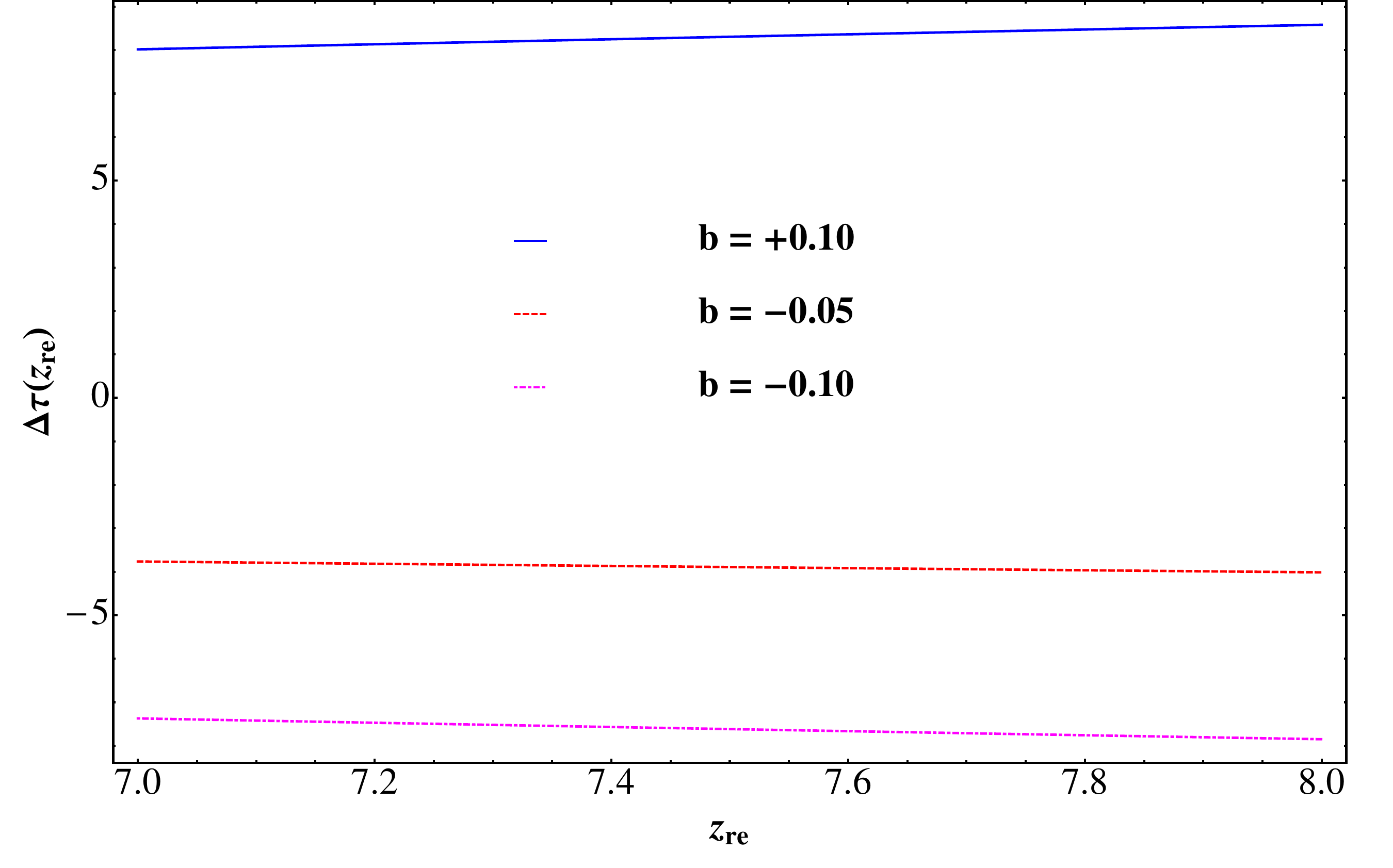}
\end{tabular}
\vspace{-0.5cm}
\caption{The optical depth to the reionization for the different values of $b$. a) The optical depth, $\tau(z_{\text{re}})$ for $b = +0.1$(solid), $0$ (dotted), $-0.05$ (dashed), and $-0.1$ (dot-dashed), respectively. b) The differences of $\tau(z_{\text{re}})$ between the meVSL model and GR (\textit{i.e.}, $b=0$).} \label{fig-4}
\vspace{1cm}
\end{figure*}

The optical depth to Thomson scattering is the integral over time of the scattering rate
\begin{align}
\tau(t) \equiv \int^{t_{0}}_{t} \Gamma_{\TT}(t) dt
\label{opticaldepth} \,.
\end{align}
For instantaneous, complete ionization at redshift $z_{\Tre}$, one can calculate the optical depth \cite{Shull:2007az}
\begin{align}
\tau(z_{\Tre}) = (1+y) \frac{(1-Y_{\Tp}) \rho_{\cro 0}}{m_{\TtH 0}} \sigma_{\TT}^{(\GR)} \frac{\tc_{0}}{H_{0}} \int_{0}^{z_{\Tre}} \frac{(1+z)^{2+b/2}}{E^{(\GR)}(z)} dz \label{tauzre}  \,,
\end{align}
where $\rho_{\cro 0}$ is the present value of critical density, $m_{\TtH 0}$ is the present mass of hydrogen. The number densities of hydrogen, helium, and electrons are given by $n_{\TtH} = [(1-Y_{\Tp}) \rho_{\cro 0} / m_{\TtH 0} ] (1+z)^3$, $n_{\text{He}} = y n_{\TtH}$, and $n_{e} = (1+y) n_{\TtH}$ if helium is singly ionized. The $y$ is related to a helium mass fraction, $Y_{\Tp}$ by $y = (Y_{\Tp}/4)/(1-Y_{\Tp})$. One can obtain the analytic solution of the above integral in Eq.~\eqref{tauzre} if one considers the late time Universe, $E(z) = \sqrt{\Omega_{\Tm 0} (1+z)^3 + \Omega_{\Lambda 0}}$ with $\Omega_{\Lambda 0} = 1 - \Omega_{\Tm 0}$
\begin{align}
\tau(z_{\Tre} \,, b) &= (1+y) \frac{(1-Y_{\Tp}) \rho_{\cro 0}}{m_{\TtH \rs}} \sigma_{\TT}^{(\GR)} \frac{\tc_{0}}{H_{0}} \frac{2\Omega_{\Tb 0}}{(6+b) \Omega_{\Lambda 0}} \nonumber \\
&\times \Biggl(- {}_2F_{1}\left[1,\frac{9+b}{6},2+\frac{b}{6},-\frac{\Omega _{\text{m0}}}{\Omega _{\text{$\Lambda $0}}}\right] \nonumber \\
&+(1+z_{\Tre})^{3+\frac{b}{2}} {}_2F_{1} \left[1,\frac{9+b}{6},2+\frac{b}{6},-\frac{(1+z_{\Tre})^3 \Omega _{\text{m0}}}{\Omega _{\text{$\Lambda $0}}}\right] \sqrt{(1+z_{\Tre})^3 \Omega _{\text{m0}}+\Omega _{\text{$\Lambda $0}}} \Biggr) \label{tauzre2} \,,
\end{align}
where ${}_2 F_{1}$ is a hypergeometric function and the reionization epoch, $z_{\Tre} = 7.5$ \cite{Aghanim:2018eyx}. Eq.~\eqref{tauzre2} provides the $\tau(z_{\Tre})$-dependence on $b$. As $b$ increases, so does $\tau(z_{\Tre})$. We show this in Fig.~\ref{fig-4}. In the left panel of Fig.~\ref{fig-4}, $\tau(z_{\Tre})$ for the different values of $b$ at the given reionization epoch $z_{\Tre}$ is depicted. The solid, dotted, dashed, and dot-dashed lines correspond to $b = +0.1, 0, -0.05$, and $-0.1$, respectively. In Planck 2018 \cite{Aghanim:2018eyx}, $\tau (7.5) = 0.0522 \pm 0.0080$ and thus its 1-$\sigma$ error is about 15\%. In the right panel of Fig.~\ref{fig-4}, we show the deviations of $\tau(z_{\Tre})$ of meVSL from that of GR, $  \Delta \tau(z_{\Tre}) =  \left( \tau(z_{\Tre}, b\neq 0) -   \tau(z_{\Tre}, b=0) \right)/\tau(z_{\Tre}, b=0) \times 100 (\%)$. The solid, dashed, and dot-dashed lines correspond to $b = +0.1, -0.05$, and $-0.1$, respectively. The differences are about 8 \%, -4 \%, and -7 \% for $b = 0.1, -0.05$, and -0.1, respectively.  These models are within the measurement errors of $\tau(z_{\Tre})$. Thus, the current observational accuracy on the optical depth might not provide a strong constraint on $b$.

The conformal time derivative of the optical depth, $\tau'$, and the visibility function are given by
\begin{align}
\tau' &\equiv \frac{d\tau}{d\eta} = - n_{\Te} \sigma_{\TT} a \tc = \tau^{'(\GR)} a^{-\frac{b}{4}} \label{tauprime} \,, \\
g(\eta) &\equiv -\tau' e^{-\tau} \quad \text{with} \quad \int_{0}^{\eta_0} d\eta g(\eta) = 1 \label{geta} \,.
\end{align}
These quantities contribute to the source terms of temperature and polarization. However, the visibility function is nonzero only during the recombination and reionization. Thus, it provides a weak constraint on $b$.

 \subsection{SZE}
 \label{subsec:SZE}

 Masses of clusters of galaxies often exceed $3 \times 10^{14} M_{\odot}$ with the effective gravitational radii, $R_{\Teff}$ of order Mpc. Any gas in hydrostatic equilibrium within a cluster's gravitational potential well has electrons with the temperature $T_e$ given by
\begin{align}
k_{\TB} T_{\Te} &\approx \frac{G M m_{\Tp}}{2 R_{\Teff}} = 6.74 \left(\frac{M}{3 \times 10^{14} M_{\odot}} \right) \left( \frac{\text{Mpc}}{R_{\Teff}} \right) \text{KeV}
\label{Te} \,,
\end{align}
where $G = 4.3 \times 10^{-3}$ pc $\text{M}_{\odot}^{-1}$ $\left( \text{km}/\text{s} \right)^{2}$ and $m_{\Tp}$ = 938.272 \text{MeV}/$c^2$. At this temperature, the X-ray part of the spectrum shows the gas's thermal emission composed of thermal bremsstrahlung and line radiation.

Among the mass of clusters of galaxies, the mass of distributed gas is about a quarter of it. Thus, clusters of galaxies are luminous X-ray sources, with the bulk of the X-rays produced as bremsstrahlung rather than line radiation due to this high mass density of the gas. However, electrons in the intracluster gas are scattered not only by ions but also by CMB photons. The cross-section of this low-energy scattering is given by the Thomson scattering cross-section, $\sigma_{\TT}$, so that the scattering optical depth becomes $\tau_{\Te} \simeq n_{\Te} \sigma_{\TT} R_{\Teff} \sim 10^{-2}$.

Due to inverse Thomson scattering with high-temperature electrons, the photon's frequency is slightly shifted, leading to more probable upscattering. On average a slight mean change of photon energy from this scattering is produced
\begin{align}
\frac{\Delta \nu}{\nu} \approx \frac{k_{\TB} T_{\Te}}{ m_{\Te} c^2} \sim \frac{6.74 KeV}{0.511 MeV} \sim 1.32 \times 10^{-2} \,. \label{Deltanuonu}
\end{align}
Thus, this inverse Compton (Thomson) scattering produces about 1 part in $10^{4}$ overall change in brightness of CMB.

Spatial distributions of clusters of galaxies determine SZE. SZE is observed towards galaxy clusters, which are large-scale structures detectable in the optical and X-ray bands, and it is localized. In addition, other observable properties of the clusters also influence the signal amplitude. However, primordial structures of the CMB are nonlocalized.  They are also not related to structures seen at different wavebands. In addition, they exhibit a nearly consistent correlation amplitude in various sky locations and are dispersed randomly throughout the entire sky. When the radiation passes through an electron population with significant energy content, its spectrum is distorted. This is called the thermal SZE.

\begin{figure*}
\centering
\vspace{1cm}
\includegraphics[width=0.8\linewidth]{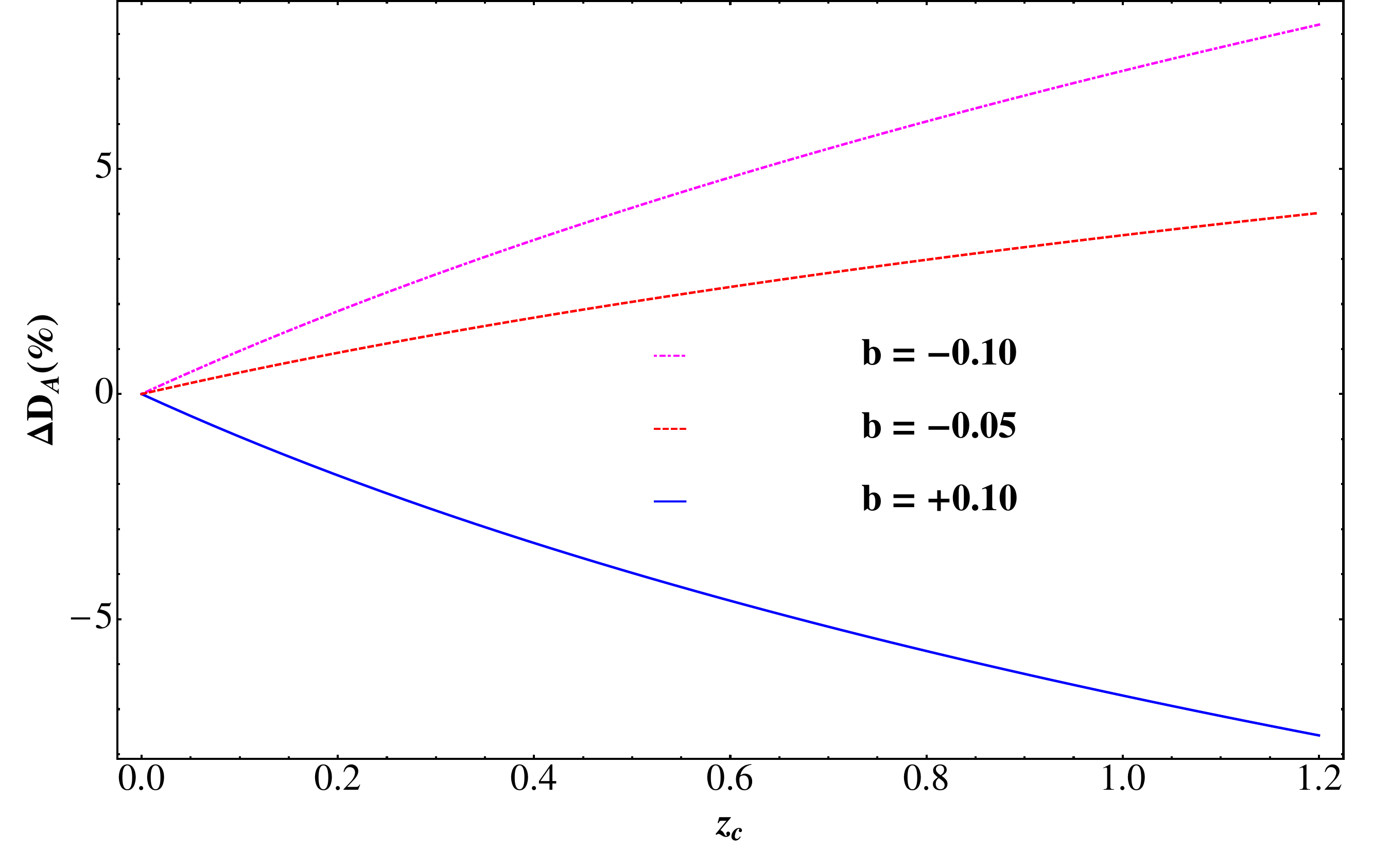} 
\vspace{-0.5cm}
\caption{The differences of angular diameter distance between the meVSL and GR, $\Delta D_{\TA}(z_c)$, as a function of $z_c$ for different values of $b$. The dot-dashed, dashed, and solid lines correspond to $b = -0.1\,,-0.05$, and $0.1$, respectively.} \label{fig-5}
\vspace{1cm}
\end{figure*}

One can express the scattering optical depth, Comptonization parameter, and X-ray spectral surface brightness along a particular line of sight with a cluster atmosphere gas of electron concentration $n_{\Te}({\bf r})$
\begin{align}
\tau_{\Te} &= \int n_{\Te} ({\bf r}) \sigma_{\TT} dl \label{tauTe} \,, \\
y &= \int n_{\Te} ({\bf r}) \sigma_{\TT} \frac{k_{\TB} T_{\Te}({\bf r})} dl \label{yTe} \,, \\
S_{X}(E) &= \frac{1}{4 \pi (1+z)^3} \int n_{\Te} ({\bf r})^2 \Lambda (E, T_{\Te}) dl \label{Sx} \,,
\end{align}
where $z$ is the redshift of the cluster and $\Lambda$ is the spectral emissivity of the gas at observed X-ray energy $E$. One obtains the factor of $4\pi$ from the assumption of the isotropic emissivity. Also, the cosmological transformations of spectral surface brightness and energy give the $(1+z)^3$ factor. Introducing a parameterized gas model in the cluster and using it to fit these parameter values to the X-ray data is convenient in many cases. One can predict the appearance of the cluster in the SZE by integrating Eq.~\eqref{yTe}. There exists a simple and popular model called the isothermal $\beta$ model. In this model, the electron temperature, $T_{\Te}$ is regarded as a constant, and the electron number density is assumed as spherically distributed
\begin{align}
n_{\Te}({\bf r}) = n_{\Te 0} \left( 1 + x^2 \right)^{-\frac{3}{2} \beta} \label{ner2} \,,
\end{align}
where $x = r/r_c$ and $r_c = \theta_{\Tc} D_{\TA}$ is the core radius of the distribution. The surface brightness profile of intracluster medium observed at the projected radius, $b_{\Tp}$, $S_{X}(b_{\Tp})$, is the projection on the sky of the plasma emissivity, $\epsilon(r)$
\begin{align}
S_{X}(b_{\Tp}) &= \frac{1}{4\pi} \frac{D_{\TA}^2}{D_{\TL}^2} \int_{b^2}^{\infty} \frac{\epsilon dr^2}{\sqrt{r^2 - b_{\Tp}^2}} \quad, \text{where} \quad \epsilon(r) = \Lambda (T_{gas}) n_{p}^2 \quad \left( \text{ergs}^{-1} \text{cm}^{-3} \right) \nonumber \\
&= \frac{1}{4 \sqrt{\pi}} \frac{1}{(1+z)^{4}} n_0^2 \Lambda(T_{gas}) r_{c} \frac{\Gamma[3\beta-\frac{1}{2}]}{\Gamma[3\beta]} \left( 1 + \frac{b_{\Tp}^2}{r_c^2} \right)^{0.5 - 3\beta}  \equiv S_{0} \left( 1 + \frac{b_{\Tp}^2}{r_c^2} \right)^{0.5 - 3\beta}  \label{SbSZ} \,,
\end{align}
where $n_{\Tp} = \rho_{\text{gas}} /(2.21 \mu m_{\Tp})$ is the proton density and the cooling function, $\Lambda (T_{\text{gas}})$ depends on the mechanism of the emission. Assuming isothermality and a $\beta$-model for the gas density, the surface brightness profile has an analytic solution.
The SZE on the temperature is given by
\begin{adjustbox}{width=\textwidth}
\begin{minipage}{\textwidth}
\begin{align}
\Delta T_{\text{SZE}}(\theta) &= f(\nu, T_{\Te}) \frac{k_{\TB} T_{\Te} T_{\gamma 0}}{m_{\Te \rm} \tc_0^2} \sigma_{\TT}^{(\GR)} (1+z)^{\frac{b}{2}} \int_{-l_{max}}^{l_{max}} n_{e} dl = f(\nu, T_{\Te}) \frac{k_{\TB} T_{\Te} T_{\gamma 0}}{m_{\Te \rm} \tc_0^2} \sigma_{\TT}^{(\GR)} (1+z)^{\frac{b}{2}} \int_{b^2}^{\infty} \frac{n_{e} dr^2}{\sqrt{r^2 - b_{\Tp}^2}}  \,, \nonumber \\
&= f(\nu, T_{\Te}) \frac{k_{\TB} T_{\Te} T_{\gamma 0}}{m_{\Te \rm} \tc_0^2} \sigma_{\TT}^{(\GR)} (1+z)^{\frac{b}{2}}  \sqrt{\pi} n_0 r_{c} \frac{\Gamma[\frac{3\beta -1}{2}]}{\Gamma[\frac{3\beta}{2}]} \left( 1 + \frac{b_{\Tp}^2}{r_c^2} \right)^{\frac{1 - 3\beta}{2}}  \equiv \Delta T_{0} \left( 1 + \frac{b_{\Tp}^2}{r_c^2} \right)^{\frac{1 - 3\beta}{2}} \label{nbSZ} \,,
\end{align}
\end{minipage}
\end{adjustbox}
where $S_0$, $\Delta T_0$, and $\theta_{\Tc}$ are observed quantities. Thus, one can solve for $n_0$ from Eqs.~\eqref{SbSZ} and \eqref{nbSZ} to obtain $D_{\TA}$
\begin{align}
D_{A}(z_{\Tc}) = \frac{1}{4 \pi \sqrt{\pi}} \left[ \frac{\Delta T_0^2}{S_{X0}} \theta_{\Tc}^{-1}  \frac{\tilde{\Gamma}[\beta]}{\tilde{\Gamma}[\beta/2]^2}  \right] \left[ \left(\frac{m_{\Te \rs} \tc_{0}^2}{ k_{\TB} T_{\Te 0}} \right) \frac{\Lambda}{f T_{\gamma 0} \sigma_{\TT}^{(\GR)} }\right]^2 \left( 1+z_{\Tc} \right)^{-4-b} \equiv D_{\TA}^{(\GR)} (z_{\Tc})(1+z_c)^{-b} \label{DASZ} \,,
\end{align}
where $\tilde{\Gamma}[\beta] = \Gamma[3\beta-1/2]/\Gamma[3\beta]$.
Thus, the observed diameter distance has an extra $(1+z_{c})^{-b}$ factor in meVSL compared to GR. The difference between GR and meVSL is given by
\begin{align}
\Delta D_{A}(z_{\Tc}) \equiv \frac{D_{A}(z_{\Tc}) - D_{A}^{(\GR)}(z_{\Tc})}{D_{A}^{(\GR)}(z_{\Tc})}  = \left( 1 + z_{\Tc} \right)^{b} - 1 \label{DeltaDASZ} \,.
\end{align}
We show the behavior of this in Fig.~\ref{fig-5}. It is obvious that $\Delta D_{A}(z_{\Tc})$ increases as $z_{\Tc}$ does. The dot-dashed, dotted, and solid lines correspond $b= -0.1\,,-0.05$, and $0.1$, respectively. The discrepancies are about 2\% and 4 (-4)\% at $z_{c} =0.5$ for $b= -0.05$ and -0.1 (+0.1), respectively. Thus, one needs to consider the time-varying speed of light effect when one interprets the cosmological parameters from the observed angular diameter obtained from the X-ray cluster.

 \subsection{BAO}
 \label{subsec:BAO}

 The oscillatory behavior of primordial plasma arises from the competition between the gravitational attraction of matter towards the primordial plasma and the outward pressure generated by the heat from photon-matter interactions. This overdense region contains dark matter (DM), baryons, and photons. The pressure results in spherical sound waves of both baryons and photons outwards from the overdensity. DM stays in the center of sound waves because it only interacts with other components through gravity. The photons and baryons moved outwards together before decoupling. However, they diffused away after the decoupling due to the lack of interactions between the photons and the baryons. That put pressure on the system, leaving behind shells of baryonic matter. The resonant shell corresponds to the first one out of all those shells representing different sound wave wavelengths, as it travels the same distance for all overdensities before decoupling. The radius of this traveling distance is called the sound horizon. There remains only the gravitational force acting on the baryons after the disappearance of the photon-baryon pressure driving the system outwards. Hence, the baryons and DM constructed a shape that comprised overdensities of matter both at the original position of the anisotropy and in the shell at the sound horizon for that anisotropy.

Such anisotropies eventually became the ripples in matter density that would form galaxies. As such, a higher number of galaxy pairs should exist at the sound horizon distance scale than at other length scales. This specific pattern of matter created overlapping ripples at each anisotropy in the early universe.

The effect of baryon loading on the CMB monopole is given by
\begin{align}
&\int_{0}^{\eta_{\Tdrag}} k c_{s} (\eta') d \eta' \equiv k r_{s}(\eta_{\Tdrag}) \nonumber \,, \\
&r_{s}(\eta_{\Tdrag}) = \int_{0}^{\eta_{\Tdrag}} \tc_s d \eta =  \int_{z_{\Tdrag}}^{\infty} \frac{\tc_s}{H} dz =  \int_{z_{\Tdrag}}^{\infty} \frac{\tc_{s}^{(\GR)}}{H^{(\GR)}} \sqrt{\frac{1+R}{1+\frac{1+b/6}{1+b/8} R} } dz  \neq r_{s}^{(\GR)}(\eta_{\Tdrag})\label{rs} \,,
\end{align}
where $r_s$ is the sound horizon evaluated at the baryon drag epoch and the speed of sound of the baryon-photon plasma, $\tc_{s}$ is given by
\begin{align}
\tc_s^2 &\equiv \frac{\partial P}{\partial \rho}  = \frac{\tc_{0}^2}{3} a^{\frac{b}{2}} \left( 1 + \frac{(3 + b/2) \rho_{b}}{(4+b/2) \rho_{\gamma}} \right)^{-1}\equiv \tc_{s}^{(GR)2 } a^{\frac{b}{2}} \frac{1+R}{1+\frac{1+b/6}{1+b/8} R} \quad \text{where} \quad R = \frac{3 \rho_b}{4 \rho_{\gamma}} \label{cs2} \,.
\end{align}
Therefore, there is a difference between the sound horizon value in the meVSL model and GR.

Since the galaxy distribution is three-dimensional, the measurement of the sound horizon must be conducted in three different directions. Two are the projected sky directions, and the other is the radial direction \cite{Salzano:2015mxk}. The former is called the tangential mode, while the latter is the radial one defined as
\begin{align}
y_{\text{t}}(z) &= \frac{D_{\text{A}}(z)}{r_{s}(z_{\Tdrag})} \quad \text{and} \quad y_{\text{r}}(z) = \frac{c}{H(z)r_{s}(z_{\Tdrag})} \label{ytyr} \,.
\end{align}
Thus, one can measure the constraint on the value of $b$ from the measurements of tangential and radial modes.
However, if we limit the measurements as the combined quantities using $D_A$ and $H$, as, for example, the cube root of the product of the radial dilation times the square of the transverse dilation, the average distance \cite{Eisenstein:2005su}
\begin{align}
D_{V} = \left[ (1+z)^2 \frac{c z}{H} D_{A}^2 \right]^{1/3} = \left[ \frac{c z}{H} D_{\TM}^2 \right]^{1/3}  = \left[ \frac{\tc_0 z}{H^{(\GR)}} D_{\TM}^{(\GR) 2} \right]^{1/3} = D_{V}^{(\GR)}  \label{DV} \,,
\end{align}
or the Alcock-Paczynski (AP) distortion parameter
\begin{align}
F = (1+z) D_{A} \frac{H}{c} = (1+z) D_{A}^{(\GR)} \frac{H^{(\GR)}}{\tc_0} = F^{(\GR)} \label{AP} \,.
\end{align}
Thus, $D_{V}$ and $F$ are the same in GR and meVSL. Therefore, it is impossible to distinguish between GR and the meVSL model using average distance or AP.

\begin{figure*}
\centering
\vspace{1cm}
\begin{tabular}{cc}
\includegraphics[width=0.5\linewidth]{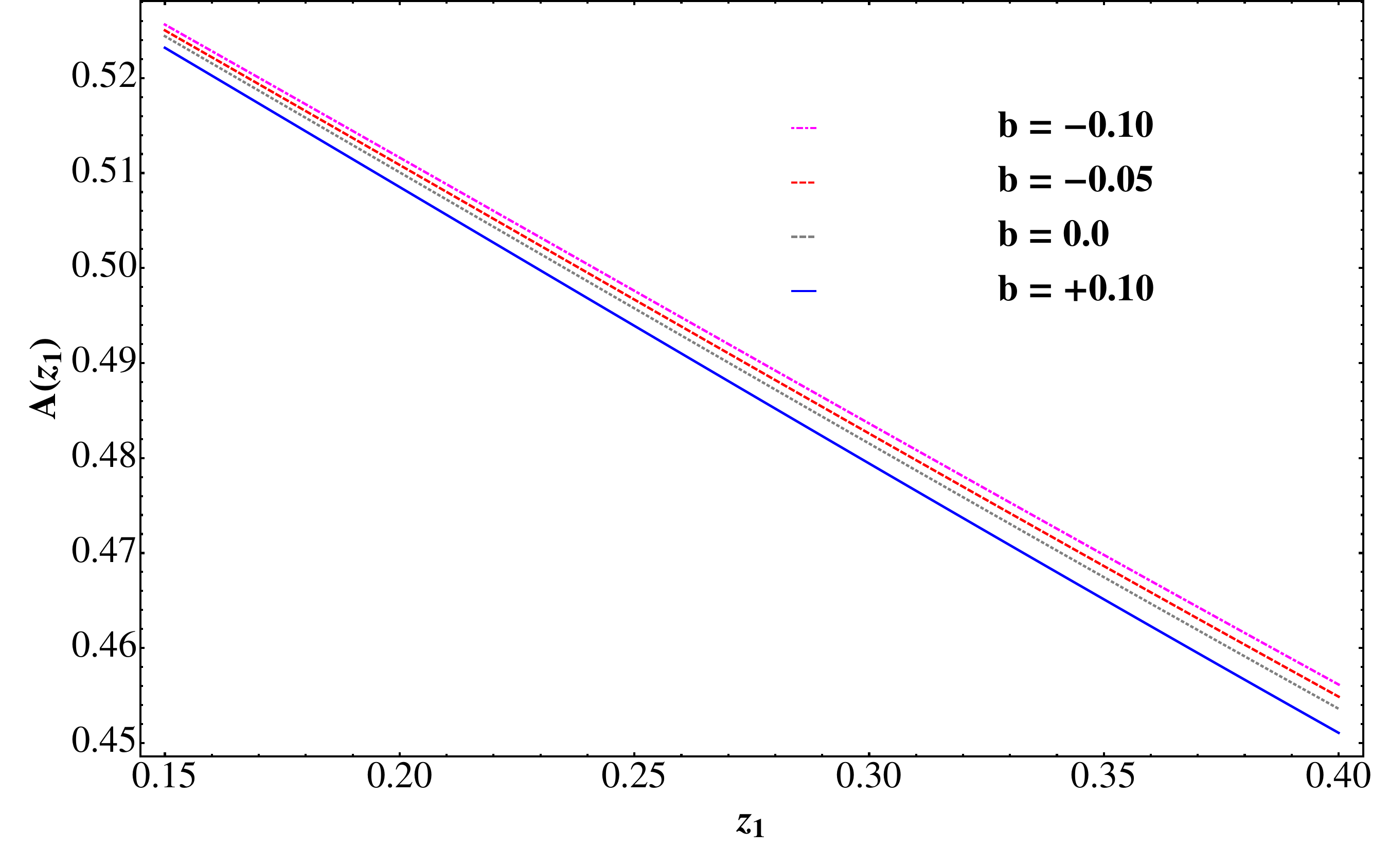} &
\includegraphics[width=0.49\linewidth]{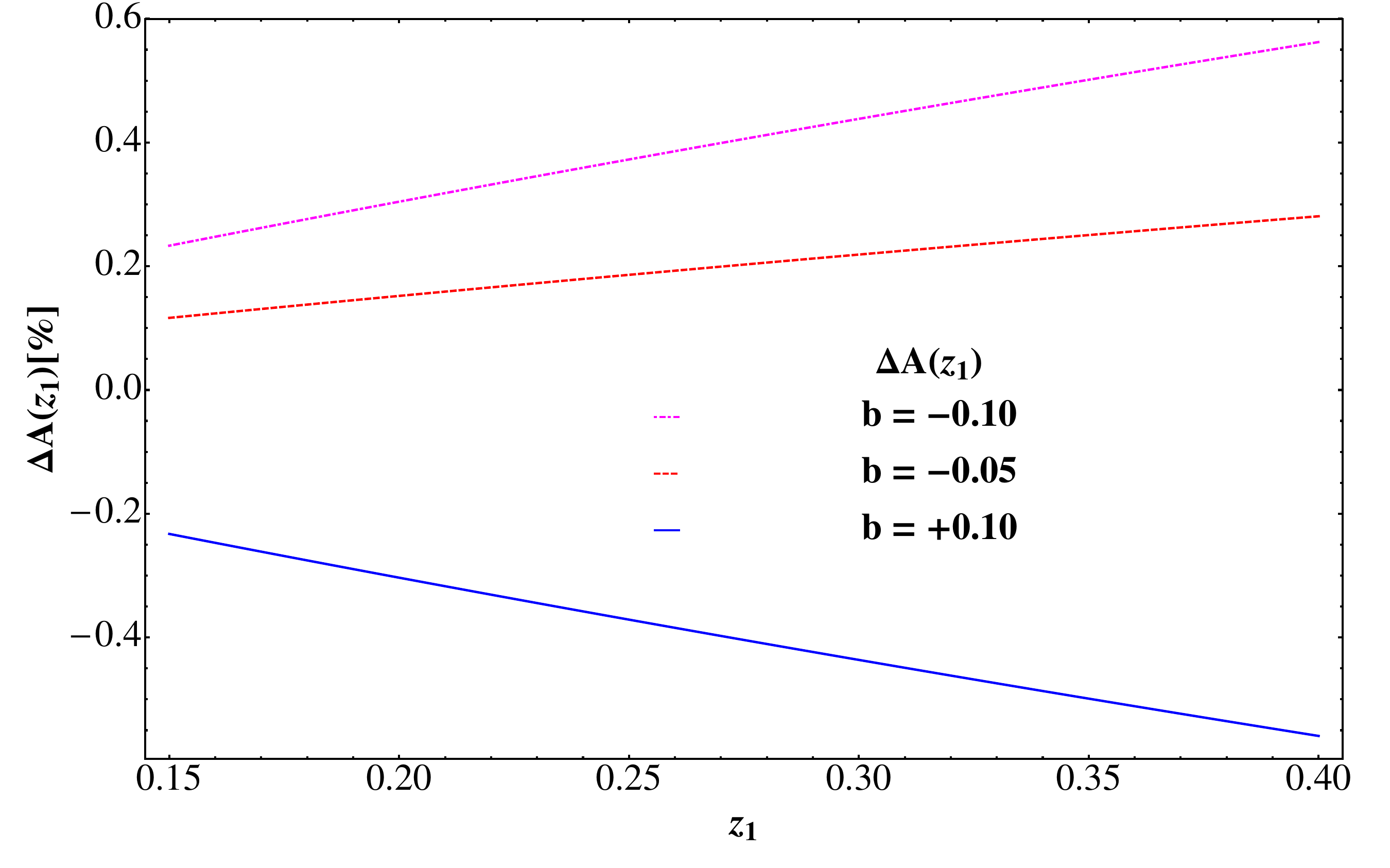}
\end{tabular}
\vspace{-0.5cm}
\caption{The BAO shift parameters for different models. a) The shift parameter, $A(z_1)$ at $z_1$ for different values of $b$. b) The difference of the BAO shift parameter, $\Delta A(z_1)$ for the different values of $b$.} \label{fig-6}
\vspace{1cm}
\end{figure*}

To investigate dark energy, the low redshift constraints on the path from today, $z=0$, to the galaxy at $z = z_1$, are investigated rather than $z_1$ to the last scattering surface. In the literature, the so-called BAO shift parameter $A(z_1)$ is often used for this purpose at a particular $z_1$ specified as \cite{Eisenstein:2005su}
\begin{align}
A(z_1) &\equiv D_{V}(z_1) \frac{\sqrt{\Omega_{m0}} H_0}{z_1 c} \equiv \left( \frac{\sqrt{\Omega_{m0}}}{E^{(\GR)} (z_1)} \right)^{\frac{1}{3}} \left[ \frac{\sqrt{\Omega_{m0}}}{z_1} \int_{0}^{z_1} \frac{dz'}{E^{(\GR)} (z')}  \right]^{\frac{2}{3}} \left( 1 + z_1 \right)^{-\frac{b}{4}} \nonumber \\
&\equiv A^{(\GR)}(z_1) (1+z_1)^{-\frac{b}{4}}  \label{AZ}  \,.
\end{align}
As shown in the above equation ~\eqref{AZ}, the BAO shift parameter of meVSL deviates from that of GR. Thus, one might need to consider the varying $c$ in this observable. We show the values of $A(z_1)$ for different models in the left panel of Fig.~\ref{fig-6}. The dot-dashed, dashed, dotted, and solid lines correspond to $b= -0.1, -0.05, 0$, and $+0.1$, respectively. The deviations of $A(z_1)$ for $b \neq 0$ from that of $b =0$ are depicted in the right panel of Fig.~\ref{fig-6}. $\Delta A(z_1) \equiv (A(z_1) - A^{(\GR)}(z_1) ) / A^{(\GR)}(z_1) \times 100  $ \% and they are all sub-percentage level for $-0.1 \leq b \leq 0.1$. Therefore, although there is a difference in $A(z_1)$ between GR and meVSL, it can be disregarded due to the given measurement accuracy \cite{Salzano:2017nvc}.

 \subsection{SNe}
 \label{subsec:SNe}

 Supernovae are promising candidates for measuring cosmic expansion. Their peak brightness appears to be quite uniform and bright enough to be visible from very great distances. The type Ia supernovae (SNe Ia) show a great uniformity both in their spectral characteristics and in their light curves that are in the way their luminosities vary as functions of time, as they reach the peak of brightness first and then fade over after around a few weeks. Therefore, they are considered standard candles.

SNe Ia are thought to be nuclear explosions of WDs in binary systems. The WD gradually cumulates matter from an evolving companion, and its mass reaches toward the Chandrasekhar limit. WDs resist gravitational collapse mainly through the electron degeneracy pressure. The Chandrasekhar limit denotes the mass above which the gravitational self-attraction of the star becomes strong enough to overcome the electron degeneracy pressure in the star’s core. Consequently, a WD with a mass greater than this limit is subject to further gravitational collapse, evolving into a different type of stellar remnant, such as a neutron star or black hole. Those with masses up to this limit remain stable as WDs. Based on the equation of state for an ideal Fermi gas, the Chandrasekhar mass limit, $M^{(\Ch)}$ is given by,
\begin{align}
M^{(\Ch)} &= \frac{\omega_{3}^{0} \sqrt{3\pi}}{2} \left( \frac{\hbar \tc}{\tG} \right)^{\frac{3}{2}} \frac{1}{\left( \mu_{e} m_{\text{H}} \right)^2} = \frac{\omega_{3}^{0} \sqrt{3\pi}}{2} \left( \frac{\hbar \tc_0}{\tG_0} \right)^{\frac{3}{2}} \frac{1}{\left( \mu_{e} m_{\text{H} 0} \right)^2}  a^{-\frac{b}{2}} \equiv M^{(\Ch)}_{0}  a^{-\frac{b}{2}}  \label{MCh} \,,
\end{align}
where $\omega_{0}^3 \approx 2.018$ is a constant related to the solution to the so-called Lane-Emden equation, $\mu_{e}$ is the average molecular weight per electron determined by the chemical composition of the star, and $m_{\text{H} 0}$ is the present value of the mass of the hydrogen atom.

\begin{figure*}
\centering
\vspace{1cm}
\begin{tabular}{cc}
\includegraphics[width=0.5\linewidth]{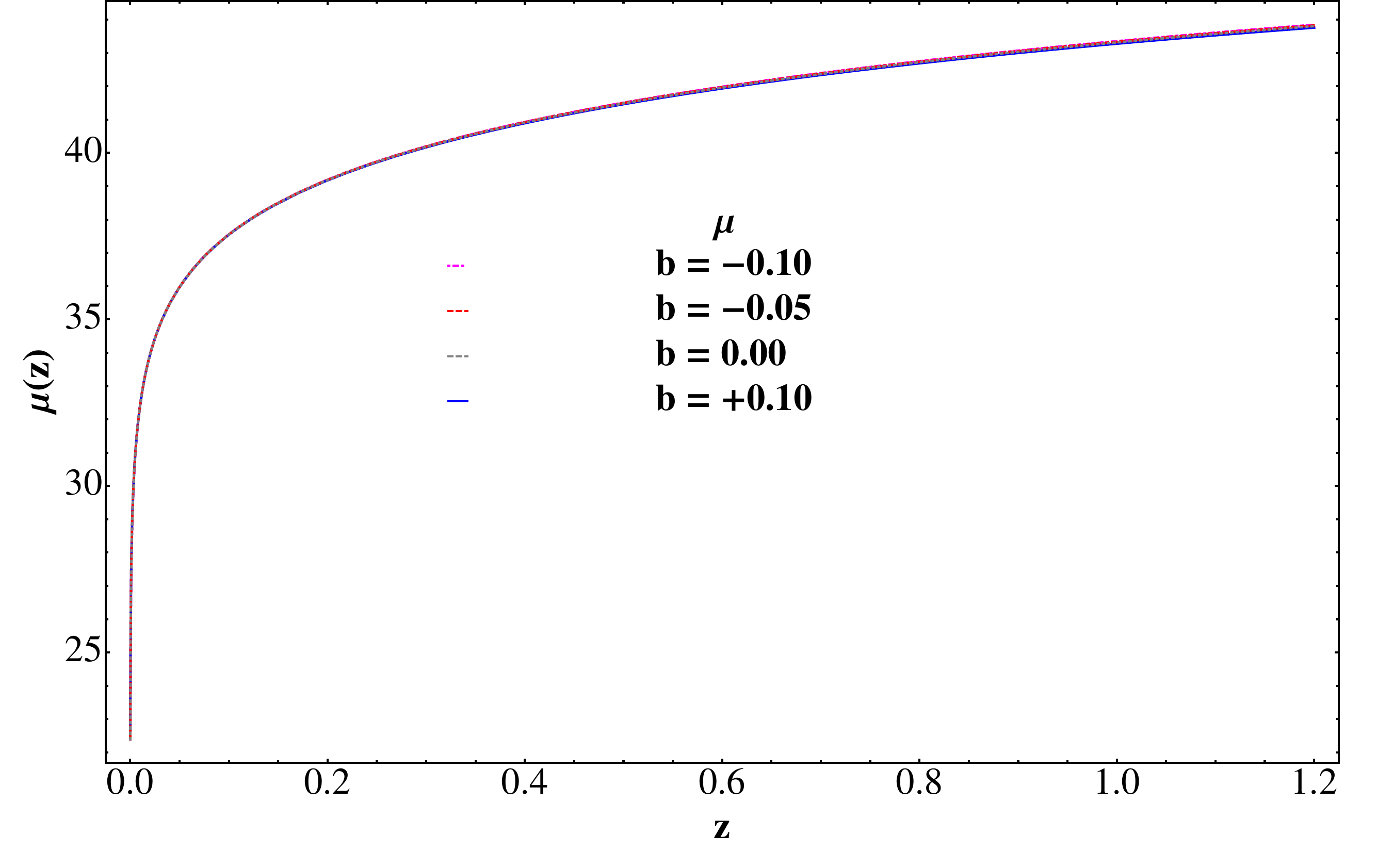} &
\includegraphics[width=0.49\linewidth]{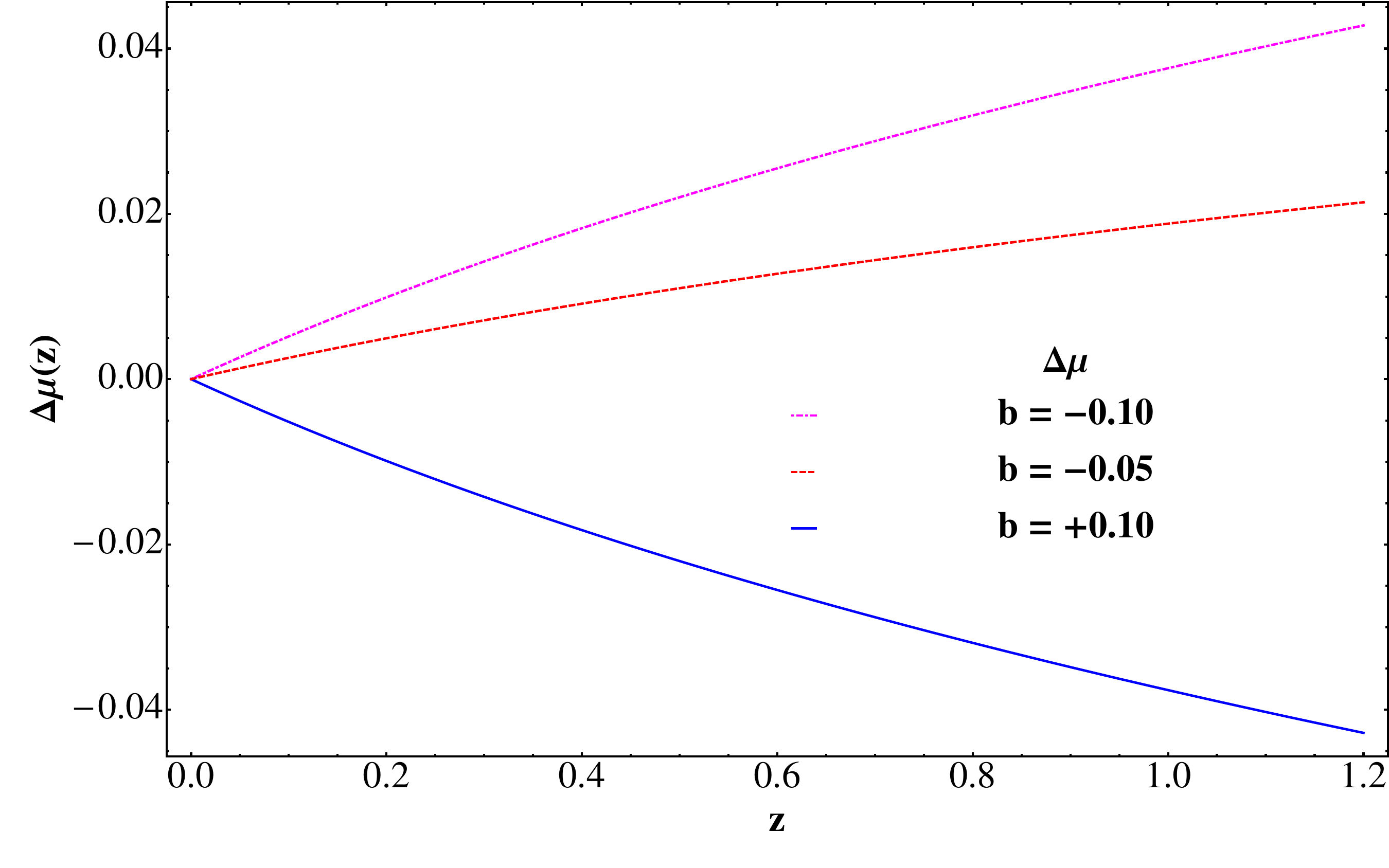}
\end{tabular}
\vspace{-0.5cm}
\caption{The distance module, $\mu$, and the difference of distance module, $\Delta \mu$, for the different values of $b$.} \label{fig-7}
\vspace{1cm}
\end{figure*}

The peak luminosity is proportional to the mass of synthesized nickel in the simple analytic light curve models. This mass is a fixed fraction of the Chandrasekhar mass to a good approximation. The actual fraction varies when different specific SNe Ia scenarios are considered, but the physical mechanisms relevant to SNe Ia naturally relate the energy yield to the Chandrasekhar mass. Thus, the peak luminosity of SNe Ia is proportional to the total amount of nickel synthesized in the SN outburst, $L \propto M^{(\Ch)}$. We define that the apparent magnitude of a star would be equal to its absolute magnitude when the star was at $10$ parsecs distance from us. Therefore, the absolute magnitude is a scale that represents the luminosity of a star as $\text{M} \propto -2.5 \log [L]$. Under this assumption, we have the modification of the absolute magnitude of SNe Ia
\begin{align}
\tM - \tM_{0} = - 2.5 \log \left[ \frac{L}{L_0} \right] = \frac{5}{4} b \log \left[ a\right] \label{MmM0} \,,
\end{align}
where the subscript $0$ refers to the local value of $M$. Thus, the distance module of meVSL, $\mu$ is written by
\begin{align}
\mu = 5 \log \left[ \frac{D_{\TL}}{1 \TMpc} \right] + 25 + \frac{5}{4} b \log \left[ a\right]  = \mu^{(\GR)} - \frac{5}{4} b \log \left[ 1+z \right]  \label{mumeVSL} \,,
\end{align}
where $D_{\TL}$ is the luminosity distance given in Eq.~\eqref{DLDA}. Thus, one might have a measurement error in the distance module when one assumes the Universe is governed by GR instead of meVSL by $- \frac{5}{4} b \log \left[ 1+z \right] $. The larger the redshift, the greater the deviation. We show this in Fig.~\ref{fig-7}. The left panel of Fig.~\ref{fig-7} shows the distance modules for the different models. The dot-dashed, dashed, dotted, and solid lines correspond $b = -0.1, -0.05, 0$, and $0.1$, respectively. The differences in distance modules, $\mu - \mu^{(\GR)}$ are depicted in the right panel of Fig.\ref{fig-7}. $\Delta \mu$ can be 0.04 (0.02, -0.04) for $b = -0.1(-0.05, 0.1)$ at $z = 1.2$. These are quite small and SNe Ia are improper observations to constrain $b$ \cite{Cao:2018rzc}. This result is rather different from that of \cite{Riazuelo:2001mg,Gaztanaga:2001fh,GarciaBerro:2005yw,Mould:2014iga,Zhao:2018gwk,Kazantzidis:2018jtb,Hanimeli:2019wrt}.

\subsection{GWs}
\label{subsec:GWs}

If it is somehow possible to learn how the source's mass quadrupole moment varies with time, then a measurement of the gravitational wave (GW) amplitude would reveal that distance by using the fact that the amplitude of a GW falls off inversely with the distance to the source, $h_{jk} = 2G /(c^4 D) \ddot{I}_{jk}$. Therefore, GWs generated by the merger of two massive compact objects provide information about the distance to the merger and offer an independent method for distance measurement. Using standard sirens for luminosity distance measurements is one of the most intriguing prospects of third-generation GW detectors.

The waveform produced by a binary inspiral is modified by the propagation across cosmological distance. In a local wave zone, where the distance to the source is sufficiently large the gravitational field already has the $1/r$ behavior characteristic of waves, but still sufficiently small, so that the expansion of the Universe is negligible. The GW produced at a distance $r_{\text{phys}} = a(t_{\text{emis}}) r$ in the local wave zone is written as
\begin{align}
h_{+} \left( t_{\Ts} \right) &= h_{\Tc}(t^{\Tret}_{\Ts}) \frac{1+ \cos^2 \imath}{2} \cos \left[ 2 \Phi \left( t^{\Tret}_{\Ts} \right) \right] \quad , \quad
h_{\times} \left( t_{\Ts} \right) =  h_{\Tc}(t^{\Tret}_{\Ts}) \cos \imath \cos \left[ 2 \Phi \left( t^{\Tret}_{\Ts} \right) \right] \label{hplusts} \, \quad , \quad \text{where} \\
h_{\Tc}(t^{\Tret}_{\Ts}) &= \frac{4}{a(t_{\Temis}) r} \left( \frac{G M_{\Tchirp}}{\tc^2} \right)^{\frac{5}{3}} \left( \frac{ \pi f^{(\Ts)}_{\Tgw} (t^{\Tret}_{\Ts}) }{\tc} \right)^{\frac{2}{3}} = \frac{4}{a(t_{\Temis}) r} \left( \frac{G^{(\GR)} M_{\Tchirp}^{(\GR)}}{\tc_0^2} \right)^{\frac{5}{3}} \left( \frac{ \pi f^{(\Ts)(\GR)}_{\Tgw} (t^{\Tret}_{\Ts}) }{\tc_0} \right)^{\frac{2}{3}} a^{-\frac{1}{16}b} \nonumber \,, \\
&= h_{\bbc}^{(\GR)}(t^{\Tret}_{\Ts}) a^{-\frac{1}{16}b} \label{hcts} \, \\
f^{(\Ts)}_{\Tgw} (\tau_{\Ts}) &= \frac{1}{\pi} \left( \frac{5}{256} \frac{1}{\tau_{\Ts}} \right)^{\frac{3}{8}} \left( \frac{G M_{\Tchirp}}{\tc^3} \right)^{-\frac{5}{8}} = \frac{1}{\pi} \left( \frac{5}{256} \frac{1}{\tau_{\Ts}} \right)^{\frac{3}{8}} \left( \frac{G^{(\GR)} M_{\Tchirp}^{(\GR)}}{\tc_0^3} \right)^{-\frac{5}{8}} a^{\frac{5}{32} b} \equiv f^{(\Ts)(\GR)}_{\Tgw} (\tau_{\Ts}) a^{\frac{5}{32} b} \label{fgws} \,, \\
M_{\Tchirp} &= \mu^{\frac{3}{5}} m^{\frac{2}{5}} = M_{\Tchirp}^{(\GR)} a^{-\frac{b}{2}} \quad \text{where} \quad \frac{1}{\mu} = \frac{1}{m_1} + \frac{1}{m_2} = \frac{1}{\mu^{(\rs)}} a^{\frac{b}{2}} \,\,, \,\, m \equiv m_1 + m_2 =  m^{(\rs)} a^{-\frac{b}{2}} \label{Mchirp} \,, \\
\Phi \left( t^{\Tret}_{\Ts} \right) &= \pi \int^{t^{\rs}_{\Ts}} dt'_{\Ts} f^{(\Ts)}_{\Tgw} (t_{\Ts}') \label{Phits} \,,
\end{align}
where $\imath$ is the angle between the observer's $z$-axis and the normal to the orbit, $\tau \equiv t_{\text{coal}} - t$ is the time to coalescence, $t_{\Ts}$ is the time measured by the clock of the source, $t^{\Tret}_{\Ts}$ is the corresponding value of retarded time, $t_{\Temis}$ is the time of emission, $h_{\Tc}$ is the amplitude of GW, $M_{\Tchirp}$ is the chirp mass, $f_{\Tgw}$ is the angular frequency of the GW which is twice the orbital frequency, and $\Phi$ is the integrated phase. So in meVSL, the amplitude and phase measured by the source's clock are modified. However, we measure GW at present and we need to replace the above results in the observer's time.

After propagation from the source to the detector, the GW amplitude is given by the above equations ~\eqref{hplusts}-\eqref{Phits} with replacing $a(t_{\Temis})$ by $a(t_0)$. Then, it is convenient to express the above results in terms of the time, $t_{\Tobs} = (1+z) t_{\Ts}$ and the GW frequency, $f^{\Tobs}_{\Tgw} = f^{\Ts}/(1+z)$ measured by the observer. Both Eqs.~\eqref{hcts} and \eqref{fgws} are rewritten by
 \begin{align}
 h_{\Tc}(t^{\Tret}_{\Tobs}) &= \frac{4}{a(t_{0}) r} (1+z)^{\frac{2}{3}} \left( \frac{G^{(\GR)} M_{\Tchirp}^{(\GR)}}{\tc_0^2} \right)^{\frac{5}{3}} \left( \frac{ \pi f^{(\Tobs)(\GR)}_{\Tgw} (t^{\Tret}_{\Tobs}) }{\tc_0} \right)^{\frac{2}{3}} \nonumber \,, \\
	&\equiv \frac{4}{D_{\TL}(z)} \left( \frac{G^{(\GR)} \mathcal M_{\Tchirp}^{(\GR)}(z)}{\tc_0^2} \right)^{\frac{5}{3}} \left( \frac{ \pi f^{(\Tobs)(\GR)}_{\Tgw} (t^{\Tret}_{\Tobs}) }{\tc_0} \right)^{\frac{2}{3}} = h_{\bbc}^{(\GR)}(t^{\Tret}_{\Ts}) \label{hctobs} \,, \\
f^{(\Tobs)}_{\Tgw} (\tau_{\Tobs}) &= \frac{1}{\pi} \left( \frac{5}{256} \frac{1}{\tau_{\Tobs}} \right)^{\frac{3}{8}} \left( \frac{G^{(\GR)} \mathcal M_{\Tchirp}^{(\GR)(z)}}{\tc_0^3} \right)^{-\frac{5}{8}} \equiv f^{(\Tobs)(\GR)}_{\Tgw} (\tau_{\Tobs}) \label{fgwobs} \,,
 \end{align}
 where $D_{\TL}$ is the luminosity distance given in Eq.~\eqref{DLDA}. We use $a_{\Tobs} = a_{0} = 1$ and define $\mathcal M_{\Tchirp} (z)= (1+z) M_{\Tchirp}$. Also we can use $\mathcal M_{\Tchirp} (z_{\Tobs}) = M_{\Tchirp}$. Thus, the observed  $h_{\Tc}$ and $f_{\Tgw}^{\Tobs}(\tau_{\Tobs})$ of meVSL are same as those of GR. 

 However, we still have the effect of meVSL on the GWs detection. As we show in the appendix.~\ref{subsec:perturbationApp}, the propagation equation of the TT gauge metric perturbations, $h$ in the FLRW background is given by Eq.~\eqref{hTTApp}
\begin{align}
h'' + 2 \mathcal H \left( 1 + \frac{b}{8} \right) h' + \tc^2 k^2 h = 0 \label{hevol} \,,
\end{align}
where $h(\eta, {\bf k})$ are the Fourier modes of the GW amplitudes, primes denote the derivative for the conformal time, and $\mathcal H= a'/a$. However, we need to replace the conformal time derivatives in the above equation.~\eqref{hevol} with the derivatives w.r.t to $\ln a$ to properly investigate the modification of GW evolutions in meVSL. Then the above equation is rewritten as
\begin{align}
\frac{d^2 h}{d \ln a^2} + \left( 2 + \frac{\mathcal H'}{\mathcal H^2} + \frac{b}{4} \right) \frac{d h}{d \ln a} + \frac{\tc^2 k^2}{\mathcal H^2} h = \frac{d^2 h}{d \ln a^2} + \left( 2 + \frac{\mathcal H'}{\mathcal H^2} + \frac{b}{4} \right) \frac{d h}{d \ln a} + \frac{\tc_0^2 k^2}{\mathcal H^{(\GR) 2}} h = 0 \label{helna} \,,
\end{align}
where $\mathcal H'/\mathcal H^2 \equiv -q = - q^{(\GR)} +b/4$ as given in Eq.~\eqref{q}. Thus, the above equation is rewritten as
\begin{align}
\frac{d^2 h}{d \ln a^2} + \left( 2 - q^{(\GR)} + \frac{b}{2} \right) \frac{d h}{d \ln a} + \frac{\tc_0^2 k^2}{\mathcal H^{(\GR) 2}} h = 0 \label{helna2} \,,
\end{align}
the source-term is the same as that of GR because $\tc = \tc_0 a^{b/4}$, $\mathcal H = \mathcal H^{(\GR)} a^{b/4}$, and the wavenumber $k$ is a constant. However, the Hubble friction term differs from GR due to the additional term $b/2$. Thus, there exists the varying-$\tc$ effect only on the friction term in the meVSL. Thus, this change affects the luminosity distance of GW which affects the amplitude of GWs, $h_{\Tc}$ as shown in Eq.~\eqref{hctobs}.
One can define $h$ as $A e^{i B}$ and insert it into Eq.~\eqref{hevol} to obtain
\begin{align}
&\frac{A''}{A} + \left(2 + \frac{b}{4} \right) \mathcal H \frac{A'}{A} - B^{'2} + \tc^2 k^2 = 0 \label{AB1} \,, \\
& 2 \frac{A'}{A} + \frac{B''}{B'} + \left( 2 + \frac{b}{4} \right) \mathcal H = 0 \label{AB2} \,.
\end{align}
Because $A$ changes slowly and the sub-horizon mode solution $k \eta \gg 1$ is a good approximation for current GWs observations, one can ignore first two terms in Eq.~\eqref{AB1} to obtain
\begin{align}
B = \pm \tc_0 k \int^{\eta} d\eta' a^{\frac{b}{4}} \label{Bh} \,.
\end{align}
By inserting Eq.~\eqref{Bh} into Eq.~\eqref{AB2}, one obtains the WKB solution
\begin{align}
h &\propto \frac{1}{\sqrt{\tc}} \exp \left[ - \int^{\eta} d \eta' \left( 1 + \frac{b}{8} \right) \mathcal H \right] \exp \left[ \pm i \tc_0 k \int^{\eta} d \eta' a^{\frac{b}{4}} \right] \nonumber \\
&\propto \frac{1}{\sqrt{\tc}} \exp \left[ - \int^{\eta} d \eta' \frac{b}{8} \mathcal H \right] \frac{\exp \left[ \pm i \tc_0 k \int^{\eta} d \eta' a^{\frac{b}{4}} \right]}{\exp \left[ \pm i \tc_0 k \int^{\eta} d \eta' \right]} \exp \left[ - \int^{\eta} d \eta' \mathcal H \right] \exp \left[ \pm i \tc_0 k \int^{\eta} d \eta' \right] \\
&\equiv \left( 1 + z \right)^{-\frac{b}{8}} e^{- i k \Delta T} h^{(\GR)} \quad \text{with} \quad \frac{\exp \left[ \pm i \tc_0 k \int^{\eta} d \eta' a^{\frac{b}{4}} \right]}{\exp \left[ \pm i \tc_0 k \int^{\eta} d \eta' \right]} \equiv  e^{- i k \Delta T}  \label{hAB2} \,,
\end{align}
where we use $\mathcal H d \eta = - dz/(1+z)$. The first term is the damping factor from the change of evolution and the second term is phase shift due to the time delay due to the change of speed of light. As a result, the lower bound interpretation of equation \eqref{hevol} moves at the effective scale $\ta = a^{1+b/8}$. The amplitude, $h$ in the propagation of cosmological distance decreases as $\ta^{-1}$ instead of $a^{-1}$. Thus, a GW luminosity distance $D_{\TL}^{\Tgw}$ can be defined as
\begin{align}
D_{\TL}^{\Tgw}(z) = \frac{a(z)}{\ta(z)} D_{\TL}(z) = (1+z)^{\frac{b}{8}} D_{\TL}(z) \label{DLgw} \,,
\end{align}
where we use $\ta = a^{1+b/8}$ and $D_{\TL}$ is the usual (electromagnetic) luminosity distance given in Eq.~\eqref{DLDA}.

\begin{figure*}
\centering
\vspace{1cm}
\begin{tabular}{cc}
\includegraphics[width=0.5\linewidth]{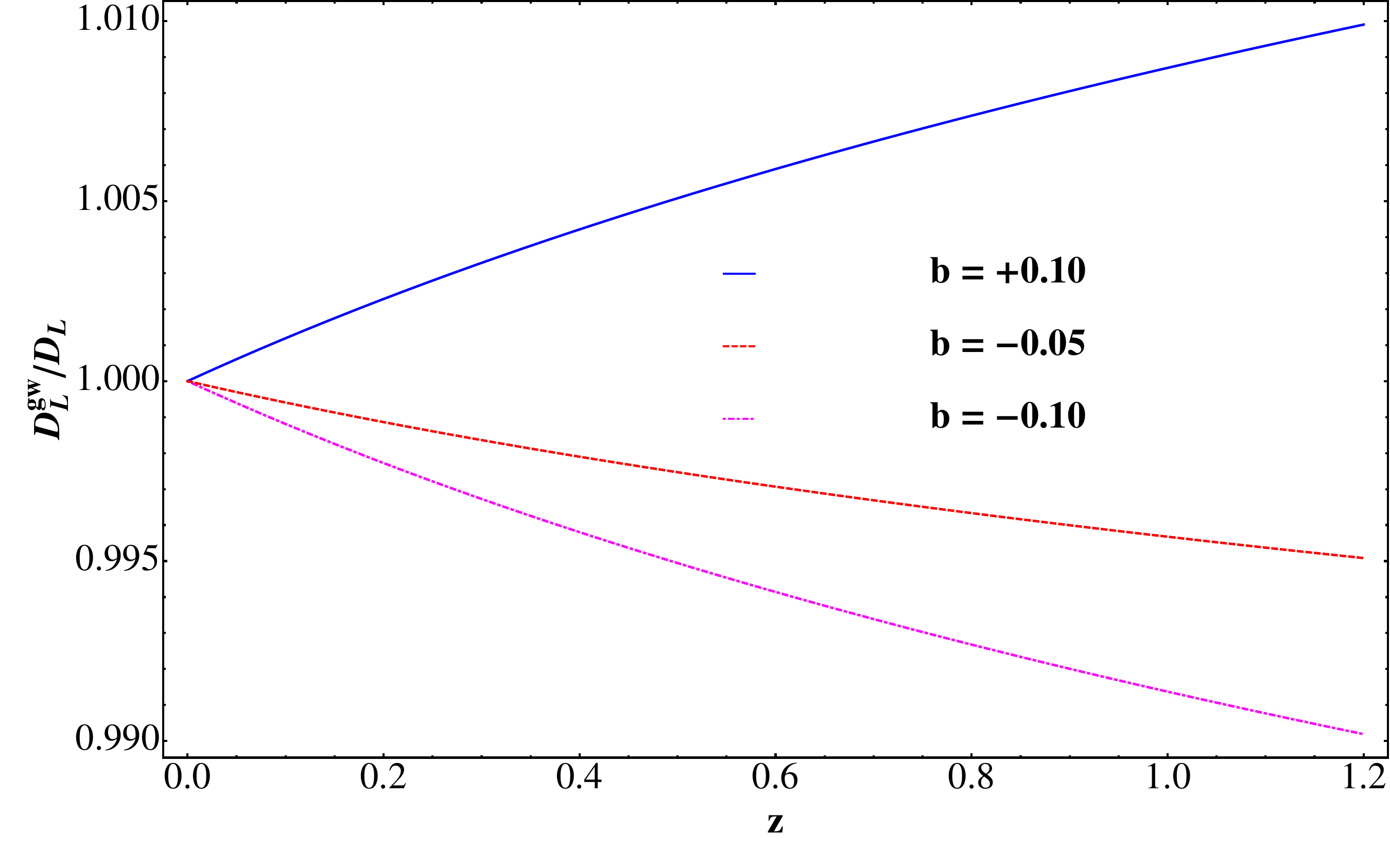} &
\includegraphics[width=0.49\linewidth]{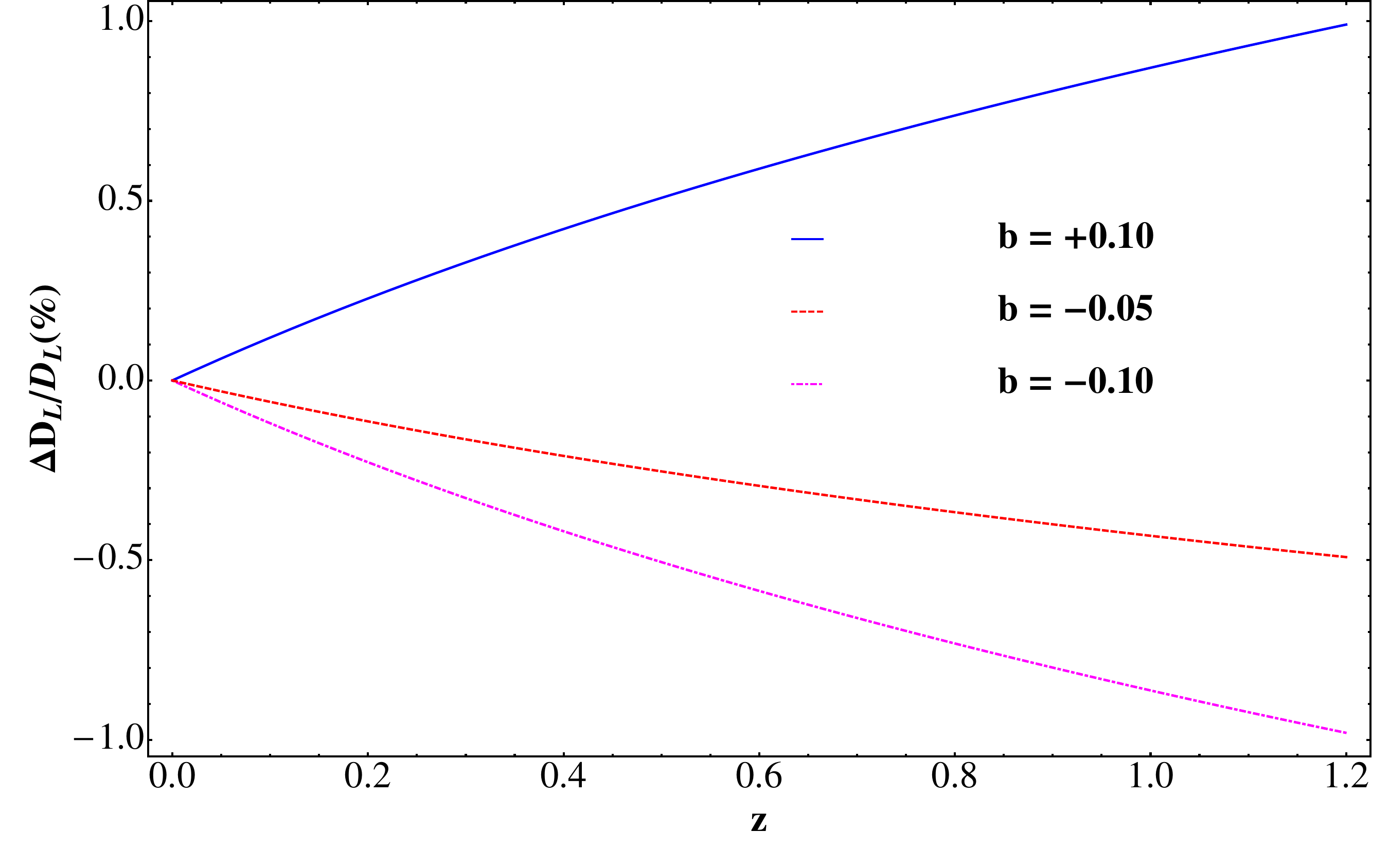}
\end{tabular}
\vspace{-0.5cm}
\caption{The GW luminosity distance $D_{\TL}^{\Tgw}$ of meVSL. a) The ratio $D_{\TL}^{\Tgw}/D_{\TL}$ for different values of $b$. b) The relative difference $\Delta D_{\TL}^{\Tgw}/D_{\TL} (\%)$ between meVSL and GR for different values of $b$. } \label{fig-8}
\vspace{1cm}
\end{figure*}

There have been similar results on the luminosity distance of GW for the modified gravity models. It uses the fact that the time variation of the gravitational constant G can be constrained from the GW observations of merging binary neutron stars. One can relate the luminosity distance to the gravitational constant G from the Friedmann equation. This equally means that the GWs give information about the value of G at the time of the merger. Thus, the measured masses of neutron stars from the GW observations can be inconsistent with the theoretically allowed range, if there exists a significant time evolution of $G$ from the merging epoch to the present epoch. One might be able to place bounds on the variation of $G$ between the merger epoch and the present one by using GWs. 

In meVSL, the ratio $D_{\TL}^{\Tgw}/D_{\TL}$ is given by $(1+z)^{b/8}$ as given in Eq.~\eqref{DLgw}. We plot this ratio in the left panel of Fig.~\ref{fig-8}. For a positive value of $b$, this ratio is greater than $1$. This ratio becomes less than 1 for the negative value of $b$. The solid, dashed, and, dot-dashed lines correspond $b= +0.1$, $-0.05$, and $-0.1$, respectively. We show the relative difference $\Delta D_{\TL}/D_{\TL} (\%) = (D_{\TL}^{\Tgw} - D_{\TL})/D_{\TL} \times 100 $ for different models ({\it i.e.}, for different values of $b$) in the right panel of Fig.~\ref{fig-8}. For standard sirens, we need to compare the GW luminosity distance of meVSL to that of GR. There is about 1\% at $z=1.2$ difference between the two models when $|b|= 0.1$. For values of $b$ smaller than $0.1$, there will be a sub-percent difference between two luminosity distances.

\subsection{Hubble parameter}
\label{subsec:H}

As we show in the subsection~\ref{subsec:FLRWsol}, the Hubble parameter of meVSL in Eq.~\eqref{H2mp} can be rewritten as
\begin{align}
H(z) &= H_0 E(z)^{(\GR)} \left( 1 + z\right)^{-\frac{b}{4}} \equiv 100 \, h(z) \, E(z)^{(\GR)} \, [\text{km/s/Mpc}]  \label{Hz} \,,
\end{align}
where $h[z] \equiv h (1+z)^{-b/4}$ with $h$ is known as 0.6688 in Planck 2018 \cite{Aghanim:2018eyx}. Therefore, while the value of $h$ remains constant in GR, we can interpret that $h[z]$ cosmologically varies in the meVSL model.

\begin{figure*}
\centering
\vspace{1cm}
\includegraphics[width=0.8\linewidth]{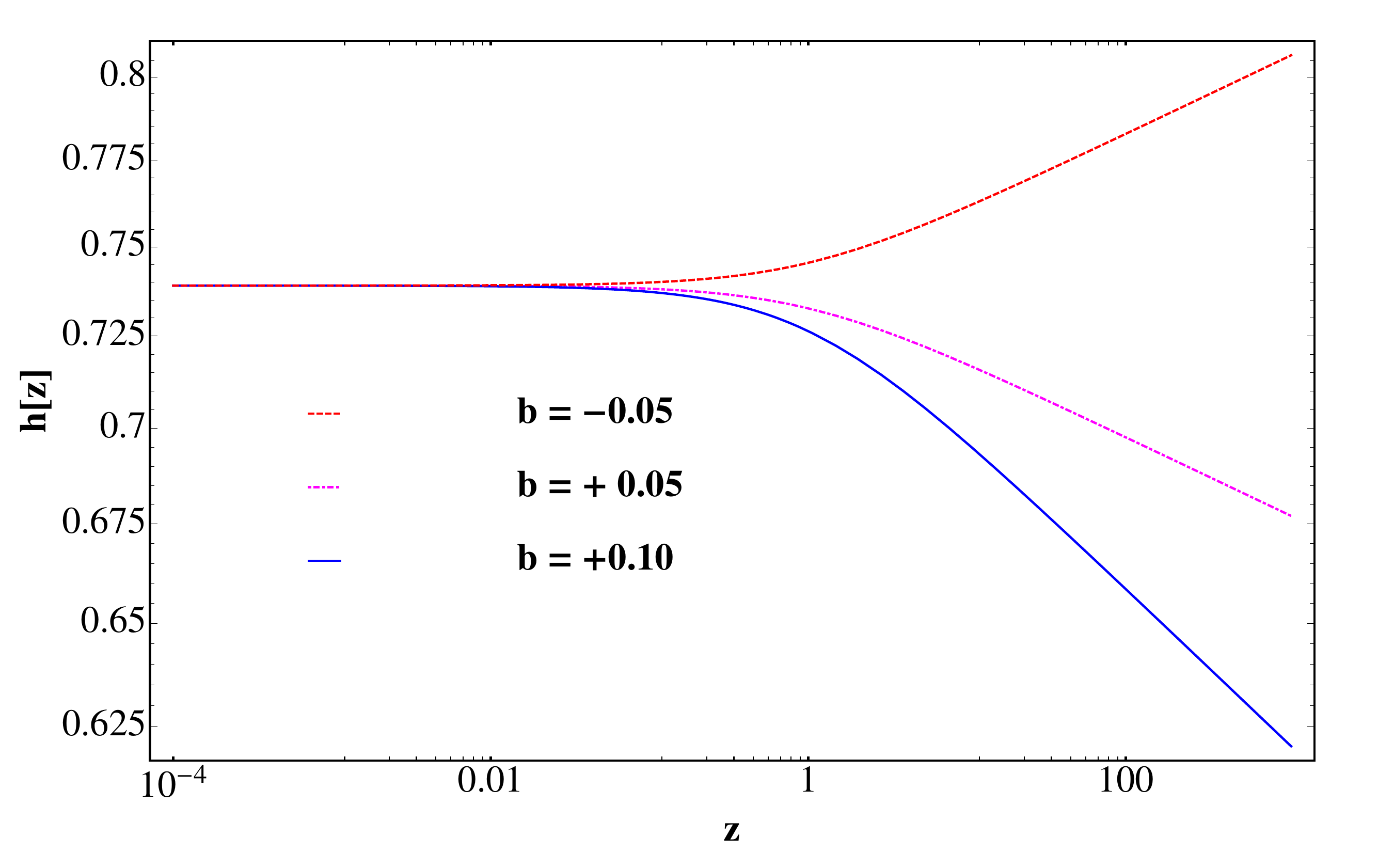} 
\vspace{-0.5cm}
\caption{The cosmological evolution of the reduced Hubble parameters, $h[z]$ for the different values of $b$.} \label{fig-9}
\vspace{1cm}
\end{figure*}

The present values of local measurements of the Hubble parameter, from the Cepheid-calibrated SN~\cite{Riess:2016jrr,Riess:2018byc,Riess:2019cxk,Reid:2019tiq} and time delays of strong lensing~\cite{Bonvin:2016crt,Birrer:2018vtm,Wong:2019kwg}, are close to $74$ km/s/Mpc. However, $H_0$-values inferred from a $\Lambda$CDM fit the CMB~\cite{Aghanim:2018eyx}, various large-scale structures~\cite{Abbott:2017smn,Cuceu:2019for,Schoneberg:2019wmt,DAmico:2019fhj,Ivanov:2019pdj,Colas:2019ret,Philcox:2020vvt}, and the local measurements based on the inverse distance ladder method \cite{Lemos:2018smw} converge to $68$ km/s/Mpc. This discrepancy (the so-called ``Hubble tension'' or ``$H_0$-tension'') is not easily explained by any obvious systematic effect in either measurement~\cite{Efstathiou:2013via,Rigault:2014kaa,Addison:2015wyg,Aghanim:2016sns,Aylor:2018drw,Kochanek:2019ruu,Blum:2020mgu} and so increasing attention is focusing on the possibility that this ``Hubble tension'' may be indicating new physics beyond the standard cosmological model~\cite{Bernal:2016gxb,Freedman:2017yms,Feeney:2017sgx,Poulin:2018cxd,Knox:2019rjx}.

If one adopts the $H_0$-tension as the evidence for the new physics, then there are two promising ways to alter early cosmology so that the tension between the CMB-inferred value and the measured value of $H_0$ is reduced. These are either changing the early time expansion history or changing the details of recombination. The second effect is changes in the details of nucleosynthesis can be captured by changes in the primordial Helium mass fraction, parametrized by $\YPBBN$. However, this is not the case for the meVSL model. Instead, the expansion history can be modified in the early Universe in the meVSL and this can be used to solve the $H_0$-tension.

We can interpret $h[z] =  h (1+z)^{-b/4}$ in Eq.~\eqref{Hz} as follows. Even though the present value of $h[z]$ is $h$, its past value was smaller (larger) than that of $h$ for the positive (negative) value of b. This is depicted in Fig.~\ref{fig-9}. The dashed, dot-dashed, and solid lines correspond to $b = -0.05, 0.05$, and $0.1$, respectively. If we choose $h = 0.739$, then $h[z=1091] = 0.6688$ when $b = 0.057$. It can explain why the local measurement $0.74$ is identical to the CMB measured value, $0.67$.

\begin{figure*}
\centering
\vspace{1cm}
\begin{tabular}{cc}
\includegraphics[width=0.5\linewidth]{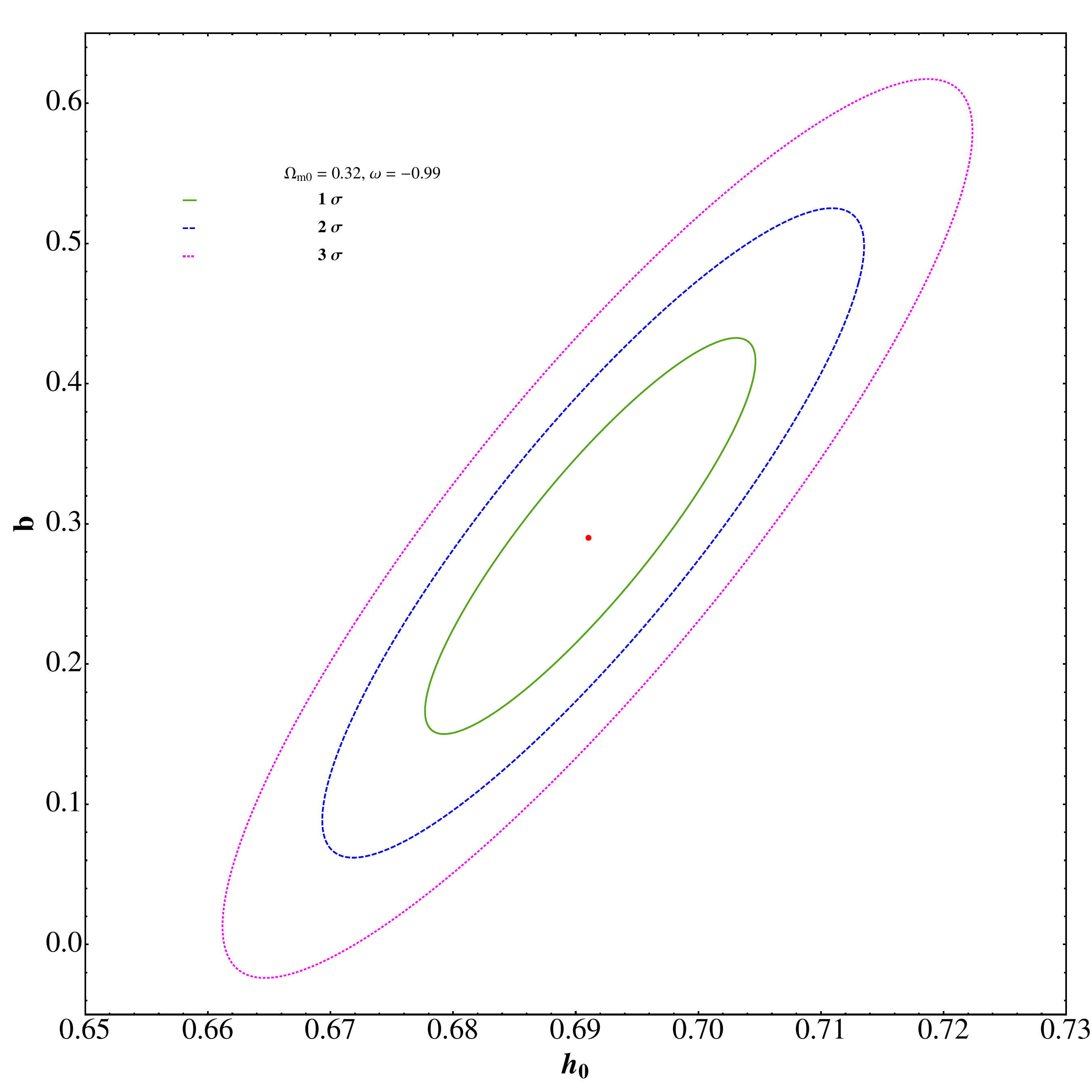} &
\includegraphics[width=0.5\linewidth]{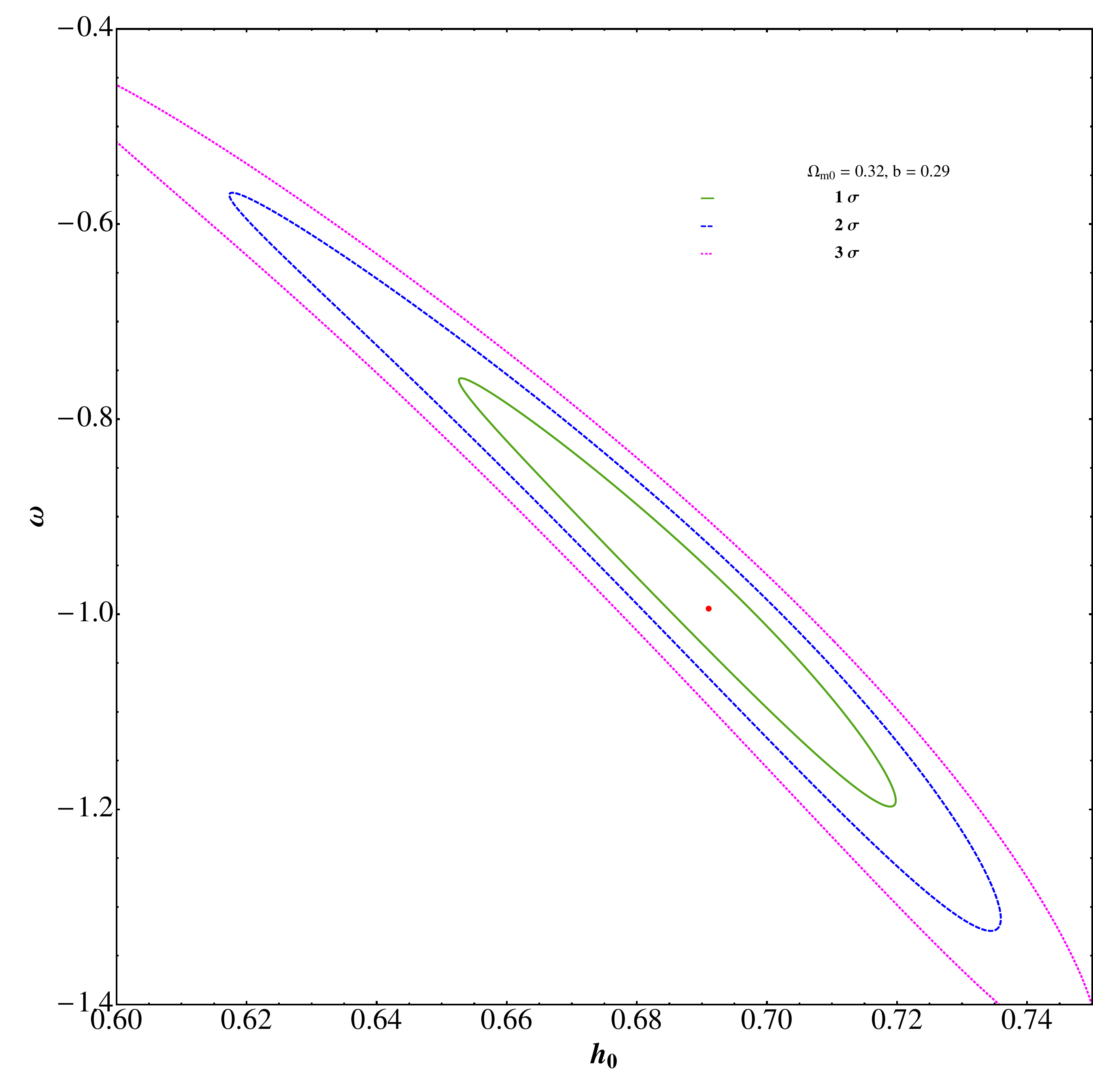}
\end{tabular}
\vspace{-0.5cm}
\caption{Contour plots of $h$ for $b$ and $\omega$. a) 1,2, and, 3-$\sigma$ contour plots for $h_0$ and $b$.  b) 1,2, and, 3-$\sigma$ contour plots for $h_0$ and $\omega$.  } \label{fig-10}
\vspace{1cm}
\end{figure*}

The Hubble parameter, which directly investigates the expansion history of the Universe, is also related to differential redshift and can be expressed as $H(z) = - (dz/dt)/(1+z)$. In this context, spectroscopic surveys provide $dz$, and the measurement of $dt$ allows for determining the Hubble parameter. Two methods generally measure the Hubble parameter values $H(z)$ at a certain redshift. They are extraction of $H(z)$ from line-of-sight BAO data and differential age (DA) method estimating $H(z)$. From the reference~\cite{Pacif:2020hai}, we use $57$ data points, of which $31$ is measured from the DA method and $26$ is obtained with BAO and other methods. Using this data, we perform a maximum likelihood analysis to obtain the optimal values for $b$, $\omega$, and $h$ in the meVSL model. The best fit values of ($ b , \omega, h$) = (0.290, -0.994, 0.691) if we adopt $\Omega_{m0} = 0.32$ and $\Omega_{\Lambda 0} = 0.68$. We obtain the 1, 2, and 3-$\sigma$ contour plots for the fixed $b$ and $\omega$, respectively. Figure~\ref{fig-10} illustrates these. In the left panel of Fig.~\ref{fig-10}, the solid, dotted, and dashed lines correspond $1$-, $2$-, and, $3$-$\sigma$ level contour plots with $\omega = -0.99$. In the right panel of Fig.~\ref{fig-10}, we also show the contour plots of $h$ verse $\omega$ for the fixed value of $b = 0.29$. Solid, dashed, and dotted lines represent contour levels at the $1$-, $2$-, and $3$-sigma levels, respectively. In this case, the $h$ value can include both the local and the CMB values.

 \subsection{Strong lensing}
 \label{subsec:StrongLensing}

 In GR, the presence of matter curves spacetime, and this causes the deflection of the path of a light ray as a result. This process is known as the gravitational lensing. The most extreme bending of light (\textit{i.e.}, strong lensing) occurs when the lens is very massive with the source of the lens close enough to it. In this case, lights can reach the observer through different paths with more than one image of the source. One can estimate the volume of space back to the sources by using the number of discovered lenses. This volume depends strongly both on cosmological parameters and on gravity theories. However, due to the bending of space, the distances each image moves differ, which can cause a time delay in the image changes. One can estimate the present value of the Hubble constant, $H_0$, from these time delays. In some cases, the source and the lens are in a special alignment to make a light will be deflected to the observer and produce an Einstein ring. The Einstein radius, $\theta_{\text{E}}$ under the singular isothermal sphere (SIS) model is given by \cite{Colaco:2020ndf}
\begin{align}
\theta_{\text{E}} = 4 \pi \frac{D_{A}^{\Tls}}{D_{A}^{\Tso}} \frac{\sigma_{\text{SIS}}^2}{\tc^2} = 4 \pi \frac{D_{A}^{\Tls}}{D_{A}^{\Tso}}  \frac{\sigma_{\text{SIS}}^2}{\tc_0^2} (1+z)^{\frac{b}{2}} = \theta_{\text{E}}^{(\GR)} (1+z)^{\frac{b}{2}} \label{thetaE} \,,
\end{align}
where $D_{A}^{\Tls}$ is the angular diameter distance from the lens to the source, $D_{A}^{\Tso}$ is that from the source to the observer, and $\sigma_{\text{SIS}}$ is the velocity dispersion due to the lens mass distribution. The angular diameter distance in both GR and the meVSL model is identical, with the only difference between the two models being the speed of light in equation~\eqref{thetaE}. $\theta_{\text{E}}$ is the observed quantity, and one can estimate cosmological parameters from the angular diameters based on the given gravity theory. Therefore, if meVSL is the true gravitational theory governing our universe, to accurately extract cosmological parameters from strong lensing signals, the Einstein radius derived from GR must be adjusted by a factor of $(1+z)^{b/2}$.

The observations of gravitationally lensed quasars are best understood in light of Fermat’s principle. Intervening mass between a source and an observer introduces an effective index of refraction, thereby increasing the light travel time. The competition between this Shapiro delay due to the gravitational field and the geometric delay from bending the ray paths leads to the formation of multiple images at the stationary points of the travel time \cite{Schneider:1992}. There exists a thin-lens approximation that applies when the optics are small compared to the distances to the source and the observer. In this approximation, we need only the effective potential, $\psi(\vec{x}) = (2/\tc^2) D_{\Tls}/D_{\text{s}} \int dz \phi$ found by integrating the 3D potential $\phi$ along the line of sight. The light-travel time is
\begin{align}
\tau (\vec{x}) = \frac{1 + z_{\text{l}}}{\tc} \frac{D_{\text{l}} D_{\text{s}}}{D_{\text{ls}}} \left( \frac{1}{2} \left( \vec{x} - \vec{\beta} \right)^2 - \psi (\vec{x}) \right) = \tau^{(\GR)} (1+z)^{-\frac{b}{4}} \label{SLtau} \,,
\end{align}
where $\vec{x}$ is the angular position of the image, $\vec{\beta}$ is the angular position of the source, $\psi (\vec{x})$ is the effective potential, $z_{\text{l}}$ is the lens redshift. $D_{\text{l}}$, $D_{\text{s}}$, and $D_{\text{ls}}$ are angular diameter distances to the lens, to the source, and from the lens to the source, respectively. $\left( \vec{x} - \vec{\beta} \right)^2/2$ is the geometric delay in the small-angle approximation. As for the Einstein radius, there is an additional factor of $(1+z)^{-b/4}$ in the light travel time.

 \subsection{Fine structure constant }
 \label{subsec:alpha}

 The strength of the electromagnetic interaction between elementary charged particles can be described by the so-called fine structure constant, $\alpha$, also known as Sommerfeld's constant. It is a dimensionless constant obtained from the combination of another physical constant as
\begin{align}
\alpha \equiv \frac{e^2}{4 \pi \tepsilon \hbar \tc} = \frac{e_0^2}{4 \pi \tepsilon_0 \hbar_0 \tc_0} (1+z)^{\frac{b}{4}} \equiv \alpha^{(\GR)} (1+z)^{\frac{b}{4}} \label{alphaEM} \,,
\end{align}
where we use $e = e_0 a^{-b/4}$, $\tepsilon = \tepsilon_0 a^{-b/4}$, $\hbar = \hbar_0 a^{-b/4}$, and $\tc = \tc_0 a^{b/4}$ as shown both in Sec. \ref{subsec:EM} and in Sec. \ref{subsec:TEQ}. In the meVSL model, we assume the conservation of particle number, and Eq.~\eqref{alphaEM} above is a result of this assumption. Compared to the $\alpha^{(\GR)}$, it has the additional factor, $(1+z)^{b/4}$ and thus $\alpha$ of meVSL in the past is larger (smaller) than $\alpha^{(\GR)}$ for the positive (negative) value of b.
Thus, a temporal variation of $\alpha$ over cosmological time periods, $\dot{\alpha}/\alpha$ and the fractional difference of it $\Delta \alpha/\alpha_0$ are given by
\begin{align}
\frac{\dot{\alpha}}{\alpha} &= - \frac{b}{4} H(z) = - \frac{b}{4} (1+z)^{-\frac{b}{4}} H^{(\GR)} (z) \, ,  \label{dotalpha} \\
\frac{\Delta \alpha}{\alpha_0} &\equiv \frac{\alpha(z) - \alpha_{0}}{\alpha_0} = (1+z)^{\frac{b}{4}} - 1 \, , \label{Deltaalphaoalpha}
\end{align}
where $\alpha_0 = \alpha^{(\GR)} \approx 1/137$ denotes the value of $\alpha$ at the present epoch ({\it i.e.}, $z=0$). Observational limits on these values from the Keck telescopes are $\dot{\alpha}/\alpha = (6.45 \pm 1.35) \times 10^{-16} \text{yr}^{-1}$ and $\Delta \alpha / \alpha = (-5.43 \pm 0.116) \times 10^{-6}$ for $0.2 < z < 3.7$ \cite{Murphy:2003hw} . If we match these results, $b$ should be negative, with a magnitude of approximately $10^{-5}$. There is a strong constraint on $b$, and adopting this value may result in weak effects predicted by the meVSL model, possibly lacking observable results comparable to those of general relativity.

The method of obtaining the observational values of $\Delta \alpha$ is called the 'many multiplets (MM)' method \cite{Kotus:2016xxb}. This method compares the relative velocity spacing between different metal ion transitions and relates it to possible variation in $\alpha$. For example, considering just a single transition,
variation in $\alpha$ is related to the velocity shift $\Delta v_i$ of a transition
\begin{align}
\frac{\Delta \alpha}{\alpha} \equiv \frac{\alpha_{\text{obs}} - \alpha_{\text{lab}}}{\alpha_{\text{lab}}} \approx -\frac{\Delta v_i}{2\tc} \frac{\omega_i}{q_i} = -\frac{\Delta v_i}{2\tc_0} \frac{\omega_i}{q_i} (1+z)^{\frac{b}{4}} \label{Deltaaoa} \,,
\end{align}
where $q_i$ is the sensitivity of the transition to $\alpha$ variation, calculated from many-body relativistic corrections to the energy levels of ions, and $\omega_i$ is its wavenumber measured in the laboratory. Thus, one should include the effect of the variation of the speed of light in the data analysis as given in the above Eq.~\eqref{Deltaaoa}. However, this effect is negligible and might not affect the observational results.

 \subsection{The gravitational constant and the speed of light}
 \label{subsec:Gdotcdot} 

Stringent constraints on violations of the equivalence principle (EP) can be obtained from LLR experiments. By varying the laser distance to the Moon, we can improve the accuracy of constraints on violations of the equivalence principle (EP). These analyses give an EP test. In addition to the SEP constraint, the PPN parameters $\gamma$ and $\beta$ affect the orbit of relativistic point masses, and $\gamma$ also influences time delay. LLR can test this orbital $\beta$ and $\gamma$ dependence, geodesic de-Sitter precession, and $\dot{G}/G$. In this LLR analysis, the limit on the temporal variation of the gravitational constant is $ \dot{G}/G = ( 4 \pm 9 ) \times 10^{-13} \text{yr}^{-1} $ and $\dot{\tc}/\tc  = (0 \pm 2) \times 10^{-12} \text{yr}^{-1} $ \cite{Williams:2004qba}.

One can use Eq.~\eqref{dotGoGmp} to obtain the expressions of time variations of those quantities
\begin{align}
\frac{\dot{G}}{G} &= \frac{d \ln G}{d t} = b H = b H^{(\GR)} (1+z)^{-b} \label{dotGoG} \,, \\
\frac{\dot{\tc}}{\tc} &= \frac{d \ln \tc}{d t} = \frac{b}{4} H = \frac{b}{4} H^{(\GR)} (1+z)^{-\frac{b}{4}}  \label{dotcoc2} \,.
\end{align}
We show these results in subsection \ref{subsec:FLRWsol}. 

\section{Conclusions}
\label{sec:conclusion}

One of the main motivations for probing previous time-varying speed of light models is to provide alternative solutions for early time problems of the Big Bang model. However, we have investigated the new kind of time-varying speed of light model as an alternative solution for various late time problems of the standard $\Lambda$CDM model.

We have proposed a new time-varying speed of light model that satisfies Lorentz invariance, including Maxwell's equations and local thermodynamic equilibrium. The cosmological evolution of the speed of light also induces the evolutions of both permittivity and permeability. As a result, charges also evolve cosmologically to satisfy charge conservation. Using local thermodynamic equilibrium and conservation of photon number density, one can also obtain the cosmological evolution of the Planck constant. It also provides the cosmological evolution of frequency. However, the Boltzmann constant is still constant, and both the wavelength and the temperature are the same as those of GR. The conservation of number density also provides the cosmological evolution of the rest mass. When we derive the Einstein field equation from the action by allowing the time variation of the speed of light, this also induces the time evolution of the gravitational constant, $G$. Table~\ref{tab:tabVSL} provides a summary of all these findings. We have called this new model of varying speed of light the minimally extended variation speed of light (meVSL).

In table~\ref{tab:table-2}, we have summarized both the consequent time variations of other physical constants and modifications of some observable quantities derived in this manuscript.

All of these changes in both the physical constants and the physical quantities induce the modifications of the Friedmann equations when we apply the meVSL to the Friedmann-Lema\^{i}tre-Robertson-Walker metric. In the meVSL model, the expansion rate (\textit{i.e.}, the Hubble parameter) and the expansion acceleration are modified compared to GR. On the other hand, the meVSL model and GR have the same cosmic redshift and geometric distances, like the angular diameter distance and the transverse comoving distance. Nevertheless, the modification of the cosmic history of the frequency results in a modification of the luminosity distance. meVSL also affects the cosmological observations, like the cosmic microwave background, the Sunyaev-Zel$^{’}$dovich effect, the baryon acoustic oscillation, the supernovae, the gravitational waves, the Hubble parameter, and the strong gravitational lensing.

Thus, it is necessary to include the effect of the time variation of the speed of light when one investigates cosmological observations. It will induce changes in cosmological parameter values compared to those obtained from GR. As an example, we have scrutinized the Hubble tension, and meVSL might be able to solve this problem, as we have shown in section \ref{subsec:H}. One can analyze the cosmological observations based on meVSL to obtain new best-fit values on cosmological parameters.

\begin{table}[]
\caption{Equivalences and differences in physically observable quantities between the meVSL model and GR. EM, TD, and KN represent electromagnetism, thermodynamics, and kinematics, respectively. Physical quantities corresponding to SR refer to their values in an expanding universe. }
\label{tab:table-2}
     \begin{adjustbox}{width=\columnwidth,center}
\begin{tabular}{|c|c|c|c|}
\hline
\multicolumn{2}{|c|}{Criteria}              &Equivalences                                                                                                                                 & Differences                                                                                                                                                                                                                           \\ \hline
\multirow{4}{*}{SR} & \multirow{2}{*}{EM}   & \multirow{2}{*}{wavelegth $\lambda = \lambda_0 a$}                                                                                                       & \multirow{2}{*}{\begin{tabular}{cc} electric charge $e = e_{0} a^{-\frac{b}{4}}$ \\ frequency $\nu = \nu_0 a^{-1+b/4}$ \end{tabular} }                                                                                 \\
                   &                       &                                                                                                                                            &                                                                                                                                                                                                                                      \\ \cline{2-4} 
                 & \multirow{2}{*}{TD}   & \multirow{2}{*}{\begin{tabular}{cc} Boltzmann constant $k_{B}$ \\ temperature $T = T_0 a^{-1}$ \end{tabular}}                                   & \multirow{2}{*}{Planck constant $h = h_{0} a^{-\frac{b}{4}}$}                                                                                                                                                                        \\
                    &                       &                                                                                                                                            &                                                                                                                                                                                                                                      \\ \hline
\multirow{4}{*}{GR} & KN                    & redshift $z$                                                                                                                               & gravitational constant $G = G_{0} a^{b}$                                                                                                                                                                                             \\ \cline{2-4} 
                    & \multirow{3}{*}{FLRW} & \multirow{3}{*} {\begin{tabular}{ccc} equation of state $\omega$ \\ Hubble radius $c/H$ \\ angular distance $D_{A}$\end{tabular}} & \multirow{3}{*}{\begin{tabular}{ccc} mass density $\rho = \rho^{(\GR)} a^{-\frac{b}{2}}$ \\ Hubble parameter $H = H^{(\GR)} a^{\frac{b}{4}}$ \\ luminosity distance $D_{L} = D_{L}^{(\GR)} a^{\frac{b}{8}}$\end{tabular}} \\
                   &                       &                                                                                                                                            &                                                                                                                                                                                                                                      \\
                    &                       &                                                                                                                                            &                                                                                                                                                                                                                                      \\ & & & \\ \hline
\end{tabular}
    \end{adjustbox}
\end{table}

\appendix
\section{Appendix}
\label{sec:App}

In the appendix, we show some detailed calculations in contents.  

\section{Special Relativity}
\label{sec:SRapp} 

In this appendix, we derive the equations concerning the impact of the meVSL model on four-acceleration, electromagnetism, and thermodynamics.

\subsection{Four acceleration}
\label{subsec:FourAccelerationApp}

A point in Minkowski spacetime is a time and spatial position called an event, or sometimes the four-position, described in some reference frame by a set of four coordinates
\begin{align}
x^{\mu} = \left( c \left(a \right) t \,, x^{i}(t) \right) = \left( c \left(a \right) \tau \,, 0 \right) \,, \label{xmuApp}
\end{align}
where $x^{i}$ is the three-dimensional space position vector being a function of the coordinate time, $t$, in the same frame. These coordinates are the components of the position four-vector for the event. In the SR, the four-position is a four-vector that transforms from one inertial frame of reference to another by Lorentz transformations. A clock fastened to a particle moving along a worldline in four-dimensional spacetime measures the particle's proper time $\tau$. This relation is shown in the second equality of Eq.~\eqref{xmuApp}. The four-velocity of a particle is defined as the rate of change of its four-position for proper time and is the tangent vector to the particle's world line
\begin{align}
&U^{\mu} \equiv \frac{d x^{\mu}}{d \tau} = \frac{dt}{d \tau} \left( \tc \,, v^{i} \right) \equiv \gamma \left( \tc \,, v^{i} \right) \equiv \left( \ttc \,, 0 \right) \quad , \quad \text{where} \, \nonumber \\
&\tc \equiv \left( \frac{d \ln c}{d \ln a} H t + 1 \right) c \quad \text{and} \quad  \ttc \equiv \left( \gamma \frac{d \ln c}{d \ln a} H \tau + 1 \right) c \label{UmuApp} \,.
\end{align}
The Lorentz factor, $\gamma$ represents the time dilation of a process since it takes a proper time $\Delta \tau$ in its rest frame and has a longer duration $\Delta t$ measured by another observer moving relative to the rest frame. The $H t$-term in the above Eq.~\eqref{UmuApp} can be obtained in the FLRW universe as
\begin{align}
&E^{2} = H^{2}/H_0^{2} = \sum_{i} \Omega_{0i} a^{-3(1+\omega_i)+b/2}  \,\, \Rightarrow \,\, E(z) = (1+z)^{-b/4}  \sqrt{\sum_{i} \Omega_{i0} (1+z)^{3(1+\omega_i)}} \label{E2InSRApp} \;, \\
&H_0 \int_0^{t} dt' = H_0 t = - \int_{\infty}^{z} \frac{dz'}{(1+z') E(z')} = \int_{z}^{\infty} \frac{dz'}{(1+z')^{1-b/4} \sqrt{\sum_{i} \Omega_{i0} (1+z')^{3(1+\omega_i)}}} \label{H0t0inSRApp} \;, \\
& H t = H_0 t E = (1+z)^{-b/4}  \sqrt{\sum_{i} \Omega_{i0} (1+z)^{3(1+\omega_i)}} \int_{z}^{\infty} \frac{dz'}{(1+z')^{1-b/4} \sqrt{\sum_{i} \Omega_{i0} (1+z')^{3(1+\omega_i)}}} \label{HtinSRApp} \;.
\end{align}
$H_0 t_0 \sim 1$ and thus if $c$ varies as a function of the scale factor $a$ then $\tc_0$ is different from $c_0$ as shown in Eq.~\eqref{UmuApp}. We emphasize that $\tc$ is a constant at the given hypersurface (for the given cosmic time). The four-acceleration, $A^{\mu}$ is defined as the rate of change in four-velocity for the particle's proper time along its worldline
\begin{align}
A^{\mu} &\equiv \frac{d U^{\mu}}{d \tau} = \frac{dt}{d \tau} \frac{d U^{\mu}}{dt} = \gamma \left[ \dot{\gamma}  \left( \, \tc  , v^{i} \, \right) + \gamma  \left( \, \dot{\tc}  , a^{i} \, \right) \right] = \gamma^2 \left[ \frac{\dot{\gamma}}{\gamma}  \left( \, \tc  , v^{i} \, \right) + \left( \, \dot{\tc}  , a^{i} \, \right) \right] \label{AmuApp} \,,
\end{align}
where we use Eq.~\eqref{UmuApp}, and dots denote the derivatives for the coordinate time, $t$. We explicitly write terms in the above equation \eqref{AmuApp}
\begin{align}
\dot{\tc} &= \frac{b}{4} H \tc \quad , \quad
\frac{\dot{\breve{r}}}{\breve{r}} = \left( \frac{\dot{\ttc}}{\ttc} - \frac{\dot{\tc}}{\tc} \right)  \quad , \quad
\frac{\dot{\tgam}}{\tgam} = \tgam^{2} \left( \frac{\vec{v} \cdot \vec{a}}{\tc^2} - \frac{v^2}{\tc^2} \frac{\dot{\tc}}{\tc} \right) \label{dottgamotgam} \,, \\
\frac{\dot{\gamma}}{\gamma}  &= \frac{\dot{\breve{r}}}{\breve{r}} + \frac{\dot{\tgam}}{\tgam} = \left( \frac{\dot{\ttc}}{\ttc} - \frac{\dot{\tc}}{\tc} \right)  + \tgam^{2} \left( \frac{\vec{v} \cdot \vec{a}}{\tc^2} - \frac{v^2}{\tc^2} \frac{\dot{\tc}}{\tc} \right) \label{dotgammaogammaApp} \,,
\end{align}
where we use the explicit form of $\tc$ given in Eq.~\eqref{dotGoGmp} to have
\begin{align}
\frac{\dot{\tc}}{\tc} = \frac{d \ln \tc}{dt} = \frac{d \ln a}{dt} \frac{d \ln \tc}{d \ln a} = \frac{b}{4} H \label{dottcptcApp} \,.
\end{align}
We inserts Eqs.~\eqref{dottgamotgam} and \eqref{dotgammaogammaApp} into Eq.~\eqref{AmuApp} to obtain
\begin{align}
A^{\mu} &= \left( \gamma \dot{\gamma} \tc + \gamma^2 \dot{\tc} \,,  \gamma \dot{\gamma} \vec{v} + \gamma^2 \vec{a} \right) = \gamma^2   \left[ \left[ \left( \frac{\dot{\ttc}}{\ttc} - \frac{\dot{\tc}}{\tc} \right)  + \tgam^{2} \left( \frac{\vec{v} \cdot \vec{a}}{\tc^2} - \frac{v^2}{\tc^2} \frac{\dot{\tc}}{\tc} \right) \right]  \left( \, \tc  , \vec{v} \, \right) + \left( \, \dot{\tc}  , \vec{a} \, \right) \right]  \nonumber \\
&= \gamma^2 \tgam^2 \left( \frac{\vec{v} \cdot \vec{a}}{\tc} - \beta^2 \dot{\tc} + \tgam^{-2} \frac{\dot{\ttc}}{\ttc} \tc  \,,  \frac{ \vec{v} \times \left( \vec{v} + \vec{a} \right) }{\tc^2 }  + \tgam^{-2}  \frac{\dot{\ttc}}{\ttc} \vec{v} - \frac{\dot{\tc}}{\tc} \vec{v} + \vec{a}  \right)    \label{4accelApp} \,,
\end{align}
where we use
\begin{align}
\vec{v} \left( \vec{v} \cdot \vec{a} \right) =  \vec{v} \times \left( \vec{v} \times \vec{a} \right) + v^2 \vec{a} \quad , \quad \frac{v^2}{\tc^2} + \tgam^{-2} = 1 \label{baccabApp} \,.
\end{align}

Even though we express the four-acceleration in Eq.~\eqref{4accelApp} by including $\dot{\tc}$ and $\dot{\ttc}$ terms, we should drop these terms from the expression to properly describe $A^{\mu}$ as a local quantity in the constant time hypersurface. And $\tc = \ttc$ in this frame. Thus, the four-acceleration is expressed by
\begin{align}
A^{\mu} &= \gamma^2 \tgam^2 \left( \frac{\vec{v} \cdot \vec{a}}{\tc} - \beta^2 \dot{\tc} + \tgam^{-2} \frac{\dot{\ttc}}{\ttc} \tc  \,,  \frac{ \vec{v} \times \left( \vec{v} + \vec{a} \right) }{\tc^2 }  + \vec{a}  \right)    \label{4accelApp2} \,.
\end{align}
It is the same as the four-acceleration of SR if $\tc = c$ and thus $\tgam = \gamma$.
The scalar product of a particle's four-velocity and its four-acceleration $U^{\mu} A_{\mu}$ is given by
\begin{align}
U^{\mu} A_{\mu} &= \gamma^3 \tgam^2 \left( - \vec{v} \cdot \vec{a} + \frac{\dot{\tc}}{\tc} v^2 - \tgam^{-2} \frac{\dot{\ttc}}{\ttc} \tc^2 + \vec{v} \cdot \vec{a} + \frac{\vec{v} \times \left( \vec{v} \times \vec{a} \right)}{\tc^2} \cdot \vec{v} - \frac{\dot{\tc}}{\tc} v^2 + \tgam^{-2} \frac{\dot{\ttc}}{\ttc} v^2  \right) \nonumber \\
&= - \gamma^3 \tgam^2 \tgam^{-2} \frac{\dot{\ttc}}{\ttc} \tc^2 \left( 1 - \frac{v^2}{\tc^2} \right) = -\gamma^3 \tgam^{-2} \frac{\dot{\ttc}}{\ttc} \tc^2 = - \ttc \frac{d \ttc}{d \tau} \,. \label{UmuAmuApp}
\end{align}
Again, in the local reference frame, one obtains constant $\tc$ and $\ttc$, and thus the inner product between the four-velocity and the four-acceleration becomes zero.

\subsection{Electromagnetism}
\label{subsec:ElectromagnetismApp}

In this appendix, we review Maxwell's equations in $4$-dimensional spacetime. The four-potential, $A^{\mu}$, fully describes the electromagnetic field
\begin{align}
A^{\mu} = \left( \frac{\phi}{\tc} \,, A^{i} \right) \;, \label{EMAmuApp}
\end{align}
where $\phi$ is the electrostatic scalar potential and $A^{i}$ is the vector potential. The Lagrangian of a charged particle and an electromagnetic field is given by
\begin{align}
L_{\text{EM}} &\equiv \int \mathcal L d^3 x = - \int \rho_{m} \tc \sqrt{ U_{\alpha} U^{\alpha}} d^3 x - \int \frac{1}{4 \tmu} F_{\alpha\beta} F^{\alpha\beta} d^3 x + \int j_{\alpha} A^{\alpha} d^3 x \label{LEMApp} \,, \\
&= -\frac{1}{2} m_{\rs} U^{\alpha} U_{\alpha} - \int \left( \frac{1}{4 \tmu} F_{\alpha\beta} F^{\alpha\beta} - j_{\alpha} A^{\alpha}  \right) d^3 x \label{LEMApp2} \,, \\
F_{\alpha\beta} F^{\alpha\beta} &= \left( \partial_{\alpha} A_{\beta} - \partial_{\beta} A_{\alpha} \right) \left( \partial^{\alpha} A^{\beta} - \partial^{\beta} A^{\alpha} \right) = 2 \left (\partial_{\alpha} A_{\beta} \partial^{\alpha} A^{\beta}  -   \partial_{\beta} A_{\alpha} \partial^{\alpha} A^{\beta}  \right) \label{FFEMApp} \,, \\
&= 2 \left (g_{\alpha\delta} g_{\beta\epsilon} \partial^{\delta} A^{\epsilon} \partial^{\alpha} A^{\beta}  -   g_{\alpha\delta} g_{\beta\epsilon} \partial^{\epsilon} A^{\delta} \partial^{\alpha} A^{\beta}  \right) \nonumber \,,
\end{align}
where $m_{\rs}$ is the rest mass of a charged particle, $F^{\alpha\beta}$ is the electromagnetic field strength tensor, a four-current density $j^{\alpha} = \rho_{\EM}^{\rs} U^{\alpha} = \rho_{\EM}^{\rs} \gamma \left( \tc \,, \vec{v} \right)$, and the rest charge density $\rho_{\EM}^{\rs} = q^{\rs} \delta (\vec{r} - \vec{s} )$.
Thus, the action of the electromagnetic field is given by
\begin{align}
S_{\text{EM}} = \int L_{\text{EM}} d \tau = \int L_{\text{EM}} \gamma^{-1} dt \label{SEMApp} \,.
\end{align}

The Euler-Lagrange equations for the electromagnetic field provide 
\begin{align}
&\frac{\partial \mathcal L}{\partial A^{\nu}} =  j_{\alpha} \frac{\partial A^{\alpha}}{\partial A^{\nu}} =  j_{\alpha} \delta^{\alpha}_{\nu} = j_{\nu} \quad , \quad
\frac{\partial \mathcal L}{\partial (\partial^{\mu} A^{\nu})} = \frac{\partial }{\partial (\partial^{\mu} A^{\nu})} \left( -\frac{1}{4 \tmu} F_{\alpha\beta} F^{\alpha\beta} \right) \quad , \quad
\frac{\partial \left( F_{\alpha\beta} F^{\alpha\beta} \right) }{\partial (\partial^{\mu} A^{\nu})} = 4 F_{\mu\nu} \,, \nonumber \\
&\partial^{\mu} \frac{\partial \mathcal L}{\partial (\partial^{\mu} A^{\nu})} = \partial^{\mu} \left( -\frac{F_{\mu\nu}}{\tmu} \right) = -\frac{1}{\tmu} \partial^{\mu} F_{\mu\nu} + \frac{\partial^{\mu} \tmu}{\tmu^2} F_{\mu\nu} \nonumber \,, \\
&\partial^{\alpha} F_{\alpha\beta} = - \tmu j_{\beta} + \frac{\partial^{\alpha} \tmu}{\tmu} F_{\alpha\beta} = - \tmu j_{\beta} + \left( \partial^{\alpha} \ln \tmu \right)F_{\alpha\beta}  \,, \label{ELEqofEMApp} \\
&\epsilon^{\alpha\beta\gamma\delta} \partial_{\gamma} F_{\alpha\beta} = 0 \,, \label{ELEqofEM2App} 
\end{align}
where $\tmu$ is the permeability and $\epsilon_{\alpha\beta\gamma\delta}$ is a four-dimensional Levi-Civita symbol. Since $\tc$ varies as a function of the scale factor in meVSL, it is reasonable to assume that so does $\tmu$. Thus, we write the permeability in the meVSL as $\tmu[a]$ and the modification in Eq.~\eqref{ELEqofEMApp} does not contribute to Maxwell's equation in the local inertial frame. 
Eq.~\eqref{ELEqofEMApp} includes inhomogeneous Maxwell's equations (Gauss's law and Amp\`{e}re's law) and Eq.~\eqref{ELEqofEM2App} implies Bianchi's identity (Gauss's law for magnetism and Maxwell-Faraday equation). We can explicitly rewrite the above equations as 
\begin{align}
F_{0i} = \partial_0 A_i - \partial_i A_0 = \frac{1}{\tc} \left( \frac{d A_i}{d t} + \frac{d \phi}{d x^i} \right) \quad , \quad J^{\alpha} = \left( \rho_{\EM} \tc \,, \vec{j} \right) \label{F0iApp} \,,				
\end{align}
where $\rho_{\EM}$ is the charge density and $\vec{j}$ is the conventional current density. 

We adopt the electric field $E_{i} = - \tc F_{0i}$ and the magnetic field $B_i = 1/2 \epsilon_{ijk} F^{jk}$ to obtain  
\begin{align}
&\partial^{i} F_{i0} = - \tmu j_{0} + \left( \partial^{i} \ln \tmu \right)  F_{i0} \quad , \quad \beta = 0 \nonumber \\
 &\partial^{i} \frac{1}{\tc} E_{i}(t) = - \tmu j_{0}(t) + \left( \partial^{i} \ln \tmu \right)  \frac{1}{\tc} E_{i}(t) \quad \Longrightarrow \quad
			  \vec{\nabla} \cdot \vec{E} (t) = - \tmu \tc j_{0}(t) = \rho \tmu \tc^2 = \frac{\rho(t)}{\tepsilon} \,, \label{GausslawApp}
\end{align}
where we use the fact that $\tmu[a]$ is a function of the scale factor $a$ and thus $\tc[a]^2 = 1/(\tmu[a] \tepsilon[a])$. We obtain $\tc = \tc_0 a^{b/4}$ and $\tepsilon = \tepsilon_0 a^{-b/4}$ and $\tmu = \tmu_0 a^{-b/4}$. Eq.~\eqref{GausslawApp} is the same as that of SR at the given cosmic epoch. We also obtain the Amp\`{e}re's law by taking $\beta$ for the spatial index,
\begin{align}
&\partial^{\alpha} F_{\alpha k} = - \mu j_{k} + \frac{\partial^{\alpha} \mu}{\mu} F_{\alpha k} \quad , \quad \beta = k \nonumber \\
&\partial^{0} F_{0 k} + \partial^{i} F_{i k} = - \mu j_{k} + \frac{\partial^{0} \mu}{\mu} F_{0 k} \nonumber \\
&\Longrightarrow \quad -\frac{1}{\tc^2} \frac{d \vec{E}(t)}{d t} + \vec{\nabla} \times \vec{B} = \tmu \vec{j} (t) - \frac{ d \ln[\tc \tmu]}{d t} \frac{\vec{E}(t)}{\tc} = \tmu \vec{j} (t) - \frac{ d \ln[\tc_0 \tmu_0]}{d t} \frac{\vec{E}(t)}{\tc} = \tmu \vec{j} \label{AmerelawApp} \,. 
\end{align}
Thus, Amp\`{e}re's law is the same as that of SR, too. 

In SR, charge conservation is that the Lorentz invariant divergence of $J^{\alpha}$ is zero
\begin{align}
\frac{\partial J^{\alpha}}{\partial x^{\alpha}} = \partial_{\alpha} J^{\alpha} =  \frac{d}{\tc d t} \left( \rho \tc \right) + \vec{\nabla} \cdot \vec{j} = 0 \label{charveconserv2} \,,
\end{align} 
Similarly, the continuity equation in GR with FLRW metric is written as
\begin{align}
\nabla_{\alpha} J^{\alpha} &= \partial_{\alpha} J^{\alpha} + \Gamma^{\alpha}_{\alpha \beta} J^{\beta} = 0 = \partial_{0} J^{0} + \partial_{i} J^{i} + \Gamma^{k}_{k 0} J^{0} = \frac{d}{\tc d t} \left( \rho \tc \right) + \vec{\nabla} \cdot \vec{j} + 3 \frac{H}{\tc} \rho \tc \,.
		 \label{charcon2} 
\end{align} 
The continuity equation can be solved 
\begin{align}
& \frac{d \rho}{dt} + \frac{d \ln \tc}{dt} \rho + 3 \frac{d \ln a}{dt} \rho = -\vec{\nabla} \cdot \vec{j} \nonumber \\
& \frac{d \ln \rho}{dt} + 3 \frac{d \ln a}{dt} + \frac{d \ln \tc}{dt} = \frac{d \ln \rho}{dt} + \left( 3 + \frac{b}{4} \right) \frac{d \ln a}{dt} = 0 \quad \text{if} \quad \vec{j} = 0 \nonumber \\
&\rho_{\text{EM}} = \rho_{\text{EM}0} a^{-3 - \frac{b}{4}} \label{rhoEM} \,,
\end{align}
where $\rho_{\text{EM}0}$ denotes the present value of the charge density. Thus, Gauss's law in Eq.~\eqref{GausslawApp} becomes
\begin{align}
\nabla \cdot \vec{E} = \frac{\rho_{\text{EM}}}{\tepsilon} = \frac{\rho_{\text{EM}0} a^{-b/4}}{\tepsilon_0 a^{-b/4}} = \frac{\rho_{\text{EM}0}}{\tepsilon_0} \label{Gausslaw02} \,.
\end{align}
This means that Gauss's law holds for any epoch in the meVSL model. 

\subsection{Thermal Equilibrium}
\label{subsec:TEQApp}

The perfect blackbody spectrum of the CMB is a piece of good observational evidence showing that the early universe was in local thermal equilibrium. The hot big bang model predicts thermal equilibrium above 100 GeV. To describe the subsequent evolution of the universe, we need to recall some basic facts of equilibrium thermodynamics, suitably generalized to apply to an expanding universe. It is convenient to describe the system of a weakly interacting particle in phase space, where it is described by the positions and momenta of all particles. In quantum mechanics, the momentum eigenstates of a particle included in a volume $V = L^3$ have a discrete spectrum. Then, the density of states in momentum space $\{{\bf p}\}$ is given by $L^3/h^3 = V/h^3$, and the state density in phase space $\{{\bf x},{\bf p}\}$ is $h^{-3}$. If $g$ denotes the internal degrees of freedom (e.g. spin) of the particle, then the density of states becomes
\begin{align}
\frac{g}{h^3} = \frac{g}{(2 \pi \hbar)^3} \label{gh3} \;.
\end{align}
The thermal equilibrium hypothesis is held until the nuclear interaction rate is not less than the expansion rate of the Universe. The distribution function is a function $f = f (\vec{x}, \vec{p}, t)$ of the position, of the proper momentum, and of the time. In other words, it is a function that takes its values in the phase space. It can be thought of as a probability density
\begin{align}
f(t, \vec{x}, \vec{p}) \frac{d^3 {\bf x} d^3 {\bf p}}{\mathcal N} = f(t, p) \frac{d^3 {\bf x} d^3 {\bf p}}{\mathcal N} \label{probabilitydensity} \;,
\end{align}
which is the probability of finding a particle at the given time $t$ in a small volume $d^3 {\bf x} d^3 {\bf p}$ of the phase space centered in $\{{\bf x},{\bf p}\}$, and $\mathcal N$ is some suitable normalization. We use the fact that distribution is homogeneous and isotropy to use $f(t, \vec{x}, \vec{p}) = f (t, p)$. In the early universe, the chemical potentials of all particles were so small that one could neglect them and thus the distribution functions are given by
\begin{align}
f(p) = \frac{1}{\exp [E/(k_{\TB} T)] \pm 1} \label{fpapp} \,,
\end{align}
where the $+$ sign and the $-$ one is for fermions and bosons, respectively. Because of the Heisenberg uncertainty principle of quantum mechanics, no particle can be localized in the phase space in a point $\{{\bf x},{\bf p}\}$, but at most in a small volume $\mathcal N = h^3$ about that point, where $h$ is Planck constant. Therefore, the probability density is given by
\begin{align}
d \mathcal P (t, \vec{x} , \vec{p}) = f(t, p) \frac{d^3 {\bf x} d^3 {\bf p}}{(2 \pi \hbar)^3} \label{dP} \;.
\end{align}
Integrating the distribution function for the momentum, one gets the particle number density
\begin{align}
n(t, \vec{x}) \equiv g \int \frac{d^3 {\bf p}}{(2 \pi \hbar)^3} f(t, p) \label{n} \;.
\end{align}
In general, one can write the energy-momentum tensor in terms of the distribution function as
\begin{align}
T^{\mu}_{\nu}(t, x) = g \int \frac{dP_1dP_2dP_3}{(2\pi \hbar)^3} \frac{1}{\sqrt{-g}} \frac{c P^{\mu} P_{\nu}}{P^0} f(t, p) \label{Tmunutherm} \;,
\end{align}
where $P^{\mu} = d x^{\mu}/ d \lambda$ is the comoving momentum.

This equation satisfies the mass-shell condition. To use thermodynamics, one can calculate the number densities, energy densities, and the pressures of the relativistic and non-relativistic particles
\begin{align}
&n= \frac{g}{2 \pi^2 \hbar^3} \int_{0}^{\infty} dp \frac{p^2}{\exp[\sqrt{p^2 c^2 + m^2 c^4}/k_{\TB} T] \pm1} \label{n2app} \,, \\
&\rho \tc^2= \frac{g}{2 \pi^2 \hbar^3} \int_{0}^{\infty} dp \frac{p^2 \sqrt{p^2 c^2 + m^2 c^4} }{\exp[\sqrt{p^2 c^2 + m^2 c^4}/k_{\TB} T] \pm1} \label{epsilon2app} \,, \\
&P = \frac{g}{2 \pi^2 \hbar^3} \int_{0}^{\infty} dp \frac{p^4 c^2/ \sqrt{p^2 c^2 + m^2 c^4} }{3 \left(\exp[\sqrt{p^2 c^2 + m^2 c^4}/k_{\TB} T] \pm1 \right)} \label{P2app} \,.
\end{align}
If we define $x \equiv \beta m c^2$, $\xi \equiv \beta pc$, and $\beta = 1/(k_{\TB} T)$, then above equations are rewritten as
\begin{align}
&n= \frac{g}{2 \pi^2} \left( \frac{k_{\TB} T}{\hbar c} \right)^3 \int_{0}^{\infty} d\xi \frac{\xi^2}{\exp[\sqrt{\xi^2 + x^2}] \pm1} \equiv \frac{g}{2 \pi^2} \left( \frac{k_{\TB} T}{\hbar c} \right)^3 I_{\pm}(x) \label{n3} \,, \\
&\rho \tc^2= \frac{g}{2 \pi^2} \left( \frac{k_{\TB}^4 T^4}{\hbar^3 c^3} \right)  \int_{0}^{\infty} d\xi \frac{\xi^2 \sqrt{\xi^2 + x^2} }{\exp[\sqrt{\xi^2 + x^2}] \pm1} \equiv \frac{g}{2 \pi^2} \left( \frac{k_{\TB}^4 T^4}{\hbar^3 c^3} \right)  J_{\pm}(x) \label{epsilon3} \,, \\
&P = \frac{g}{2 \pi^2} \left( \frac{k_{\TB}^4 T^4}{\hbar^3 c^3} \right) \int_{0}^{\infty} d\xi \frac{\xi^4/\sqrt{\xi^2 + x^2}}{3 \left( \exp[\sqrt{\xi^2 + x^2}] \pm1 \right)} \equiv \frac{g}{2 \pi^2} \left( \frac{k_{\TB}^4 T^4}{\hbar^3 c^3} \right)  K_{\pm}(x)  \label{P3} \,.
\end{align}
Thus, the number densities, energy densities, and the pressures of relativistic and non-relativistic particles are given by
\begin{align}
n &= \begin{cases} \frac{g}{\pi^2} \left( \frac{k_{\TB} T}{\thbar \tc} \right)^3 \frac{3}{4} \zeta(3) & \text{fermion} \\
\frac{g}{\pi^2} \left( \frac{k_{\TB} T}{\thbar \tc} \right)^3 \zeta(3) & \text{boson} \\
g \left( \frac{1}{2 \pi} \frac{m_{\rs} \tc^2}{\thbar \tc} \frac{k_{\TB}T}{\thbar \tc} \right)^{\frac{3}{2}}  e^{-\frac{m_{\rs} \tc^2}{k_{\TB} T}} & \text{non-relativistic} \end{cases} \label{nthapp} \,, \\
\rho \tc^2 &= \begin{cases} \frac{g \pi^2}{30} \frac{ \left( k_{\TB} T \right)^4}{\left( \thbar \tc \right)^3} \frac{7}{8} & \text{fermion} \\
\frac{g \pi^2}{30} \frac{ \left( k_{\TB} T \right)^4}{\left( \thbar \tc \right)^3}  & \text{boson} \\
n \left( m_{\rs} \tc^2 + \frac{3}{2} k_{\TB} T \right) \approx n m_{\rs} \tc^2 & \text{non-relativistic} \end{cases} \,, \quad
P = \begin{cases} \frac{1}{3} \rho \tc^2 & \text{fermion} \\
\frac{1}{3} \rho \tc^2  & \text{boson} \\
n T \approx 0 & \text{non-relativistic} \end{cases} \label{rhoc2Pthapp} \,.
\end{align}

\subsection{Lorentz Transformation}
\label{subsec:LTApp} 

From the translational symmetry of space and time, a transformation of the coordinates $x$ and $t$ from the inertial reference frame $\mathcal{O}$ to $x'$ and $t'$ in another reference frame $\mathcal{O}'$ should be linear functions. This fact is written by 
\begin{align}
\begin{pmatrix} t' \\ x' \end{pmatrix} = \begin{pmatrix} A & B \\ C & D \end{pmatrix} \begin{pmatrix} t \\ x \end{pmatrix} \quad , \quad
t' = At + Bx \quad , \quad x' = Ct + Dx \label{xptp2App} \,.
\end{align}
If we set $x' = 0$ as the origin of $\mathcal{O}'$ and it moves with velocity $v$ relative to $\mathcal{O}$, so that $x = vt$
\begin{align}
x' &= C t  + D x \Rightarrow 0 = C t + D v t = (C + D v) t \Rightarrow C = - v D \,, \nonumber \\ 
x' &= D ( - v t + x ) \label{xpApp} \,.
\end{align} 
Now $x = 0$ is the origin of $\mathcal{O}$ and it moves with velocity $-v$ relative to $\mathcal{O}'$, so that $x' = -vt'$
\begin{align}
x' &= D (- v t + x ) \Rightarrow - v t' =  -  D v t + 0  \Rightarrow t' = D t  \nonumber \,,  \\
t' &= At + B x \Rightarrow t' = At + 0 \Rightarrow t' = At \nonumber \,, \\
A &= D \label{tpApp3} \,.
\end{align} 
One can rewrite Eq.~\eqref{xptp2App} as
\begin{align}
t' &= At + Bx = A (t + F x) \quad \text{where} \quad F = B/A \,. \label{tpApp4} 
\end{align}

If one changes the notation $A$ as $\gamma$, then the above equation~\eqref{tpApp4} becomes
\begin{align}
t' &= \gamma \left( t + F x \right) \quad , \quad x' = \gamma \left( - v t + x \right) \quad , \quad
\begin{pmatrix} t' \\ x' \end{pmatrix} = \gamma[v] \begin{pmatrix} 1 & F[v] \\ - v & 1 \end{pmatrix} \begin{pmatrix} t \\ x \end{pmatrix} \label{xptp4App} \,.
\end{align}
A combination of two Lorentz transformations also must be a Lorentz transformation (form a group). If a reference frame $\mathcal{O}'$ moving relative to $\mathcal{O}$ with velocity $v_1$ and a reference frame $\mathcal{O}''$ moving relative to $\mathcal{O}'$ with velocity $v_2$ then
\begin{align}
\begin{pmatrix} t'' \\ x'' \end{pmatrix} &= \gamma[v_2] \begin{pmatrix} 1 & F[v_2] \\ -v_2 & 1 \end{pmatrix} \begin{pmatrix} t' \\ x' \end{pmatrix} = \gamma[v_2] \begin{pmatrix} 1 & F[v_2] \\ -v_2 & 1 \end{pmatrix}  \gamma[v_1] \begin{pmatrix} 1 & F[v_1] \\ -v_1 & 1 \end{pmatrix} \begin{pmatrix} t \\ x \end{pmatrix}  \nonumber \\
&= \gamma[v_2] \gamma[v_1] \begin{pmatrix} 1 - F[v_2] v_1 & F[v_1] + F[v_2] \\ -v_2 - v_1 & 1 - F[v_1] v_2 \end{pmatrix} \begin{pmatrix} t \\ x \end{pmatrix}  \label{DoubleLorentzApp} \,.
\end{align}
One compares the coefficients in Eqs.~\eqref{xptp4App} and \eqref{DoubleLorentzApp} to obtain
\begin{align}
1 - F[v_2] v_1 = 1 - F[v_1] v_2 \Rightarrow \frac{F[v_1]}{v_1} = \frac{F[v_2]}{v_2} \equiv \frac{1}{\alpha} = \text{constant} \label{alphaApp} \,.
\end{align}
If one puts Eq.~\eqref{alphaApp} into Eq.~\eqref{xptp4App}, then one obtains
\begin{align}
t' &= \gamma \left( t + \frac{v}{\alpha} x \right) \quad , \quad x'= \gamma \left( - v t + x \right) \quad , \quad
\begin{pmatrix} t' \\ x' \end{pmatrix} = \gamma[v] \begin{pmatrix} 1 & \frac{v}{\alpha} \\ - v & 1 \end{pmatrix} \begin{pmatrix} t \\ x \end{pmatrix} \,.\label{xptp6App}
\end{align}

If one makes the Lorentz transformation from the reference frame $\mathcal{O}$ to $\mathcal{O}'$ and then from $\mathcal{O}'$ to $\mathcal{O}$ back, then one can use Eq.~\eqref{DoubleLorentzApp}  
\begin{align}
\begin{pmatrix} t \\ x \end{pmatrix} &= \gamma[-v] \gamma[v] \begin{pmatrix} 1 - F[-v] v & F[v] + F[-v] \\  v - v & 1 + F[v] v \end{pmatrix} \begin{pmatrix} t \\ x \end{pmatrix} =  \gamma[-v] \gamma[v] \begin{pmatrix} 1 + \frac{v^2}{\alpha}  & 0 \\  0 & 1 + \frac{v^2}{\alpha} \end{pmatrix} \begin{pmatrix} t \\ x \end{pmatrix} \nonumber \,, \\
\Rightarrow & \quad \gamma[-v] \gamma[v] = 1/ \left(1 + \frac{v^2}{\alpha} \right) \equiv \gamma[v]^2 \quad \text{because of space symmetry} \label{DoubleLorentz3App} \,.
\end{align}
Thus, the Lorentz transformation is given by
\begin{align}
\begin{pmatrix} t' \\ x' \end{pmatrix} &=  \frac{1}{\sqrt{1 + \frac{v^2}{\alpha}}} \begin{pmatrix} 1 & \frac{v}{\alpha} \\ - v & 1 \end{pmatrix} \begin{pmatrix} t \\ x \end{pmatrix} \label{xptp7App} \,.
\end{align}
Finally, if one puts $\alpha = -\tc^2$, then Eq.~\eqref{xptp7App} becomes
\begin{align}
& t' =  \frac{t - \frac{v}{\tc^2} x}{\sqrt{1 - \frac{v^2}{\tc^2}}} \quad , \quad  x' =  \frac{- v t +  x}{\sqrt{1 - \frac{v^2}{\tc^2}}} \quad , \quad 
\tc t' =  \frac{\tc t - \beta x}{\sqrt{1 - \beta^2} } \quad , \quad  x' =  \frac{- \beta \tc t +  x}{\sqrt{1 - \beta^2}} \label{xptp9App} \,, \\
&\begin{pmatrix} \tc t' \\ x' \end{pmatrix} =  \frac{1}{\sqrt{1 - \beta^2}} \begin{pmatrix} 1 & - \beta \\ -\beta & 1 \end{pmatrix} \begin{pmatrix} \tc t \\ x \end{pmatrix} \quad , \quad \text{where} \quad \beta = \frac{v}{\tc} \,. \label{xptp10App}
\end{align}

\section{General Relativity}
\label{sec:GRApp}

We now expand the meVSL model into the curved spacetime. In general relativity, a geodesic is a "straight line" in curved spacetime. It is the shortest path, determined by the metric of the specified spacetime, that a physical item takes when moving freely. This indicates that a particle in free fall or motion always follows a geodesic. In this part, we examine the meVSL model's geodesic equation and its geodesic deviation equation.

\subsection{Geodesic equation}
\label{subsec:GEApp}

As per the considerations of the Equivalence Principle, if we were to describe the movement of an object in the Earth's gravitational field, we would then have to follow the following steps: First, Describe the movement in a local inertial free-falling referential. Second, operate a coordinate transformation from this local inertial referential to the Earth referential, it is seen as accelerated upwards. One can derive the geodesic equation directly from the equivalence principle. A free-falling particle does not accelerate in the neighborhood of a point-event for a freely falling coordinate system, $X^{\mu}$. Setting $X^0 \equiv T \equiv c \tau$, one has the following equation that is locally applicable in free fall
\begin{align}
\frac{d^2 X^{\mu}}{d T^2} = 0 \label{geod1App} \,.
\end{align}
We emphasize that $\tau$ refers to the time measured by an observer at rest in her rest referential and is called the proper time. One can express the above equation in general non-inertial referential coordinate $x^{\mu}$ by using the chain-rule
\begin{align}
\frac{d X^{\mu}}{dT} &= \frac{d x^{\alpha}}{d T} \frac{\partial X^{\mu}}{\partial x^{\alpha}} = \frac{1}{\ttc} \frac{d x^{\alpha}}{d \tau} \frac{\partial X^{\mu}}{\partial x^{\alpha}}  \label{dXmudTApp} \,, \\
\frac{d^2 X^{\mu}}{dT^2} &= \frac{d}{\ttc d\tau} \left[ \frac{d x^{\alpha}}{\ttc d \tau} \frac{\partial X^{\mu}}{\partial x^{\alpha}}  \right] = \frac{1}{\ttc^2} \left[ \frac{d^2 x^{\alpha}}{ d \tau^2} - \frac{1}{\ttc} \frac{d \ttc}{d\tau} \frac{dx^{\alpha}}{d\tau} \right] \frac{\partial X^{\mu}}{\partial x^{\alpha}} + \frac{1}{\ttc^2} \left[ \frac{\partial^2 X^{\mu}}{\partial x^{\alpha} \partial x^{\beta}} \right] \frac{d x^{\alpha}}{d \tau} \frac{d x^{\beta}}{d \tau} \nonumber \\
&= \frac{1}{\ttc^2} \left[ \frac{d^2 x^{\lambda}}{ d \tau^2} + \left( \frac{\partial^2 X^{\mu}}{\partial x^{\alpha} \partial x^{\beta}} \frac{\partial x^{\lambda}}{\partial X^{\mu} } \right) \frac{d x^{\alpha}}{d \tau} \frac{d x^{\beta}}{d \tau}  - \frac{1}{\ttc} \frac{d \ttc}{d\tau} \frac{dx^{\lambda}}{d\tau} \right] \nonumber \,, \\
&= \frac{1}{\ttc^2} \left[ \frac{d^2 x^{\lambda}}{ d \tau^2} + \Gamma^{\lambda}_{\alpha\beta} \frac{d x^{\alpha}}{d \tau} \frac{d x^{\beta}}{d \tau}  - \frac{d \ln \ttc}{d \ln a} \frac{d \ln a}{d \tau} \frac{dx^{\lambda}}{d\tau} \right] = 0 \label{geod2App} \,.
\end{align}

Thus, the geodesic equation in the locally free-falling coordinate in Eq.~\eqref{geod1App} can be rewritten in the general coordinate w.r.t the proper time, $\tau$. Thus, the inertial force in the fixed laboratory reference frame is given by
\begin{align}
F^{\lambda} &\equiv m \frac{d^2 x^{\lambda}}{ d \tau^2} \equiv m A^{\lambda} = - m \Gamma^{\lambda}_{\alpha\beta} \frac{d x^{\alpha}}{d \tau} \frac{d x^{\beta}}{d \tau}  + m \frac{dx^{\lambda}}{d\tau} \frac{d \ln \ttc}{d \ln a} \frac{d \ln a}{d \tau} \nonumber \\
&\equiv - m \Gamma^{\lambda}_{\alpha\beta} U^{\alpha} U^{\beta} + m U^{\lambda} \frac{d \ln \ttc}{d \ln a} \frac{d \ln a}{d \tau} \label{FlambdaApp} \,,
\end{align}
and this is the gravitational force. The last term of Eq.~\eqref{FlambdaApp} is the correction term in the meVSL model compared to GR. This term is turned on, only when one considers the cosmological evolution of the inertial force. Also, one can rewrite the above geodesic equation in Eq.~\eqref{geod2App} w.r.t the cosmic time $t$ by using the chain-rule again
\begin{align}
\frac{d x^{\lambda}}{d \tau} &= \frac{d t}{d \tau} \frac{d x^{\lambda}}{d t} \quad , \quad
\frac{d^2 x^{\lambda}}{d \tau^2} = \left( \frac{d t}{d \tau} \right)^2 \frac{d^2 x^{\lambda}}{d t^2} + \frac{d^2 t}{d \tau^2} \frac{d x^{\lambda}}{d t}  \label{d2xldtau2App} \,.
\end{align}
One inserts Eq.~\eqref{d2xldtau2App} into Eq.~\eqref{geod2App} to obtain
\begin{align}
\frac{d^2 x^{\lambda}}{d t^2} &= -  \Gamma^{\lambda}_{\alpha\beta} \frac{d x^{\alpha}}{d t} \frac{d x^{\beta}}{d t} - \left( \frac{d \tau}{d t} \right)^{2} \frac{d^2 t}{d \tau^2} \frac{d x^{\lambda}}{d t} + \frac{d \ln \ttc}{d \ln a} \frac{d \ln a}{dt} \frac{dx^{\lambda}}{dt} \label{geod3App} \,.
\end{align}

Therefore, one obtains the modification due to the varying $\ttc$ in the last term of the above equation~\eqref{geod3App} compared to the GR.
Applying $\lambda = 0$ in the above Eq.~\eqref{geod3App} to obtain
\begin{align}
\frac{d^2 t}{d \tau^2} \left( \frac{d \tau}{d t} \right)^{2} &= - \Gamma^{0}_{\alpha\beta} \frac{1}{\tc} \frac{d x^{\alpha}}{d t} \frac{d x^{\beta}}{d t}  + \frac{d \ln \ttc}{dt} - \frac{d \ln \tc}{dt} \label{d2tdtau2App} \,.
\end{align}
Thus, the geodesic equation~\eqref{geod3App} using the coordinate time $t$ becomes
\begin{align}
\frac{d^2 x^{\lambda}}{d t^2}	&= -  \Gamma^{\lambda}_{\alpha\beta} \frac{d x^{\alpha}}{d t} \frac{d x^{\beta}}{d t} + \left[ \Gamma^{0}_{\alpha\beta} \frac{1}{\tc} \frac{d x^{\alpha}}{d t} \frac{d x^{\beta}}{d t}  + \frac{d \ln \tc}{d \ln a} H \right] \frac{dx^{\lambda}}{dt} \label{d2xdt2App}  \,.
\end{align}
One can see that the last term in the above Eq.~\eqref{d2xdt2App} contains $(d \ln \tc/d \ln a)$-term, which gives the explicit correction on the geodesic equation compared to that of GR. Both $\Gamma^{0}_{\alpha\beta}$ and $\tc$ are also different from those of GR. This equation contains four components. For $\lambda = 0$, the above equation shows the consistency $\dot{\tc} = \dot{\tc}$. One can estimate the magnitude of the correction by comparing the magnitude of the third term to the second one in the above Eq.~\eqref{d2xdt2App} for $\lambda = i$ with some results of meVSL, $\tc = \tc_0 a^{b/4}$ and $H = H^{(\GR)} a^{b/4}$
\begin{align}
\frac{d^2 x^{i}}{d t^2} &= -  H \frac{d x^{i}}{d t} + \left[ a^2 H \frac{v^2}{\tc^2} + \frac{d \ln \tc}{d \ln a} H \right] \frac{dx^{i}}{dt} = \left[ - 1 + a^2 \frac{v^2}{\tc^2} + \frac{d \ln \tc}{d \ln a} \right] H \frac{dx^{i}}{dt}  \nonumber \, \\
&=\left[ - 1 + a^{2-\frac{b}{2}} \frac{v^2}{\tc_0^2} + \frac{b}{4} \right] H \frac{dx^{i}}{dt} \,. \label{d2xidt2App}
\end{align}
We show that the constraint on $b$ is rough $b \leq 0.1$, and so the correction on the geodesic equation due to the varying $c$ is less than the percent level.

If we want to apply the above equation to Newtonian dynamics, then we should remember the so-called ``Newtonian limit'' which is based on three assumptions, the particle is moving relatively slowly (compared to the speed of light), the gravitational field is weak, and the field does not change with time, ({\it i.e.}, it is static). As the geodesic equation describes the worldline of a particle acted only upon by gravity, we need to show that in the context of the Newtonian limit, the geodesic equation reduces to the first Newton's gravity equation. From the geodesic equation, the second term hides a sum in $\alpha$ and $\beta$ overall indices. But the particle in question is moving slowly, the time component dominates the other spatial components, and every term containing one or two spatial four-velocity components will be then dwarfed by the term containing two time components. We can therefore take the approximation
\begin{align}
\frac{d x^i}{d \tau} &\ll \frac{\tc dt}{d \tau} \quad , \quad
\frac{d^2 x^{\lambda}}{ d \tau^2} \approx - \Gamma^{\lambda}_{00} \left( \frac{\tc dt }{d \tau} \right)^2 + \frac{1}{\ttc} \frac{d \ttc}{d\tau} \frac{dx^{\lambda}}{d\tau} \label{geod2NewApp} \,.
\end{align}
One obtains from Eq.~\eqref{geod2NewApp} for $\lambda = 0$,
\begin{align}
\frac{d^2 t}{d \tau^2} &\simeq - \Gamma^{0}_{00} \tc \left( \frac{dt }{d \tau} \right)^2 + \left( \frac{d \ln \ttc}{d \tau} - \frac{d \ln \tc}{d \tau} \right) \frac{dt}{d\tau} \equiv  - \Gamma^{0}_{00} \tc \left( \frac{dt }{d \tau} \right)^2 + \frac{d \ln \bbc}{d\tau} \frac{dt}{d\tau} \label{d2tdtau2appApp} \,,
\end{align}
where $\bar{c} = \ttc/\tc$. Now, if the gravitational field is weak enough, then spacetime will be only slightly deformed from the gravity-free Minkowski space of SR, and we can consider the spacetime metric as a small perturbation from the Minkowski metric $\eta_{\mu\nu}$
\begin{align}
g_{\mu\nu} &= \eta_{\mu} + h_{\mu\nu} \quad , \quad |h_{\mu\nu} | \ll 1 \quad , \quad
g_{00,i} = h_{00,i} \,. \label{gmunuapprox}
\end{align}

We restrict ourselves to the Newtonian $3$-D space, meaning that we assign $\beta$ to spatial dimensions only, we can then replace $\lambda$ by the Latin letter ($i = x, y, z$), giving
\begin{align}
\frac{d^2 x^i}{ d \tau^2} &\approx - \Gamma^{i}_{00} \left( \frac{\tc dt }{d \tau} \right)^2 + \frac{1}{\ttc} \frac{d \ttc}{d\tau} \frac{dx^{i}}{d\tau} \quad \text{where} \quad \Gamma^{i}_{00} = - \frac{1}{2} g^{ii} g_{00,i} \nonumber \\
&= \frac{1}{2} h_{00,i} \left( \frac{\tc dt }{d \tau} \right)^2 + \frac{1}{\ttc} \frac{d \ttc}{d\tau} \frac{dx^{i}}{d\tau} \label{geod2New2App} \,. 
\end{align}
Again, one can repeat the above process for the cosmic time $t$ to obtain
\begin{align}
\frac{d^2 x^i}{dt^2} &\approx \frac{\tc^2}{2} h_{00,i} + v^{i} \frac{d \ln \tc}{dt} \quad \Rightarrow \quad \frac{d^2 \vec{x}}{dt^2} = - \vec{\nabla} \Phi + \vec{v} \frac{d \ln \tc}{dt} = - \vec{\nabla} \Phi + \frac{b}{4} H \vec{v}  \nonumber \,, \\
h_{00} &= 2 \frac{\Phi}{\tc^2} = \frac{2}{\tc^2} \frac{G M_{\text{Earth}}}{R_{\text{Earth}}} = \frac{2}{(3 \times 10^{8} \text{m/s})^2} \frac{6.67 \times 10^{-11} \text{$m^3$/kg/$s^2$} 6 \times 10^{24} \text{kg}}{6.4 \times 10^6 \text{m}} \approx 1.39 \times 10^{-9} \label{h00} \,, \\
\frac{H_0}{\tc_0} \frac{v}{\tc_0} &= \frac{100 h \text{km/s/Mpc}}{(3 \times 10^{5} \text{km/s})} \frac{v}{\tc_0} = \frac{h}{(3 \times 10^{5}) (3.09 \times 10^{22} \text{m})} \frac{v}{\tc_0} \approx 10^{-26} h \frac{v}{\tc_0} \frac{1}{\text{m}} \,.
\end{align}
Thus, the contribution to the geodesic equation from the varying $\tc$ can be negligible even compared to the contribution from the Earth.

\subsection{Geodesic deviation equation}
\label{subsec:geodesicdeviationApp}

We show how the evolution of the separation measured between two adjacent geodesics, also known as geodesic deviation can be related to a non-zero curvature of the spacetime, or to use a Newtonian language, to the presence of tidal force. So let us pick out two particles following two very close geodesics. Their respective path could be described by the functions $x^{\mu}(\tau)$ (reference particle) and $y^{\mu}(\tau) = x^{\mu}(\tau)+ \xi^{\mu}(\tau)$ (second particle) where $\tau$ is the proper time along the reference particle's worldline and where $\xi^{\mu}$ refers to the deviation four-vector joining one particle to the other at each given time $\tau$ ($\xi^{\mu} \ll x^{\mu}$). The relative acceleration $A^{\mu}$ of the two objects is defined, roughly, as the second derivative of the separation vector $\xi^{\mu}$ as the objects advance along their respective geodesics. As each particle follows a geodesic, the equation of their respective coordinate is by using Eq.~\eqref{geod2App} 
\begin{align}
&\frac{d^2 x^{\alpha}}{d \tau^2} + \Gamma^{\alpha}_{\mu\nu} (x^{\alpha}) \frac{d x^{\mu}}{d \tau} \frac{d x^{\nu}}{d \tau} = \frac{d \ln \ttc}{d \tau} \frac{d x^{\alpha}}{d \tau} \label{d2xaodt2App} \,, \\
&\frac{d^2 x^{\mu}}{d \tau^2} + \frac{d^2 \xi^{\mu}}{d \tau^2} + \left( \Gamma^{\alpha}_{\mu\nu} (x^{\alpha} )  + \partial_{\sigma} \Gamma^{\alpha}_{\mu\nu} \xi^{\sigma} \right) \left( \frac{d x^{\mu}}{d \tau} + \frac{d \xi^{\mu}}{d \tau}\right) \left( \frac{d x^{\nu}}{d \tau} + \frac{d \xi^{\nu}}{d \tau}\right) = \frac{d \ln \ttc}{d \tau} \left( \frac{d x^{\alpha}}{d \tau} + \frac{d \xi^{\alpha}}{d \tau}\right)  \label{d2xxiaodt2App} \,.
\end{align}

If one subtracts Eq.~\eqref{d2xaodt2App} from Eq.~\eqref{d2xxiaodt2App}, then one obtains for the linear order of $\xi$ ({\it i.e.}, $\mathcal{O}(\xi)$)
\begin{align}
\frac{d^2 \xi^{\alpha}}{d \tau^2} + 2 \Gamma^{\alpha}_{\mu\nu} U^{\mu} \frac{d \xi^{\nu}}{d \tau} + \partial_{\sigma} \Gamma^{\alpha}_{\mu\nu} U^{\mu} U^{\nu} \xi^{\sigma} = \frac{d \ln \ttc}{d \tau} \frac{d \xi^{\alpha}}{d \tau} \label{d2xiodt2App} \,,
\end{align}
where we use the torsion-free condition $\Gamma^{\alpha}_{\mu\nu} = \Gamma^{\alpha}_{\nu\mu}$. We now have an expression for $d \xi^{\mu}/ d \tau$ as usual, it is not the total derivative of the four-vector $\xi^{\mu}$ since its derivative could also get a contribution from the change of the basis vectors $e_{\alpha}$ as the object moves along its geodesic. To get the total derivative, we have
\begin{align}
\frac{d \xi}{d\tau} &= \frac{d}{d \tau} \left( \xi^{\alpha} e_{\alpha} \right) = \frac{d \xi^{\alpha}}{d \tau} e_{\alpha} + \xi^{\alpha} \frac{d e_{\alpha}}{d \tau} = \left( \frac{d \xi}{d \tau} \right)^{\alpha} e_{\alpha} \label{dxiodtauApp} \,, \\
\frac{d e_{\alpha}}{d \tau} &= \frac{d x^{\mu}}{d \tau} \frac{\partial e_{\alpha}}{\partial x^{\mu}} \equiv  \frac{d x^{\mu}}{d \tau}  \Gamma^{\sigma}_{\mu\alpha} e_{\sigma} = \Gamma^{\sigma}_{\mu\alpha} U^{\mu} e_{\sigma} \label{deodtApp} \,, \\
\left( \frac{d \xi}{d \tau} \right)^{\alpha} &= \frac{d \xi^{\alpha}}{d \tau} + \Gamma^{\alpha}_{\mu\sigma} U^{\mu} \xi^{\sigma} \label{dxiodtaualphaApp} \,.
\end{align}

Since ${\bf \xi}$ is a four-vector, its derivative for proper time is also a four-vector, so we can find the second absolute derivative by using the same development as for the first-order derivative
\begin{align}
\left( \frac{d^2 \xi}{d \tau^2} \right)^{\alpha} &= \left( \frac{d}{d \tau} \left[ \frac{d \xi}{d \tau} \right] \right)^{\alpha} = \frac{d}{d \tau} \left( \frac{d \xi^{\alpha}}{d \tau} + \Gamma^{\alpha}_{\mu\sigma} U^{\mu} \xi^{\sigma} \right) + \Gamma^{\alpha}_{\mu\sigma} U^{\mu} \left( \frac{d \xi^{\sigma}}{d \tau} + \Gamma^{\sigma}_{\beta\gamma} U^{\beta} \xi^{\gamma} \right) \nonumber \\
&= - \left( \partial_{\sigma} \Gamma^{\alpha}_{\mu\nu} - \partial_{\nu} \Gamma^{\alpha}_{\mu\sigma} + \Gamma^{\alpha}_{\gamma\sigma} \Gamma^{\gamma}_{\nu\mu} - \Gamma^{\alpha}_{\nu\gamma} \Gamma^{\gamma}_{\mu\sigma} \right) U^{\mu} U^{\nu} \xi^{\sigma} + \frac{d \ln \ttc}{d \tau} \left( \frac{d \xi}{d \tau} \right)^{\alpha} \nonumber \\
&\equiv - \tensor{R}{^\alpha_\mu_\sigma_\nu} U^{\mu} U^{\nu} \xi^{\sigma} + \frac{d \ln \ttc}{d \tau} \left( \frac{d \xi}{d \tau} \right)^{\alpha}\label{d2xiodt2alphaApp} \,,
\end{align}
where we use Eqs.~\eqref{d2xiodt2App} and \eqref{dxiodtaualphaApp} to obtain the above equation. This is the geodesic deviation equation of the meVSL model. Compared to GR, we obtain the additional term related to the derivatives of $\ttc$ w.r.t the scale factor $a$. One can rewrite the above equation w.r.t the cosmic time $t$
\begin{align}
\frac{d^2 \xi^{\alpha} }{d t^2} &= - \tensor{R}{^\alpha_\mu_\sigma_\nu} \frac{d x^{\mu}}{dt} \frac{d x^{\nu}}{dt} \xi^{\sigma} + \left[ \Gamma^{0}_{\mu\nu} \frac{1}{\tc} \frac{d x^{\mu}}{dt} \frac{d x^{\nu}}{dt}  + \frac{d \ln \tc}{d \ln a} H \right] \frac{d \xi^{\alpha}}{dt} \,, \label{d2xiApp1} \\
&=  - \tensor{R}{^\alpha_\mu_\sigma_\nu} \frac{d x^{\mu}}{dt} \frac{d x^{\nu}}{dt} \xi^{\sigma} + \left[ a^2 \frac{v^2}{\tc^2} + \frac{b}{4} \right] H \frac{d \xi^{\alpha}}{dt} \label{d2xiApp2}
\end{align}
where we use Eq.~\eqref{d2tdtau2App} to obtain the above equation~\eqref{d2xiApp1}. The modification in the geodesic deviation equation of the meVSL model is indicated in the last term in Eq.~\eqref{d2xiApp2}. In addition, there is another contribution from the Riemann curvature tensor as shown in Eq.~\eqref{RijApp}. This modifies the geodesic deviation equations and it affects the GWs detections.

\subsection{FLRW metric}
\label{subsec:FLRWmppp}

We now derive the Einstein equation of the meVSL model based on the FLRW metric. The FLRW metric is a spatially homogenous and isotropic spacetime given by
\begin{align}
ds^2 = g_{\mu\nu} dx^{\mu} dx^{\nu} = -\tc^2 dt^2 + a^2 \gamma_{ij} dx^i dx^j = -\tc^2 dt^2 + a^2 \left( \frac{dr^2}{1-k r^2} + r^2 d \Omega^2 \right)\label{ds2App} \,.
\end{align} 
The Christoffel symbols $\Gamma^{\mu}_{\nu\lambda}$ for the FLRW metric in Eq.~\eqref{ds2App} are given by
\begin{align}
&\Gamma^{\mu}_{\nu\lambda} \equiv \frac{1}{2} g^{\mu\alpha} \left( g_{\alpha\nu,\lambda} + g_{\alpha\lambda,\nu} - g_{\nu\lambda,\alpha} \right) \label{GammaApp} \, \\
&\Gamma^{0}_{ij} = \frac{a\dot{a}}{\tc} \gamma_{ij} \quad , \quad \Gamma^{i}_{0j} = \frac{1}{\tc}  \frac{\dot{a}}{a} \delta^i_j \quad , \quad \Gamma^{i}_{jk} = ^{s}\Gamma^{i}_{jk}  \label{GammacompApp} \,,
\end{align}
where $^{s}\Gamma^{i}_{jk}$ denotes the Christoffel symbols for the spatial metric $\gamma_{ij}$. As shown in the above Eq.~\eqref{GammacompApp}, the Christoffel symbols of meVSL are the same forms as those of GR. However, $\tc$ varies as a function of the scale factor.

The Riemann curvature tensors express the curvature of the Riemann manifold given by 
\begin{align}
&\tensor{R}{^\alpha_\beta_\mu_\nu} = \Gamma^{\alpha}_{\beta\nu,\mu} - \Gamma^{\alpha}_{\beta\mu,\nu} + \Gamma^{\alpha}_{\lambda\mu} \Gamma^{\lambda}_{\beta\nu} - \Gamma^{\alpha}_{\lambda\nu} \Gamma^{\lambda}_{\beta\mu} \label{RabmuApp} \,, \\
&\tensor{R}{^0_i_0_j} = \frac{\gamma_{ij}}{\tc^2} \left( a \ddot{a} - \dot{a}^2 \frac{d \ln \tc}{d \ln a} \right) \quad , \quad \tensor{R}{^i_0_0_j} = \frac{\delta^{i}_{j}}{\tc^2} \left( \frac{\ddot{a}}{a} - \frac{\dot{a}^2}{a^2} \frac{d \ln \tc}{ d \ln a}  \right) \,, \label{R0i0jApp} \\
&\tensor{R}{^i_j_k_m} = \frac{\dot{a}^2}{\tc^2} \left( \delta^{i}_{k} \gamma_{jm} - \delta^i_m \gamma_{jk} \right) + \tensor[^s]{R}{^i_j_k_m} \quad , \quad \tensor[^s]{R}{^i_j_k_m} = k \left( \delta^i_k \gamma_{jm} - \delta^i_m \gamma_{jk} \right) \,. \label{RijkmApp}
\end{align}
Even though the Christoffel symbols of the FLRW metric of the meVSL model are the same form as those of GR, the Riemann curvature tensors of the meVSL are different from those of GR. The reason for this is that the derivatives of the Christoffel symbols, which include the time-varying speed of light for the cosmic time $t$ (\textit{i.e.}, the scale factor, $a$), yield the Riemann curvature tensors. Thus, one obtains the correction term ($H^2 d \ln \tc/ d \ln a$) in both $\tensor{R}{^0_i_0_j}$ and $\tensor{R}{^i_0_0_j}$.

The Ricci curvature tensors measuring how a shape is deformed as one moves along geodesics in the space are obtained from the contraction of the Riemann curvature tensors given in Eqs.~\eqref{R0i0jApp} and \eqref{RijkmApp}
\begin{align}
R_{\mu\nu} &= \Gamma^{\lambda}_{\mu\nu,\lambda} - \Gamma^{\lambda}_{\mu\lambda,\nu} + \Gamma^{\lambda}_{\mu\nu} \Gamma^{\sigma}_{\lambda\sigma} - \Gamma^{\sigma}_{\mu\lambda} \Gamma^{\lambda}_{\nu \sigma} \label{RmnApp} \,, \\
R_{00} &= -\frac{3}{\tc^2} \left( \frac{\ddot{a}}{a} - \frac{\dot{a}^2}{a^2} \frac{d \ln \tc}{ d \ln a}  \right) \quad , \quad
R_{ij} = \frac{\gamma_{ij}}{\tc^2} a^2 \left( 2 \frac{\dot{a}^2}{a^2} + \frac{\ddot{a}}{a} + 2 k \frac{\tc^2}{a^2} - \frac{\dot{a}^2}{a^2} \frac{d \ln \tc}{ d \ln a}  \right) \label{RijApp} \,.
\end{align}
Again, there are correction terms in both $R_{00}$ and $R_{ij}$ due to the time-variation of the speed of light.

Finally, one can also obtain the Ricci scalar by taking the trace of the Ricci tensors
\begin{align}
R &= \frac{6}{\tc^2} \left( \frac{\ddot{a}}{a} + \frac{\dot{a}^2}{a^2} + k \frac{\tc^2}{a^2} - \frac{\dot{a}^2}{a^2} \frac{d \ln \tc}{ d \ln a}  \right) \label{RmpApp} \,,
\end{align}
where the time-varying speed of light effect is shown in the last term.

\subsection{Hilbert Einstein action}
\label{subsec:HEaction}

We adopt the Einstein-Hilbert (EH) action to obtain EFEs through the principle of least action from the action. In the mVSL model, the only difference from GR is that the speed of light varies as a function of scale factor, $a$. However, that causes the problem in recovering the EFE due to the Palatini identity term acting on the varying speed of light. Thus, one also should allow the gravitational constant to vary as a function of the scale factor to obtain the EFE from the EH action with the varying speed of light in the way that the combination of them (\textit{i.e.}, $\tkapp \equiv 8 \pi \tG/\tc^4$) in the EH does not depend on the cosmic time. The EH action of the meVSL is given by
\begin{align}
S &\equiv \int \Biggl[ \frac{1}{2 \tkapp} \left( R - 2 \Lambda \right) + \mathcal L_{m} \Biggr] \sqrt{-g} dt d^3x \label{SHmpApp} \,,
\end{align}
where $\tkapp$ is sometimes called the Einstein gravitational constant. As we will show shortly, not only the speed of light but also the gravitational constant should cosmologically evolve to recover the EFE of GR from the EH action in the meVSL. The variation of action for the inverse metric should be zero
\begin{align}
\delta S &= \int \left( \left[ \frac{\left( R - 2 \Lambda \right)}{2 \tkapp} \right] \delta \left( \sqrt{-g} \right)  + \frac{1}{2\tkapp} \sqrt{-g} \delta R \right) dt d^3 x + \int \delta \left( \sqrt{-g} \mathcal L_{m} \right) dt d^3 x \nonumber \, \\
&= \int \frac{\sqrt{-g}}{2 \tkapp} \left[ R_{\mu\nu} - \frac{1}{2} g_{\mu\nu} \left( R - 2 \Lambda \right) - \tkapp T_{\mu\nu} \right] \delta g^{\mu\nu} dtd^3 x + \int \frac{\sqrt{-g}}{2 \tkapp} \left[ \nabla_{\mu} \nabla_{\nu} - g_{\mu\nu} \Box \right] \delta g^{\mu\nu} dtd^3 x \label{deltaSHmp2App} \,.
\end{align}

In order not to spoil the EFE, the second term (\textit{i.e.}, the Palatini identity term) in the above Eq.~\eqref{deltaSHmp2App} should vanish. It means that $\tkapp$ should be constant even though both $\tc$ and $\tG$ cosmologically evolve.
\begin{align}
&\tkapp = \text{const} \quad \Rightarrow \quad \tc^4 \propto \tG \nonumber \,, \\
& \tc = \tc_0 a^{b/4} \quad , \quad \tG = \tG_{0} a^{b} \label{tkappaconstmpApp} \,,
\end{align}
where we put $a_0 = 1$ and $\tc_0$ and $\tG_0$ denote present values of the speed of light and the gravitational constant, respectively. From the above two equations \eqref{deltaSHmp2App} and \eqref{tkappaconstmpApp}, one can obtain EFE including the cosmological constant
\begin{align}
&R_{\mu\nu} - \frac{1}{2} g_{\mu\nu} R + \Lambda g_{\mu\nu} \equiv G_{\mu\nu} + \Lambda g_{\mu\nu}  = \frac{8 \pi \tG}{\tc^4} T_{\mu\nu} \label{tEFEmpApp} \;,
\end{align}
where $G_{\mu\nu}$ is the so-called Einstein tensor. The above EFE is the same form as that of GR.

\subsection{FLRW universe}
\label{subsec:FLRWUniv}

To solve the EFE given in Eq.~\eqref{tEFEmpApp}, one needs the stress-energy tensor in addition to the geometric terms in Eqs.~\eqref{RijApp} and \eqref{RmpApp}. The source of the spacetime curvature is the symmetric stress-energy tensor of a perfect fluid in thermodynamic equilibrium, as provided by 
\begin{align}
T_{\mu\nu} = \left( \rho + \frac{P}{\tc^2} \right) U_{\mu} U_{\nu} + P g_{\mu\nu} \label{TmunumpApp} \;,
\end{align}
where $\rho$ is the mass density and $P$ is the hydrostatic pressure. The Bianchi identity states that the covariant derivatives of the metric $g_{\mu\nu}$ and the Einstein tensor $G_{\mu\nu}$ are both zero. From the Bianchi identity and the constancy of the Einstein gravitational constant $\kappa$, one can obtain the local conservation of energy and momentum as
\begin{align}
&\tensor{T}{^\mu_\nu_;_\mu} = 0 \quad \Rightarrow \quad \frac{\partial \rho_i}{\partial t} + 3 H \left( \rho_i + \frac{P_i}{\tc^2} \right) + 2 \rho_i H \frac{d \ln \tc}{d \ln a} = 0 \label{BI1mpApp} \,.
\end{align}
One can solve for this equation~\eqref{BI1mpApp} to obtain
\begin{align}
&\rho_i \tc^{2} = \rho_{i0} \tc_0^2 a^{-3 (1 + \omega_i)} \quad , \quad
\rho_{i} = \rho_{i0} a^{-3(1 + \omega_i) -\frac{b}{2}} \label{rhompApp} \;,
\end{align}
where we use the equation of state $\omega_i = P_i / (\rho_i \tc^2)$ for the $i$-component. The mass density of $i$-component redshifts slower (faster) than that of GR for the negative (positive) value of $b$. Alternatively, one might interpret Eq.~\eqref{rhompApp} that the rest mass evolves cosmologically as $a^{-b/2}$. By inserting Eqs.~\eqref{RijApp}, ~\eqref{RmpApp}, ~\eqref{TmunumpApp}, and ~\eqref{rhompApp} into Eq.~\eqref{tEFEmpApp}, one can obtain the components of EFE as
\begin{align}
&\frac{\dot{a}^2}{a^2} + k \frac{\tc^2}{a^2}  -\frac{ \Lambda \tc^2}{3} = \frac{8 \pi \tG}{3} \sum_i \rho_i \label{tG00mpApp} \,, \\
&\frac{\dot{a}^2}{a^2} + 2 \frac{\ddot{a}}{a} +  k \frac{\tc^2}{a^2} - \Lambda \tc^2 - 2 \frac{\dot{a}^2}{a^2} \frac{d \ln \tc}{d \ln a} = -\frac{8 \pi \tG}{\tc^2} \sum_{i} P_i  \label{tG11mpApp} \;.
\end{align}

One also obtains the expansion acceleration from Eqs.~\eqref{tG00mpApp} and \eqref{tG11mpApp} as
\begin{align}
&\frac{\ddot{a}}{a} = -\frac{4\pi \tG}{3} \sum_i \left( 1 + 3 \omega_i \right) \rho_i  + \frac{\Lambda \tc^2}{3} + \frac{\dot{a}^2}{a^2} \frac{d \ln \tc}{d \ln a}\label{t3G11mG00mpApp} \,.
\end{align}
One can rewrite the Hubble parameter $H$ and the acceleration $\ddot{a}/a$ by using Eqs.~\eqref{tkappaconstmpApp} and \eqref{rhompApp} to obtain
\begin{align}
H^2 &= \left[ \frac{8 \pi \tG^0}{3} \sum_{i} \rho_{0i} a^{-3(1+\omega_i)} + \frac{ \Lambda \tc_0^2}{3} - k \frac{\tc_0^2}{a^2} \right] a^{\frac{b}{2}} \equiv H^{(\GR)2} a^{\frac{b}{2}} \label{H2meApp} \;, \\
\frac{\ddot{a}}{a} &= \left[ -\frac{4\pi \tG_0}{3} \sum_i \left( 1 + 3 \omega_i \right) \rho_{0i} a^{-3(1+\omega_i)} + \frac{\Lambda \tc_0^2}{3} \right] a^{\frac{b}{2}}  + H^2 \frac{d \ln \tc}{d \ln a} = \left[ \left( \frac{\ddot{a}}{a} \right)^{(\GR)} + \frac{b}{4} H^{(\GR)2} \right] a^{\frac{b}{2}} \,.
\end{align}
These equations are background evolutions of the FLRW universe of the meVSL model. The expansion speed of the Universe in meVSL, $H$ has the extra factor $(1+z)^{-b/4}$ compared to that of GR, $H^{(\GR)}$. The present values of the Hubble parameter of GR and meVSL are the same because $a_0 =1$. However, the value of the Hubble parameter of meVSL in the past is $(1+z)^{-b/4}$-factor larger (smaller) than that of GR if $b$ is negative (positive). It is possible to solve the Hubble tension with this basic information \cite{Bernal:2016gxb,Knox:2019rjx}. In the meantime, there are two changes to the meVSL expansion's acceleration compared to GR. One is the extra additional factor comes from $ \frac{b}{4} H^{(\GR)2}$ compared to the acceleration of GR. It depends on the sign of $b$ whether this term gives the additional contribution or the subtraction. The second effect is the same as the one in the Hubble parameter. Compared to the acceleration of GR, there is an overall factor $(1+z)^{-b/4}$.

\subsection{Luminosity distance}
\label{subsec:dLapp} 

Eq.~\eqref{ds2App} provides the line element of the FLRW metric, which can be rewritten as 
\begin{align}
ds^2 = - \tc^2 dt^2 + a^{2}(t) \left[ \frac{dr^2}{1 - kr^2} + r^2 d \Omega^2 \right] = - \tc^2 dt^2 + a^2(t) \left[ d \chi^2 + f_{k}^2(\chi) d \Omega^2 \right] \,,
\end{align}
where $\chi = D_{\Tc}$ is the comoving distance given in Eq.~\eqref{Dcmp} and $f_{k}(\chi) = \sinh (\sqrt{-k} \chi )/\sqrt{-k} = D_{\TM}$ is the transverse comoving distance given in Eq.~\eqref{DMmp}. Unlike in other VSL models, both $D_{\Tc}$ and $D_{\TM}$ are the same in both GR and the meVSL as shown in Sec.~\ref{subsection:cosDistance}. To obtain the luminosity distance in meVSL, one needs to re-examine it from its definition. We define the observed luminosity as $L_0$ detected at the present epoch, which differs from the absolute luminosity $L_{s}$ of the source emitted at the redshift $z$. We can write the conservation of flux $\mathcal F$ from the source point to the observed point as
\begin{align}
\mathcal F = \frac{L_s}{4 \pi D_{\TL}^2(z)} = \frac{L_0}{4 \pi D_{\TM}^2(z_0)} \label{mathF} \,.
\end{align}
The absolute luminosity, $L_s \equiv \Delta E_1/\Delta t_1$, is defined by the energy ratio of the emitted light $\Delta E_1$ to the time interval of that emission $\Delta t_1$. And the observed luminosity also can be written as $L_0 = \Delta E_0 / \Delta t_0$. Thus, one can rewrite the luminosity distance by using Eq.~\eqref{mathF} as
\begin{align}
D_{\TL}^2 (z) = \frac{L_s}{L_0} D_{\TM}^2 (z_0) = \frac{\Delta E_1}{\Delta E_0} \frac{\Delta t_0}{\Delta t_1} D_{\TM}^2 (z_0) = \left(1 + z \right)^{2-\frac{b}{4}} D_{\TM}^2 (z_0) \,, \label{DL2app}
\end{align}
where we use
\begin{align}
\frac{\Delta E_1}{\Delta E_0} &= \frac{\tih_1 \tnu_1}{\tih_0 \tnu_0} = \frac{\tnu_1^{(\GR)}}{\tnu_0^{(\GR)}} = (1+z) \quad , \quad
\frac{\Delta t_0}{\Delta t_1} = \frac{\tnu_1}{\tnu_0} = \frac{\tnu_1^{(\GR)} (1+z)^{-b/4}}{\tnu_0^{(\GR)}} = (1+z)^{1-\frac{b}{4}}  \label{t0ot1} \,.
\end{align}
Thus, we obtain the relation between the luminosity distance and the transverse co-moving distance in meVSL (also for the angular diameter distance, $D_{\TA}$)
\begin{align}
D_{\TL}(z) = \left( 1 + z \right)^{1 - \frac{b}{8}} D_{\TM}(z) = \left( 1 + z \right)^{2 - \frac{b}{8}} D_{\TA}(z) \label{CDDRmeVSL} \,.
\end{align}
It is the meVSL model's cosmic distance duality relation (CDDR), and it is rewriteable as 
\begin{align}
\frac{(1+z)^2 D_{\TA}}{D_{\TL}} = (1+z)^{\frac{b}{8}} \label{z2DAoDL} \,.
\end{align}

\subsection{Perturbation}
\label{subsec:perturbationApp}

The spacetime geometry is embedded in the metric tensor, and the homogeneous, isotropic, and expanding universe is described by the background FLRW metric, $\bar{g}_{\mu\nu}$, given in Eq.~\eqref{ds2App}. This is written by the conformal time $\eta$ as
\begin{equation}\label{FRWmetetaApp}
\bar{g}_{\mu\nu} = a^{2}(\eta)(-d\eta^2 + \gamma_{ij}dx^idx^j) \;.
\end{equation}
This metric describes the background spacetime (\textit{i.e.}, manifold). The background spacetime is artificial, though, in that we have to account for deviations from homogeneity and isotropy, which are the differences between the background and the real physical spacetime
\begin{equation}\label{metricdecApp}
\delta g_{\mu\nu}(x^{\sigma}) = g_{\mu\nu}(x^{\sigma}) - \bar{g}_{\mu\nu}(x^{\sigma}) \,,
\end{equation}
at a certain spacetime coordinate $x^{\sigma}$. But $x^{\sigma}$ is an ill-posed statement because $g_{\mu\nu}$ and $\bar{g}_{\mu\nu}$ are tensors defined on different manifolds and $x^{\sigma}$ is a coordinate defined through different charts. The background spacetime is artificial, though, in that we have to account for deviations from homogeneity and isotropy, which are the differences between the background and the real physical spacetime. The difference between two tensors assessed at different places is an ill-defined operation, even if we embed the two manifolds in a single one. Therefore, to make Eq.~\eqref{metricdecApp} meaningful, we need a map that identifies points of the background manifold with those of the physical one. This map is called a gauge.

Gauge is arbitrary and allows us to use a fixed coordinate system in the background manifold for the points in the physical one. We shall still use conformal or cosmic time plus comoving spatial coordinates even when describing perturbative quantities. This property leads to the so-called gauge problem. We use the following relations
\begin{align}
	g^{\mu\rho}g_{\rho\nu} = \delta^\mu{}_\nu \quad, \quad \bar{g}^{\mu\rho}\bar{g}_{\rho\nu} = \delta^\mu{}_\nu \quad, \quad
	\delta g^{\mu\nu} = - \bar{g}^{\mu\rho}\delta g_{\rho\sigma}\bar{g}^{\nu\sigma} \label{metricdifApp} \,.
\end{align}
In particular, we consider the spatially flat metric using the conformal time
\begin{equation}
	g_{\mu\nu} = a^2(\eta_{\mu\nu} + h_{\mu\nu}) \,. \label{flatmeticApp}
\end{equation}
Hence, the perturbed contravariant metric is given by
\begin{equation}
	\delta g^{00} = -\frac{1}{a^2}h_{00}\;, \quad \delta g^{0i} = \frac{1}{a^2}\delta^{il}h_{0l} = \frac{1}{a^2}h_{0i}\;, \quad \delta g^{ij} = -\frac{1}{a^2}\delta^{il}h_{lm}\delta^{mj} = -\frac{1}{a^2}h_{ij}\;,
\end{equation}
where we have used the hypothesis that the indices of $h_{ij}$ are raised by $\delta^{ij}$ and the property $h^{ij} = h_{ij}$. $\delta g_{\mu\nu}$ is the perturbed covariant metric but $\delta g^{\mu\nu}$ is not the contravariant perturbed metric.

The affine connection can be decomposed as
\begin{align}\label{ChristdecApp}
\Gamma^\mu_{\nu\rho} &= \bar{\Gamma}^\mu_{\nu\rho} + \delta \Gamma^\mu_{\nu\rho}\;, \\
\delta \Gamma^\mu_{\nu\rho} &= \frac{1}{2}\bar{g}^{\mu\sigma}\left(\delta g_{\sigma\nu,\rho} + \delta g_{\sigma\rho,\nu} - \delta g_{\nu\rho,\sigma} - 2\delta g_{\sigma\alpha}\bar{\Gamma}^\alpha_{\nu\rho}\right) \,, \label{delChirstApp}
\end{align}
where the barred one is computed from the background metric only as given in Eq.~\eqref{GammaApp}. The background Christoffel symbols are given by
\begin{equation}
\bar\Gamma^0_{00} = \frac{1}{\tc} \frac{a'}{a}\;, \quad \bar\Gamma^0_{ij} = \frac{1}{\tc}  \frac{a'}{a}\delta_{ij}\;, \quad \bar\Gamma^i_{0j} = \frac{1}{\tc}  \frac{a'}{a}\delta^i{}_j\;, \label{barGammaApp}
\end{equation}
where the prime denotes derivation w.r.t the conformal time. The perturbed Christoffel symbols are also given by
\begin{align}
\delta \Gamma^0_{00} &= -\frac{1}{2 \tc}h_{00}'\;, \quad \delta \Gamma^0_{i0} =  -\frac{1}{2}\left(h_{00,i} - 2\frac{\mathcal H}{\tc} h_{0i}\right)\;,\\
\delta \Gamma^i_{00} &= \left( \frac{1}{\tc} h_{j0}' + \frac{\mathcal H}{\tc} h_{j0} - \frac{1}{2}h_{00,j} \right) \delta^{ij} \;,\\
\delta\Gamma^0_{ij} &= -\frac{1}{2}\left(h_{0i,j} + h_{0j,i} - \frac{1}{\tc} h'_{ij} - 2 \frac{\mathcal H}{\tc} h_{ij} - 2 \frac{\mathcal H}{\tc} \delta_{ij}h_{00}\right)\;,\\
\delta\Gamma^i_{j0} &= \frac{1}{2} \left( \frac{1}{\tc} h_{kj}' + h_{k0,j} - h_{0j,k} \right) \delta^{ki} \;,\\
\delta\Gamma^i_{jk} &= \frac{1}{2}\left(h_{lj,k} + h_{lk,j} - h_{jk,l} - 2 \frac{\mathcal H}{\tc} \delta_{jk}h_{l0}\right) \delta^{li} \;.
\end{align} \index{FLRW metric!Perturbed Christoffel symbols}
The indices might seem unbalanced, but we have used the fact that $h_{i0}$ and $h_{ij}$ are 3-tensors for the metric $\delta_{ij}$ and hence, for example, $h^{i}{}_0 = h_{i0}$.
With this result, we compute the components of both the background and the perturbed Ricci tensor
\begin{align}
R_{\mu\nu} &\equiv \bar{R}_{\mu\nu} + \delta R_{\mu\nu} = \bar\Gamma^\rho_{\mu\nu,\rho} - \bar\Gamma^\rho_{\mu\rho,\nu}  + \bar\Gamma^\rho_{\mu\nu}\bar\Gamma^\sigma_{\rho\sigma} - \bar\Gamma^\rho_{\mu\sigma}\bar\Gamma^\sigma_{\nu\rho}\nonumber \\
&+ \delta\Gamma^\rho_{\mu\nu,\rho} - \delta\Gamma^\rho_{\mu\rho,\nu} + \bar\Gamma^\rho_{\mu\nu}\delta\Gamma^\sigma_{\rho\sigma} + \delta\Gamma^\rho_{\mu\nu}\bar\Gamma^\sigma_{\rho\sigma} - \bar\Gamma^\rho_{\mu\sigma}\delta\Gamma^\sigma_{\nu\rho} - \delta\Gamma^\rho_{\mu\sigma}\bar\Gamma^\sigma_{\nu\rho}\;,
\end{align}\index{Ricci tensor!Perturbation}
by neglecting second-order terms in the connection. Both the background and the perturbed Ricci tensors are given by
\begin{align}
\bar{R}_{00} &= \frac{3}{\tc^2} \left(\mathcal H^2 - \frac{a''}{a} + \frac{\tc'}{\tc} \mathcal H \right) = \bar{R}_{00}^{(\GR)} + \frac{1}{\tc^2} \frac{d \ln \tc}{d \ln a} \mathcal H^2  \quad , \quad \text{where} \,\,\, \mathcal H \equiv \frac{a'}{a} \;, \label{barR00App} \\
\bar{R}_{ij} &= \frac{1}{\tc^2} \left(\mathcal H^2 + \frac{a''}{a} -  \frac{\tc'}{\tc} \mathcal H \right) \delta_{ij} = \bar{R}_{ij}^{(\GR)} - \frac{3}{\tc^2} \frac{d \ln \tc}{d \ln a} \mathcal H^2 \delta _{ij} \label{barRijApp} \;, \\
\delta R_{00} &= -\frac{1}{2}\nabla^2h_{00} - \frac{3}{2} \frac{\mathcal H}{\tc^2} h_{00}' + \frac{1}{\tc^2} \left[ \tc h_{k0,l}' + \tc \mathcal Hh_{k0,l} - \frac{1}{2}\left(h_{kl}'' + \mathcal Hh_{kl}'\right) + \frac{1}{2} \frac{\tc'}{\tc} h_{kl}' \right] \delta^{lk} \; \nonumber \\
&\equiv \delta R_{00}^{(GR)} + \frac{1}{2\tc^2} \frac{d \ln \tc}{d \ln a} \mathcal H h_{kl}' \delta^{kl} = \delta R_{00}^{(GR)} + \frac{1}{2\tc^2} \frac{d \ln \tc}{d \ln a} \mathcal H h' \;, \label{deltaR00App} \\
\delta R_{0i} &= -\frac{\mathcal H}{\tc} h_{00,i} - \frac{1}{2} \nabla^2h_{0i} + \frac{1}{\tc^2} \left(\frac{a''}{a} + \mathcal H^2\right)h_{0i} + \frac{1}{2 \tc}\left( \tc h_{k0,il} + h_{ki,l}'- h_{kl,i}' \right) \delta^{lk} - \frac{\mathcal H}{\tc^2} \frac{\tc'}{\tc} h_{0i} \; \nonumber \\
&\equiv \delta R_{0i}^{(GR)} - \frac{\mathcal H^2}{\tc^2} \frac{d \ln \tc}{d \ln a} h_{0i} \;, \label{deltaRi0App} \\
\delta R_{ij} &= \frac{1}{2}h_{00,ij} + \frac{1}{\tc^2} \left[ \frac{\mathcal H}{2}h_{00}'+ \left(\mathcal H^2 + \frac{a''}{a}\right)h_{00} \right] \delta_{ij} \nonumber\\
&-\frac{1}{2} \nabla^2 h_{ij} + \frac{1}{\tc^2} \left[ \frac{1}{2}h_{ij}'' + \mathcal Hh_{ij}' + \left(\mathcal H^2 + \frac{a''}{a}\right)h_{ij} \right] + \frac{1}{2}\left( h_{ki,lj} + h_{kj,li} - h_{kl,ij}\right)  \delta^{lk} \nonumber\\
&+ \left( \frac{\mathcal H}{2\tc^2}h_{kl}'- \frac{\mathcal H}{\tc} h_{k0,l} \right) \delta^{lk} \delta_{ij} - \frac{1}{2 \tc}(h_{0i,j}' + h_{0j,i}') - \frac{\mathcal H}{\tc} (h_{0i,j} + h_{0j,i}) \nonumber \\ &-\frac{1}{\tc^2} \frac{\tc'}{\tc} \left( \frac{1}{2} h_{ij}' + \mathcal H h_{ij} + \mathcal H \delta_{ij} h_{00} \right) \nonumber \\
&\equiv \delta R_{ij}^{(GR)} -\frac{1}{\tc^2} \frac{d \ln \tc}{d \ln a} \mathcal H \left( \frac{1}{2} h_{ij}' + \mathcal H h_{ij} + \mathcal H \delta_{ij} h_{00} \right) \;, \label{deltaRijApp}
\end{align}
where $h \equiv h_{i}^{i}$ denotes the trace of $h_{ij}$. Both the background Ricci scalar and the perturbed Ricci scalar are obtained by contracting the Ricci tensor
\begin{align}
R &= \bar{R} + \delta R = \bar{g}^{\mu\nu} \bar{R}_{\mu\nu} + \bar{g}^{\mu\nu}\delta R_{\mu\nu} + \delta g^{\mu\nu}\bar{R}_{\mu\nu} \;, \label{RApp} \\
\delta R &= -\frac{1}{a^2}\delta R_{00} + \frac{1}{a^2}\delta^{ij}\delta R_{ij} - a^2h_{\rho\sigma}\bar{g}^{\rho\mu}\bar{g}^{\sigma\nu}\bar{R}_{\mu\nu}\;. \label{deltaRApp}
\end{align}
One can write the Ricci scalar for FLRW as
\begin{align}
\bar{R} &= \frac{6}{\tc^2 a^2} \left( \frac{a''}{a} - \frac{\tc'}{\tc} \mathcal H \right)  \;, \label{barRcomApp} \\
a^2\delta R &= \nabla^2h_{00} + \frac{3}{\tc^2} \left( \mathcal H h_{00}' + 2\frac{a''}{a}h_{00} \right) - \frac{2}{\tc} \left( h_{k0,l}' + 3\mathcal Hh_{k0,l} \right) \delta^{lk} \nonumber \\
&+ \frac{1}{\tc^2} \left( h_{kl}'' + 3\mathcal Hh_{kl}' \right) \delta^{lk} - \left( \nabla^2h_{kl} - h_{kj,lj} \right) \delta^{lk} + \frac{1}{\tc^2} \frac{\tc'}{\tc} h_{kl}' \delta^{lk} \; \nonumber \\
&\equiv \nabla^2h_{00} + \frac{3}{\tc^2} \left( \mathcal H h_{00}' + 2\frac{a''}{a}h_{00} \right) - \frac{2}{\tc} \left( h_{k0,l}' + 3\mathcal Hh_{k0,l} \right) \delta^{lk} + a^2 \delta R^{(3)} + \frac{1}{\tc^2} \frac{\tc'}{\tc} h_{kl}' \delta^{lk} \nonumber \; \\
&\equiv a^2 \delta R^{(\GR)} + \frac{1}{\tc^2} \frac{d \ln \tc}{d \ln a} \mathcal H h'  \label{deltaRcomApp} \;,
\end{align}
where $a^2\delta R^{(3)}$ denotes the intrinsic spatial perturbed curvature scalar.

It is convenient to work with mixed indices when one solves for Einstein's equations
\begin{align}
G^{\mu}{}_{\nu} &= g^{\mu\rho}R_{\rho\nu} - \frac{1}{2}\delta^{\mu}{}_{\nu}R = \bar{g}^{\mu\rho}\bar{R}_{\rho\nu} - \frac{1}{2}\delta^{\mu}{}_{\nu}\bar{R} + \bar{g}^{\mu\rho}\delta R_{\rho\nu} + \delta g^{\mu\rho}\bar{R}_{\rho\nu} - \frac{1}{2}\delta^{\mu}{}_{\nu}\delta{R} \nonumber \\
&\equiv \bar{G}^{\mu}{}_{\nu} + \delta G^{\mu}{}_{\nu}  \label{GdecApp} \;,
\end{align}
where $\bar{G}^{\mu}{}_{\nu}$ is the background Einstein tensor whereas $\delta G^{\mu}{}_{\nu}$ is the linearly perturbed Einstein tensor, which depends on both $\bar{g}_{\mu\nu}$ and $h_{\mu\nu}$.
\begin{align}
2a^2\delta G^0{}_0 &= -6 \frac{\mathcal H^2}{\tc^2} h_{00} + 4 \frac{\mathcal H}{\tc} h_{k0,k} - 2 \frac{\mathcal H}{\tc^2} h_{kk}' + \nabla^2h_{kk} - h_{kl,kl} -\frac{1}{\tc^2} \frac{\tc'}{\tc} h_{kk}' + \frac{\tc'}{\tc} \mathcal H h_{00} \; \nonumber \\
&\equiv 2a^2\delta G^{0(\GR)}_0 - \frac{d \ln \tc}{d \ln a} \mathcal H \left( \frac{1}{\tc^2} h_{kk}' - \mathcal H h_{00} \right) \label{deltaG00App} \;, \\
2a^2\delta G^0{}_i &= 2\mathcal Hh_{00,i} + \nabla^2h_{0i} - h_{k0,ki} + h_{kk,i}' - h_{ki,k}' = 2a^2\delta G^{0(\GR)}_i \label{deltaG0igen} \;, \\
2a^2\delta G^i{}_j &= \left[ \frac{1}{\tc^2} \left( -4\frac{a''}{a}h_{00} - 2\mathcal Hh_{00}'  + 2\mathcal H^2 h_{00} - 2\mathcal H h_{kk}' - h_{kk}'' \right) - \nabla^2h_{00} \right.\nonumber\\
&\left. + \nabla^2h_{kk} - h_{kl,kl} + \frac{1}{\tc} \left( 2h_{k0,k}' + 4\mathcal Hh_{k0,k} \right) \right]\delta^i{}_j + h_{00,ij} - \nabla^2h_{ij} + h_{ki,kj} \nonumber\\ &+ h_{kj,ki} - h_{kk,ij} + \frac{1}{\tc^2} \left( h_{ij}'' + 2\mathcal Hh_{ij}' \right) -  \frac{1}{\tc} \left( (h_{0i,j}' + h_{0j,i}') + 2\mathcal H(h_{0i,j} + h_{0j,i}) \right) \nonumber \\ &- \frac{1}{\tc^2} \frac{\tc'}{\tc} \left( \frac{1}{2} h^{i'}_{j} + \frac{1}{2} h^{k'}_{k} \delta ^{i}_{j} - \mathcal H h_{00} \delta^{i}_j \right) \; \nonumber \\
&\equiv  2a^2\delta G^{i(\GR)}_j  - \frac{1}{\tc^2} \frac{d \ln \tc}{d \ln a} \mathcal H \left( \frac{1}{2} h^{i'}_{j} + \frac{1}{2} h^{k'}_{k} \delta ^{i}_{j} - \mathcal H h_{00} \delta^{i}_j \right) \;. \label{deltaGijgen}
\end{align}

Regarding the gravitational waves, $h_{00} = h_{0i} = 0$ can be utilized. Also for the tracelss-transpose spatial components, $h^{\text{TT}}_{ij}$, $h^{k(\text{TT}) }{}_{k} =0$ and $h^{(\text{TT})}_{i,i} = 0$. From this TT component, $\delta G^{i}{}_{j}$ becomes
\begin{align}
0 &= \frac{1}{\tc^2} h^{(\text{TT})''}_{ij} + 2 \frac{\mathcal H}{\tc^2} h^{(\text{TT})'}_{ij} - \frac{1}{\tc^2} \frac{\tc'}{\tc} h^{(\text{TT})'}_{ij} - \nabla^2 h_{ij}^{(\text{TT})}  \nonumber \\
&\equiv \frac{1}{\tc^2} \hTT^{''}_{ij} + 2 \frac{\mathcal H}{\tc^2} \left( 1+ \frac{b}{8} \right) \hTT^{'}_{ij} - \nabla^2 \hTT_{ij}  \label{hTTApp} \,,
\end{align}
where $\hTT$ denotes the tracelss-transpose component $h^{\text{TT}}$. One can also replace the derivatives w.r.t the conformal time with the derivatives w.r.t $\ln a$, and the above equation becomes
\begin{align}
&h' = \mathcal H \frac{d h}{d \ln a} \quad , \quad h'' = \mathcal H^2 \frac{d^2 h}{d \ln a^2} + \mathcal H' \frac{d h}{d \ln a} \label{hphppApp} \;, \\
&\frac{d^2 \hTT_{ij}}{d \ln a^2} + \left( 2 + \frac{b}{4} + \frac{\mathcal H'}{\mathcal H^2} \right) \frac{d \hTT_{ij}}{d \ln a} - \frac{\tc^2 \nabla^2}{\mathcal H^2} \hTT_{ij} = \frac{d^2 \hTT_{ij}}{d \ln a^2} + \left( 2 + \frac{b}{2} + \frac{\left( \frac{\ddot{a}}{a} \right)^{(\GR)}}{ H^{(\GR)2} }  \right) \frac{d \hTT_{ij}}{d \ln a} \nonumber \\ 
&- \frac{\tc_0^2 \nabla^2}{\mathcal H^{(\GR)2}} \hTT_{ij} = 0 \label{hTTAppdlna} \;,
\end{align}
where we use
\begin{align}
\frac{\mathcal H'}{\mathcal H^2} = \frac{\left( \frac{\ddot{a}}{a} \right)^{(\GR)}}{ H^{(\GR)2} }+ \frac{d \ln \tc}{d \ln a} =  \frac{\left( \frac{\ddot{a}}{a} \right)^{(\GR)}}{ H^{(\GR)2} }+ \frac{b}{4} \label{HpoH2App} \;.
\end{align}
Thus, the difference of GW between meVSL and GR only appears in the friction term as $b/2$.

\bibliographystyle{JHEP}


\end{document}